\documentclass[english,onecolumn,11pt]{IEEEtran}
\usepackage[T1]{fontenc}
\usepackage[latin9]{inputenc}
\usepackage{color}
\usepackage{babel}
\usepackage{array}
\usepackage{multirow}
\usepackage{amsmath}
\usepackage{amsthm}
\usepackage{amssymb}
\usepackage{wasysym}
\usepackage[unicode=true,pdfusetitle,
 bookmarks=true,bookmarksnumbered=true,bookmarksopen=false,
 breaklinks=true,pdfborder={0 0 1},backref=false,colorlinks=true]
 {hyperref}
\hypersetup{
 pdfborderstyle=}

\makeatletter

\providecommand{\tabularnewline}{\\}

\theoremstyle{plain}
\newtheorem{thm}{\protect\theoremname}
\theoremstyle{remark}
\newtheorem{rem}{\protect\remarkname}
\theoremstyle{plain}
\newtheorem{prop}{\protect\propositionname}
\theoremstyle{plain}
\newtheorem{cor}{\protect\corollaryname}
\theoremstyle{plain}
\newtheorem{lem}{\protect\lemmaname}
\theoremstyle{remark}
\newtheorem{claim}{\protect\claimname}
\theoremstyle{plain}
\newtheorem{fact}{\protect\factname}

\usepackage{babel}

\usepackage[mathscr]{eucal}
\usepackage{epsfig,epsf,psfrag}
\usepackage{amssymb,amsmath,amsthm,latexsym}
\usepackage{amsmath,graphicx,xcolor,url}
\usepackage[caption=false]{subfig} 
\usepackage{fixltx2e}%ordering of single and double column floats
\usepackage{array}%array and tabular environments
\usepackage{verbatim}
\usepackage{bm}
\usepackage{algorithmic}
\usepackage{algorithm}
\usepackage{verbatim}
\usepackage{textcomp}
\usepackage{mathrsfs,overpic}
\usepackage{epstopdf}
\usepackage{amsfonts}
\usepackage[numbers]{natbib}
\usepackage{mathtools}
\mathtoolsset{showonlyrefs}

\usepackage{tikz}
\usepackage{float}
\usepackage{tabularx}
\usepackage{multirow}
\usetikzlibrary{patterns}

\newcommand{\Unif}{\mathrm{Unif}}

\newcommand{\e}{ e}

\def\D{\mathsf{D}}
\def\DD{\D}

\def\Var{\operatorname{Var}}

\def\DSBS{\mathrm{DSBS}}
\def\BSC{\mathrm{BSC}}

\def\1{\mathbf{1}}

\def\d{{\text {\rm d}}}
\catcode`~=11 \def\UrlSpecials{\do\~{\kern -.15em\lower .7ex\hbox{~}\kern .04em}} \catcode`~=13 

\allowdisplaybreaks[1]

\newcommand{\calK}{\mathcal{K}}

\newcommand{\calM}{\mathcal{M}}

\newcommand{\calP}{\mathcal{P}}

\newcommand{\calX}{\mathcal{X}}
\newcommand{\calY}{\mathcal{Y}}

\newcommand{\rmM}{\mathrm{M}}

\newcommand{\bbE}{\mathbb{E}}

\newcommand{\bbN}{\mathbb{N}}

\newcommand{\bbR}{\mathbb{R}}

\DeclareMathAlphabet{\mathbsf}{OT1}{cmss}{bx}{n}
\DeclareMathAlphabet{\mathssf}{OT1}{cmss}{m}{sl}% slanted sans serif

\newcommand{\rvD}{\mathsf{D}}

\DeclareSymbolFont{bsfletters}{OT1}{cmss}{bx}{n}  
\DeclareSymbolFont{ssfletters}{OT1}{cmss}{m}{n}
\DeclareMathSymbol{\bsfGamma}{0}{bsfletters}{'000}
\DeclareMathSymbol{\ssfGamma}{0}{ssfletters}{'000}
\DeclareMathSymbol{\bsfDelta}{0}{bsfletters}{'001}
\DeclareMathSymbol{\ssfDelta}{0}{ssfletters}{'001}
\DeclareMathSymbol{\bsfTheta}{0}{bsfletters}{'002}
\DeclareMathSymbol{\ssfTheta}{0}{ssfletters}{'002}
\DeclareMathSymbol{\bsfLambda}{0}{bsfletters}{'003}
\DeclareMathSymbol{\ssfLambda}{0}{ssfletters}{'003}
\DeclareMathSymbol{\bsfXi}{0}{bsfletters}{'004}
\DeclareMathSymbol{\ssfXi}{0}{ssfletters}{'004}
\DeclareMathSymbol{\bsfPi}{0}{bsfletters}{'005}
\DeclareMathSymbol{\ssfPi}{0}{ssfletters}{'005}
\DeclareMathSymbol{\bsfSigma}{0}{bsfletters}{'006}
\DeclareMathSymbol{\ssfSigma}{0}{ssfletters}{'006}
\DeclareMathSymbol{\bsfUpsilon}{0}{bsfletters}{'007}
\DeclareMathSymbol{\ssfUpsilon}{0}{ssfletters}{'007}
\DeclareMathSymbol{\bsfPhi}{0}{bsfletters}{'010}
\DeclareMathSymbol{\ssfPhi}{0}{ssfletters}{'010}
\DeclareMathSymbol{\bsfPsi}{0}{bsfletters}{'011}
\DeclareMathSymbol{\ssfPsi}{0}{ssfletters}{'011}
\DeclareMathSymbol{\bsfOmega}{0}{bsfletters}{'012}
\DeclareMathSymbol{\ssfOmega}{0}{ssfletters}{'012}

\newcommand{\Bern}{\mathrm{Bern}}

\newcommand{\dotleq}{\stackrel{.}{\leq}}

\newcommand{\dotgeq}{\stackrel{.}{\geq}}

\newcommand{\qednew}{\nobreak \ifvmode \relax \else
      \ifdim\lastskip<1.5em \hskip-\lastskip
      \hskip1.5em plus0em minus0.5em \fi \nobreak
      \vrule height0.75em width0.5em depth0.25em\fi}

\usepackage{bm,bbm}

\newcommand{\bone}{\mathbbm{1}}

\allowdisplaybreaks

\providecommand{\corollaryname}{Corollary}

\providecommand{\factname}{Fact}
\providecommand{\lemmaname}{Lemma}
\providecommand{\propositionname}{Proposition}
\providecommand{\remarkname}{Remark}
\providecommand{\theoremname}{Theorem}

\providecommand{\claimname}{Claim}

\makeatother

\providecommand{\claimname}{Claim}
\providecommand{\corollaryname}{Corollary}
\providecommand{\factname}{Fact}
\providecommand{\lemmaname}{Lemma}
\providecommand{\propositionname}{Proposition}
\providecommand{\remarkname}{Remark}
\providecommand{\theoremname}{Theorem}

\begin{document}
\title{Rényi Resolvability, Noise Stability, and Anti-contractivity}
\author{Lei Yu\thanks{L. Yu is with the School of Statistics and Data Science, LPMC, KLMDASR,
and LEBPS, Nankai University, Tianjin 300071, China (e-mail: leiyu@nankai.edu.cn).
This work was supported by the National Key Research and Development
Program of China under grant 2023YFA1009604, the NSFC under grant
62101286, and the Fundamental Research Funds for the Central Universities
of China (Nankai University) under grant 054-63253112.}}
\maketitle
\begin{abstract}
This paper investigates three closely related topics---Rényi resolvability,
noise stability, and anti-contractivity. The Rényi resolvability problem
refers to approximating a target output distribution of a given channel
in the Rényi divergence when the input is set to a function of a given
uniform random variable. This problem for the Rényi parameter in $(0,2]\cup\{\infty\}$
was first studied by the present author and Tan in 2019. In the present
paper, we provide a complete solution to this problem for the Rényi
parameter in the entire range $\mathbb{R}\cup\{\pm\infty\}$. We then
connect the Rényi resolvability problem to the noise stability problem,
by observing that maximizing or minimizing the $q$-stability of a
set is equivalent to  a variant of the Rényi resolvability problem.
By such a connection, we provide sharp dimension-free bounds on the
$q$-stability. We lastly relate the noise stability problem to the
anti-contractivity of a Markov operator (i.e., conditional expectation
operator), where the terminology ``anti-contractivity'' introduced
by us refers to as the opposite property of the well-known contractivity/hyercontractivity.
We derive sharp dimension-free anti-contractivity inequalities. All
of the results in this paper are evaluated for binary distributions.
Our proofs in this paper are mainly based on the method of types,
especially strengthened versions of packing-covering lemmas. 
\end{abstract}


\tableofcontents{}

\section{\label{sec:Introduction}Introduction }

\subsection{Rényi Resolvability }

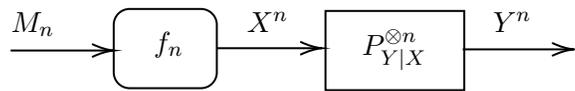
\begin{figure}[t]
\centering \tikzset{every picture/.style={line width=0.75pt}}
%set default line width to 0.75pt        
\begin{tikzpicture}[x=0.75pt,y=0.75pt,yscale=-1,xscale=1] %uncomment if require: \path (0,300); %set diagram left start at 0, and has height of 300
%Straight Lines [id:da5062303514766782] 
\draw    (107.2,78.98) -- (157.2,78.98) ;
\draw [shift={(159.2,78.98)}, rotate = 180] [color={rgb, 255:red, 0; green, 0; blue, 0 }  ][line width=0.75]    (10.93,-3.29) .. controls (6.95,-1.4) and (3.31,-0.3) .. (0,0) .. controls (3.31,0.3) and (6.95,1.4) .. (10.93,3.29)   ; %Straight Lines [id:da34105045268552203] 
\draw    (211.2,77.82) -- (262.2,77.82) ;
\draw [shift={(264.2,77.82)}, rotate = 180] [color={rgb, 255:red, 0; green, 0; blue, 0 }  ][line width=0.75]    (10.93,-3.29) .. controls (6.95,-1.4) and (3.31,-0.3) .. (0,0) .. controls (3.31,0.3) and (6.95,1.4) .. (10.93,3.29)   ; %Straight Lines [id:da12512838415449656] 
\draw    (335.2,78.4) -- (389.2,78.4) ;
\draw [shift={(391.2,78.4)}, rotate = 180] [color={rgb, 255:red, 0; green, 0; blue, 0 }  ][line width=0.75]    (10.93,-3.29) .. controls (6.95,-1.4) and (3.31,-0.3) .. (0,0) .. controls (3.31,0.3) and (6.95,1.4) .. (10.93,3.29)   ; %Shape: Rectangle [id:dp4228819541283999] 
\draw   (266,59) -- (336,59) -- (336,99) -- (266,99) -- cycle ; %Rounded Rect [id:dp7070214805095945] 
\draw   (159.2,66) .. controls (159.2,61.58) and (162.78,58) .. (167.2,58) -- (203.2,58) .. controls (207.62,58) and (211.2,61.58) .. (211.2,66) -- (211.2,90) .. controls (211.2,94.42) and (207.62,98) .. (203.2,98) -- (167.2,98) .. controls (162.78,98) and (159.2,94.42) .. (159.2,90) -- cycle ;
% Text Node
\draw (106,58) node [anchor=north west][inner sep=0.75pt]  [font=\normalsize]  {$M_{n}$}; % Text Node
\draw (283,67) node [anchor=north west][inner sep=0.75pt]    {$P_{Y|X}^{\otimes n}$}; % Text Node
\draw (225,58) node [anchor=north west][inner sep=0.75pt]  [font=\normalsize]  {$X^{n}$}; % Text Node
\draw (349,58) node [anchor=north west][inner sep=0.75pt]  [font=\normalsize]  {$Y^{n}$}; % Text Node
\draw (178,68) node [anchor=north west][inner sep=0.75pt]    {$f_{n}$};
\end{tikzpicture}\caption{Channel resolvability problem.}
\label{fig:Resolvability}
\end{figure}

The {\em channel resolvability problem} concerns how much information
required to simulate a random process via a given channel, as illustrated
in Fig.~\ref{fig:Resolvability}. Let $M_{n}$ be a random variable
uniformly distributed over\footnote{For simplicity, we assume that $\e^{nR}$ and similar expressions
(such as $\e^{R}$) are integers.} $[\e^{nR}]:=\{1,\ldots,\e^{nR}\}$, where $R$ is a positive number
known as the {\em rate}. Let $P_{Y|X}$ be a conditional distribution
(also known as a channel) and $P_{Y|X}^{\otimes n}$ its product version.
For a deterministic function $f_{n}:[\e^{nR}]\to\mathcal{X}^{n}$,
the output distribution of the memoryless channel $P_{Y|X}$ is 
\begin{align}
Q_{Y^{n}}(y^{n}) & :=\mathbb{E}[P_{Y|X}^{\otimes n}(y^{n}|f_{n}(M_{n}))]\label{eq:-113-1}\\
 & =\e^{-nR}\sum_{m\in[\e^{nR}]}P_{Y|X}^{\otimes n}(y^{n}|f_{n}(m)).
\end{align}
Here $f_{n}$ is known as a {\em resolvability code}. Given the
rate $R$, the channel $P_{Y|X}$, and a target distribution $P_{Y}$,
we wish to minimize the discrepancy between the true distribution
$Q_{Y^{n}}$ and the target product distribution $P_{Y}^{\otimes n}$
over all codes $f_{n}$. 

In this paper, we employ the Rényi divergence, a generalization of
the relative entropy, to measure the discrepancy between $Q_{Y^{n}}$
and $P_{Y}^{\otimes n}$. For two distributions $P,Q$ on the same
space $\calX$, the {\em Rényi divergence of order $q\in(-\infty,0)\cup(0,1)\cup(1,\infty)$}
is defined as 
\begin{align}
D_{q}(Q\|P) & :=\frac{1}{q-1}\log\sum_{x\in\calX}Q(x)^{q}P(x)^{1-q}.\label{eq:Renyidivergence}
\end{align}
Throughout the paper, $\log$ is to the natural base. The Rényi divergence
of orders $0,1,\pm\infty$ is defined by continuous extension. In
particular, the Rényi divergence of order $1$ reduces to the {\em
relative entropy} 
\begin{align}
D_{1}(Q\|P)=D(Q\|P) & :=\sum_{x\in\calX}Q(x)\log\frac{Q(x)}{P(x)}.
\end{align}
Based on the definitions above, we aim at minimizing $D_{q}(Q_{Y^{n}}\|P_{Y}^{\otimes n}),q\in[0,\infty]$
and maximizing $D_{q}(Q_{Y^{n}}\|P_{Y}^{\otimes n}),q\in[-\infty,0)$
over all functions $f_{n}:[\e^{nR}]\to\mathcal{X}^{n}$.

In the literature, the resolvability problems with the total variation
(TV) distance and the relative entropy (Kullback-Leibler divergence)
normalized by the dimension $n$ were studied by Han and Verdú \cite{Han93}.
The resolvability problem with the unnormalized relative entropy was
studied by Hayashi \cite{Hayashi06,Hayashi11}. In these works, it
was shown that the minimum rates of randomness needed for simulating
a channel output under the TV distance and normalized/unnormalized
relative entropy are the same, and are all equal to the minimum mutual
information over all input distributions that induce the target output
distribution. The channel resolvability problem was generalized to
the Rényi divergence setting by the present author and Tan, and the
resulting problem is coined by us the Rényi resolvability problem
\cite{yu2019renyi}. The achievability part for the Rényi resolvability
problem with the Rényi parameter in $(0,2]\cup\{\infty\}$ and the
converse part with the Rényi parameter in $(0,\infty]$ were provided
in \cite{yu2019renyi}. That is, in \cite{yu2019renyi} we completely
solved the Rényi resolvability problem when the Rényi parameter taking
values in $(0,2]\cup\{\infty\}$. When specialized to the binary setting,
Samorodnitsky \cite{samorodnitsky2022some} complemented the achievability
results for the Rényi parameter in $(2,\infty)\cap\mathbb{N}$, and
Pathegama and Barg \cite{pathegama2023smoothing} independently complemented
the case with the Rényi parameter in $(2,\infty)$. Moreover, the
asymptotics of Rényi resolvability for non-product channels was also
considered by Pathegama and Barg \cite{pathegama2023smoothing}. (Note
that the existing one-shot bounds, e.g., the ones in \cite{Hayashi06,Hayashi11,yu2019renyi},
also imply certain bounds for non-product channels.) In this work,
we aim at providing a complete solution to this problem for arbitrary
product target distributions and product channels defined on arbitrary
finite alphabets with the Rényi parameter taking values in the all
range $\mathbb{R}\cup\{\pm\infty\}$.

Liu, Cuff, and Verdú \cite{Liu} extended the theory of resolvability
by using $E_{\gamma}$ metric with $\gamma\geq1$ to measure the level
of approximation. The $E_{\gamma}$ metric reduces to the TV distance
when $\gamma=1$, but it is weaker than the TV distance when $\gamma>1$.
Furthermore, simulating a random variable from another given random
variable under Rényi divergence measures of all orders in $\mathbb{R}\cup\{\pm\infty\}$
was studied by the present author and Tan in \cite{yu2019simulation}.
This random variable simulation problem reduces to the resolvability
problem with the identity channel, known as the source resolvability
problem, when the seed random variable is uniform. The exact channel
resolvability problem was studied by the present authors in \cite{yu2018asymptotic},
in which the output distribution is required to be exactly equal to
the target distribution, and meanwhile, the rate of the input process
is instead measured by the normalized entropy.

Lastly, it was worth noting that while the term ``channel resolvability''
was coined by Han and Verdú in \cite{Han93}, the problem of approximating
a given product measure was first investigated by Wyner \cite{WynerCI}
in the study of common information. Wyner's framework of common information
was generalized to the Rényi divergence by the present author and
Tan in \cite{YuTan2018,yu2020corrections}. The achievability parts
of the channel resolvability and common information problems usually
rely on so-called soft-covering lemmas, for which the resolvability
code chooses a codeword from a random codebook according to the uniform
random variable. The convergence exponents in the soft-covering lemma
under the TV distance, the relative entropy, and the Rényi divergence
of orders in $(0,2]$ were respectively characterized in \cite{yagli2019exact},
\cite{Parizi}, and \cite{yu2019renyi}.

\subsection{Noise Stability }

The Rényi resolvability is closely related to the $q$-stability problem.
Consider a joint distribution $P_{XY}$. The $q$-stability problem
concerns the following question: For a measurable event $A$, if the
probability $P_{X}^{\otimes n}(A)$ is given, then how large and how
small could the $q$-norm of the noisy version, $\Vert P_{X|Y}^{\otimes n}(A|\cdot)\Vert_{q}$,
be? Here, $\Vert g\Vert_{q}:=\mathbb{E}_{P_{Y}^{\otimes n}}[g(Y^{n})^{q}]^{1/q}$
for $q\neq0$, and $\Vert g\Vert_{0}:=e^{\mathbb{E}_{P_{Y}^{\otimes n}}[\log g(Y^{n})]}$
for $q=0$.

The connection between Rényi resolvability and $q$-stability is as
follows. Denote $Q_{X^{n}}:=P_{X}^{\otimes n}(\cdot|A)$ as the conditional
distribution of $P_{X}^{\otimes n}$ given $A$, and denote $Q_{Y^{n}}$
as the output of the channel $P_{Y|X}^{\otimes n}$ when the input
is $Q_{X^{n}}$. Then, 
\begin{align}
\log\Vert P_{X|Y}^{\otimes n}(A|\cdot)\Vert_{q} & =\log P_{X}^{\otimes n}(A)+\frac{1}{q'}D_{q}(Q_{Y^{n}}\|P_{Y}^{\otimes n}),\label{eq:-36}
\end{align}
where $q'=\frac{q}{q-1}$ is the Hölder conjugate of $q$. Hence,
the $q$-stability problem is equivalent to a variant of Rényi resolvability
problem, in which we minimize or maximize the Rényi divergence of
the output distributions under the condition that the input distribution
is set to $P_{X}^{\otimes n}(\cdot|A)$ for some set $A$ of probability
$P_{X}^{\otimes n}(A)=e^{-n\alpha}$.

In this paper, we focus on the part of minimizing $\Vert P_{X|Y}^{\otimes n}(A|\cdot)\Vert_{q}$
for $q>1$ or maximizing $\Vert P_{X|Y}^{\otimes n}(A|\cdot)\Vert_{q}$
for $q<1$, and derive sharp dimension-free bounds for this part.
Obviously, if $P_{X}$ is set to the uniform distribution on $\mathcal{X}$
in this case, then the $q$-stability problem is exactly identical
to the original Rényi resolvability problem. As for the other part,
i.e., maximizing $\Vert P_{X|Y}^{\otimes n}(A|\cdot)\Vert_{q}$ for
$q>1$ or minimizing $\Vert P_{X|Y}^{\otimes n}(A|\cdot)\Vert_{q}$
for $q<1$, it was shown that this part is in fact closely related
to the Brascamp--Lieb (BL) inequalities, and by this relation, sharp
dimension-free bounds were derived; see, e.g., \cite{yu2021strong}.

The $q$-stability problem for $q=2$ is known as the noise stability
problem. The study of the noise stability problem dates back to Gács
and Körner's and Witsenhausen's seminal papers \cite{gacs1973common,witsenhausen1975sequences},
as a key ingredient in investigating the Gács--Körner--Witsenhausen
common information, another kind of common information different from
Wyner's one. Most of the existing works on this topic focus on doubly
symmetric binary sources (DSBSes). For the DSBS, by utilizing the
tensorization property of maximal correlation, Witsenhausen proved
sharp bounds on $P_{XY}^{\otimes n}(A\times B)$ for the case $P_{X}^{\otimes n}(A)=P_{Y}^{\otimes n}(B)=\frac{1}{2}$,
where the upper and lower bounds are respectively attained by symmetric
$(n-1)$-subcubes (e.g., $A=B=\{x^{n}:x_{1}=1\}$) and anti-symmetric
$(n-1)$-subcubes (e.g., $A=-B=\{x^{n}:x_{1}=1\}$). Combining Fourier
analysis with a coding-theoretic result, the first author and Tan
\cite{yu2021non} derived the sharp upper bound for the case $P_{X}^{\otimes n}(A)=P_{Y}^{\otimes n}(B)=\frac{1}{4}$,
where the upper bound is attained by symmetric $(n-2)$-subcubes (e.g.,
$A=B=\{x^{n}:x_{1}=x_{2}=1\}$). Kahn, Kalai, and Linial \cite{kahn1988influence}
first applied the single-function version of (forward) hypercontractivity
inequalities to obtain bounds for the noise stability problem, by
replacing nonnegative functions in the hypercontractivity inequalities
with Boolean functions. Mossel and O'Donnell \cite{mossel2006non,ODonnell14analysisof},
as well as Kamath and Anantharam \cite{kamath2016non}, applied the
two-function version of hypercontractivity inequalities to obtain
bounds in a similar way. Furthermore, the noise stability with $P_{X}^{\otimes n}(A_{n}),P_{Y}^{\otimes n}(B_{n})$
vanishing exponentially fast was investigated in \cite{ordentlich2020note,kirshner2021moment,yu2021strong}.

The $q$-stability problem for general $q>1$ was completely solved
for Gaussian measures by Borell \cite{borell1985geometric} in 1985.
In fact, he proved that a more general quantity, known as $\Phi$-stability,
is maximized by the indicators of half-spaces over all measurable
Boolean functions $f:\mathbb{R}^{n}\to\{0,1\}$ of the same measure;
see a more explicit statement of this result in \cite{kindler2015remarks}.
Such a result is known as Borell's isoperimetric theorem. As a special
$\Phi$-stability, the $q$-stability is also maximized by the same
functions. For doubly symmetric binary distributions, the $q$-stability
problem was investigated in \cite{mossel2005coin,courtade2014boolean,li2020boolean,yu2023phi},
but still remains open. Interestingly, the $q$-stability problem
for $1<q<2$ was shown by Li and Médard \cite{li2020boolean} to be
equivalent to the ``most informative Boolean function'' conjecture
posed by Courtade and Kumar \cite{courtade2014boolean}, one of the
most important conjectures in information theory at the time of the
writing of this paper. The first dimension-independent bound was derived
in \cite{samorodnitsky2016entropy} and the currently best known bound
was provided in \cite{yu2023phi}. We refer readers to the monograph
\cite{yu2022common} for a comprehensive introduction to this topic. 

\subsection{Anti-contractivity}

It will be shown in the next section that the part of the $q$-stability
problem considered in this paper is equivalent to a new kind of inequalities,
called anti-contractivity inequalities.

The {\em Markov operator} or {\em conditional expectation operator}
induced by a (regular) conditional distribution $P_{X|Y}$ is the
operator that maps a function $f:\mathcal{X}\to\mathbb{R}$ to the
function 
\begin{equation}
y\in\calY\mapsto P_{X|Y=y}(f):=\bbE\big[f(X)\,\big|\,Y=y\big].\label{eqn:cond_expectation_fi}
\end{equation}
Then, given a joint distribution $P_{XY}$, the (forward) contractivity-type
inequality (or the single function version of BL inequality) read
for $q\in[1,\infty]$, 
\begin{align}
\Vert P_{X|Y}(f)\Vert_{q} & \le e^{-\underline{C}}\Vert f\Vert_{p},\forall f\ge0,\label{eq:FIFBL-2}
\end{align}
and its reverse version read for $q\in[-\infty,1]$, 
\begin{align}
\Vert P_{X|Y}(f)\Vert_{q} & \geq e^{-\overline{C}}\Vert f\Vert_{p},\forall f\ge0,\label{eq:FIRBL-3}
\end{align}
where $p\in[-\infty,\infty]$ for both the two inequalities, ``$\forall f\ge0$''
denotes ``for all functions $f:\mathcal{X}\to[0,\infty)$'', and
$\underline{C}$ and $\overline{C}$ are two constants independent
of $f$, but possibly dependent on $(P_{XY},p,q)$.

Inspired by forward and reverse contractivity-type inequalities, we
are interested in investigating the forward and reverse anti-contractivity
inequalities: for $q\in[1,\infty]$, 
\begin{align}
\Vert P_{X|Y}(f)\Vert_{q} & \geq e^{-\overline{C}}\Vert f\Vert_{p},\forall f\ge0,\label{eq:FAC}
\end{align}
and for $q\in[-\infty,1]$, 
\begin{align}
\Vert P_{X|Y}(f)\Vert_{q} & \le e^{-\underline{C}}\Vert f\Vert_{p},\forall f\ge0,\label{eq:RAC}
\end{align}
where $p\in[-\infty,\infty]$. Samorodnitsky first investigated the
inequality in \eqref{eq:FAC} for binary distributions and $p=q\ge1$
in \cite{samorodnitsky2022some}. 

The intuitive difference between contractivity-type and anti-contractivity
inequalities is as follows. It is well known that the Markov operator
plays a smoothing role. The contractivity-type inequality concerns
how unsmooth a function could be after the Markov operator acting
on, while the anti-contractivity inequality concerns how smooth a
function could be after the Markov operator acting on. So, the extremizers
in contractivity-type inequalities are mainly concentrated on sets
with good isoperimetric properties (i.e., in which points are as close
to each other as possible), while the extremizers in anti-contractivity
inequalities are mainly concentrated on sets with good packing/covering
properties (i.e., in which points are as far away from each other
as possible).

If we restrict $f$ to a Boolean-valued function, i.e., $f=\bone_{A}$,
then the contractivity-type and anti-contractivity inequalities reduce
to the $q$-stability problem. So, any contractivity-type and anti-contractivity
inequalities can yield bounds on the $q$-stability problem. In fact,
the implication in the opposite direction is also true.

Hypercontractivity inequalities were investigated in \cite{bonami1968ensembles,kiener1969uber,schreiber1969fermeture,bonami1970etude,beckner1975inequalities,gross1975logarithmic,ahlswede1976spreading,borell1982positivity,bakry1994hypercontractivite,mossel2013reverse}
among others. Information-theoretic characterizations of the hypercontractivity
inequalities can be traced back to Ahlswede and Gács's seminal work
\cite{ahlswede1976spreading}, where a related quantity, known as
the hypercontractivity constant, was expressed in terms of relative
entropies. The information-theoretic characterization for the forward
hypercontractivity (in fact, a general version, BL inequalities) on
Euclidean spaces was given in \cite{carlen2009subadditivity}; this
was independently discovered later \cite{nair2014equivalent} in the
case of finite alphabets. An information-theoretic characterization
of the reverse BL inequalities for finite alphabets was provided in
\cite{kamath2015reverse,beigi2016equivalent,liu2016brascamp}. By
using Fenchel duality, the extension of the characterization to Polish
spaces was studied in \cite{liu2018forward}. 

\subsection{\label{subsec:Main-Contributions}Main Contributions}

Our main contributions are as follows: 
\begin{enumerate}
\item We provide a complete solution to the Rényi resolvability problem
on finite alphabets for the Rényi parameter in the all range $\mathbb{R}\cup\{\pm\infty\}$,
complementing the work \cite{yu2019renyi}. Specifically, we first
provide one-shot (i.e., finite-length) bounds and asymptotic expressions
for the Rényi divergence between the simulated and target output distributions.
We then characterize the Rényi resolvability rates, which is defined
as the minimum rate needed to guarantee that the (normalized or unnormalized)
Rényi divergence vanishes asymptotically. We lastly prove that the
optimal Rényi divergence between the simulated and target output distributions
vanishes (at least) exponentially fast as long as the code rate is
strictly larger than the Rényi resolvability rate. We also characterize
the optimal (ensemble tight) exponential decay rate for the ensemble
of i.i.d.\ random codes for positive Rényi orders, i.e., the convergence
exponent for the soft-covering problem, which complements the works
\cite{yagli2019exact}, \cite{Parizi}, and \cite{yu2019renyi}.
\item We then connect the Rényi resolvability problem to the noise stability
problem. In fact, the $q$-stability of a set can be expressed in
terms of the Rényi divergence between the true output distribution
and the target distribution in a variant of the Rényi resolvability
problem. Using such a connection and applying our results on Rényi
resolvability, we derive sharp dimension-free bounds on the $q$-stability
for a certain range of $q$, which yields the optimal exponential
decay rates of the maximal and minimal $q$-stability when the measure
of the set vanishes exponentially. This part of results complements
the work \cite{yu2021strong} which maximizes or minimizes the $q$-stability
for $q$ in the complement regions. 
\item We lastly relate the noise stability problem to anti-contractivity
of a Markov operator, where anti-contractivity refers to as an opposite
property of the well-known contractivity/hyercontractivity. We derive
sharp dimension-free anti-contractivity inequalities. Particularly,
in the binary setting with crossover probability $\epsilon\in(0,1/2)$
(i.e., for the Bonami--Beckner operator $P_{X|Y=y}(f)=(1-\epsilon)f(y)+\epsilon f(1-y)$),
it holds that for $1\le p\le q$, 
\begin{align}
\Vert P_{X|Y}^{\otimes n}(f)\Vert_{q} & \geq e^{-nH_{q}(\epsilon)/p'}\Vert f\Vert_{p},\forall f\ge0,\label{eq:FAHC}
\end{align}
and for $0\le p\le\!1$ and $q\le p$, 
\begin{align}
\Vert P_{X|Y}^{\otimes n}(f)\Vert_{q} & \le e^{-nH_{p}(\epsilon)/p'}\Vert f\Vert_{p},\forall f\ge0,\label{eq:RAHC}
\end{align}
where $H_{q}(\epsilon):=\frac{1}{1-q}\log(\epsilon^{q}+(1-\epsilon)^{q})$
is the binary Rényi entropy function. The exponents $H_{q}(\epsilon)/p'$
and $H_{p}(\epsilon)/p'$ of the factors in \eqref{eq:FAHC} and \eqref{eq:RAHC}
cannot be further improved. This is because, the exponent in \eqref{eq:FAHC}
is with high probability asymptotically attained by the indicator
of a random subset of $\{0,1\}^{n}$ of size $2^{n(1-H_{q}(\epsilon))}$,
and the exponent in \eqref{eq:RAHC} is with high probability asymptotically
attained by the indicator of a random subset of $\{0,1\}^{n}$ of
size $2^{n(1-H(\hat{\epsilon}))}$ with $\hat{\epsilon}=\frac{\epsilon^{p}}{\epsilon^{p}+(1-\epsilon)^{p}}$.
 The special case of \eqref{eq:FAHC} with $p=q$ was first shown
by Samorodnitsky \cite{samorodnitsky2022some}. The inequalities in
\eqref{eq:FAHC} and \eqref{eq:RAHC} are the opposite part of the
classic hyercontractivity inequalities (or more generally, the single
function version of BL inequalities) \cite{bonami1968ensembles,kiener1969uber,schreiber1969fermeture,bonami1970etude,beckner1975inequalities,gross1975logarithmic,borell1982positivity}:
for $q\ge1$ and $p\ge1+(1-2\epsilon)^{2}(q-1)$, 
\begin{align}
\Vert P_{X|Y}^{\otimes n}(f)\Vert_{q} & \le\Vert f\Vert_{p},\forall f\ge0,\label{eq:FIRBL-1}
\end{align}
and for $q\le1$ and $p\le1+(1-2\epsilon)^{2}(q-1)$, 
\begin{align}
\Vert P_{X|Y}^{\otimes n}(f)\Vert_{q} & \ge\Vert f\Vert_{p},\forall f\ge0.\label{eq:FIFBL-1}
\end{align}
\item The main ingredients in our proofs are strengthened versions of packing-covering
lemmas, which estimate how many ``balls'' each point is covered
by for random codes (more precisely, constant composition codes or
typical codes). We expect these lemmas to find more applications. 
\end{enumerate}

\subsection{Structure of the paper}

We introduce our results on Rényi resolvability, $q$-stability, and
anti-contractivity respectively in Sections \ref{sec:Resolvability}-\ref{subsec:Single-Function-Version}.
We conclude this paper in Section \ref{sec:concl}. The proofs of
our results are provided in appendices. In particular, we introduce
strengthened versions of packing-covering lemma in Section \ref{sec:usefullemmas}
that play a key role in our proofs.

\subsection{Notation }

\label{sec:notation}

In this paper, we use $P_{X}(x)$ to denote the probability distribution
of a random variable $X$, which is also shortly denoted as $P(x)$
(when the random variable $X$ is clear from the context). We also
use $Q_{X}$, $R_{X}$, and $S_{X}$ to denote another three arbitrary
probability distributions on the alphabet $\mathcal{X}$. All alphabets
considered in the sequel are finite, unless otherwise expressly stated.
The set of probability distributions on $\mathcal{X}$ is denoted
as $\mathcal{P}(\mathcal{X})$, and the set of conditional probability
distributions on $\mathcal{Y}$ given a variable in $\mathcal{X}$
is denoted as $\mathcal{P}(\mathcal{Y}|\mathcal{X}):=\{P_{Y|X}:P_{Y|X}(\cdot|x)\in\mathcal{P}(\mathcal{Y}),x\in\mathcal{X}\}$.
Given $P_{X}$ and $P_{Y|X}$, we write $P_{XY}=P_{X}P_{Y|X}$ as
the joint distribution, and $P_{Y}$ or $P_{X}\circ P_{Y|X}$ as the
$Y$-marginal distribution, i.e., $P_{Y}(y)=P_{X}\circ P_{Y|X}(y)=\sum_{x}P_{X}(x)P_{Y|X}(y|x)$.
Throughout this paper, $P_{XY}$ is a given joint distribution. Without
loss of generality, we assume that $\mathcal{X}$ and $\mathcal{Y}$
are respectively the supports of $P_{X}$ and $P_{Y}$. 

We use $T_{x^{n}}(x):=\frac{1}{n}\sum_{i=1}^{n}\bone\{x_{i}=x\}$
to denote the type (i.e., empirical distribution) of a sequence $x^{n}$,
$T_{X}$ and $T_{Y|X}$ to respectively denote an $n$-type and a
conditional $n$-type, i.e., a type of sequences in $\mathcal{X}^{n}$
and a conditional type of sequences in $\mathcal{Y}^{n}$ given a
sequence in $\calX^{n}$. For a type $T_{X}$, the type class (the
set of sequences having the same type $T_{X}$) is denoted by $\mathcal{T}_{T_{X}}^{(n)}$
or shortly $\mathcal{T}_{T_{X}}$. For a conditional type $T_{Y|X}$
and a sequence $x^{n}$, the conditional type class of $x^{n}$ (the
set of $y^{n}$ sequences having the same conditional type $T_{Y|X}$
given $x^{n}$) is denoted by $\mathcal{T}_{T_{Y|X}}^{(n)}(x^{n})$
or $\mathcal{T}_{T_{Y|X}}(x^{n})$. The set of types of sequences
in $\mathcal{X}^{n}$ is denoted as 
\begin{align*}
\mathcal{P}_{n}(\mathcal{X}) & :=\{T_{x^{n}}:x^{n}\in\mathcal{X}^{n}\}\\
 & =\{P\in\mathcal{P}(\mathcal{X}):nP(x)\in\mathbb{Z},\forall x\in\mathcal{X}\}.
\end{align*}
The set of conditional types of sequences in $\mathcal{Y}^{n}$ given
a sequence in $\mathcal{X}^{n}$ with the type $T_{X}$ is denoted
as 
\[
\mathcal{P}_{n}(\mathcal{Y}|T_{X}):=\{T_{Y|X}\in\mathcal{P}(\mathcal{Y}|\mathcal{X}):T_{X}T_{Y|X}\in\mathcal{P}_{n}(\mathcal{X}\times\mathcal{Y})\}.
\]

The $\epsilon$-typical set relative to $Q_{X}$ is denoted as 
\[
\mathcal{T}_{\epsilon}^{(n)}(Q_{X}):=\{x^{n}\in\mathcal{X}^{n}:\left|T_{x^{n}}(x)-Q_{X}(x)\right|\leq\epsilon Q_{X}(x),\forall x\in\mathcal{X}\}.
\]
The conditionally $\epsilon$-typical set relative to $Q_{XY}$ is
denoted as 
\[
\mathcal{T}_{\epsilon}^{(n)}(Q_{XY}|x^{n}):=\{y^{n}\in\mathcal{X}^{n}:(x^{n},y^{n})\in\mathcal{T}_{\epsilon}^{(n)}(Q_{X})\}.
\]
For brevity, we sometimes write $\mathcal{T}_{\epsilon}^{(n)}(Q_{X})$
and $\mathcal{T}_{\epsilon}^{(n)}(Q_{XY}|x^{n})$ as $\mathcal{T}_{\epsilon}(Q_{X})$
and $\mathcal{T}_{\epsilon}(Q_{XY}|x^{n})$, or $\mathcal{T}_{\epsilon}$
and $\mathcal{T}_{\epsilon}(x^{n})$ respectively. Other notation
generally follows the book by Csiszár and Körner~\cite{Csi97}.

The total variation distance between two probability mass functions
$P$ and $Q$ with a common alphabet $\calX$ is defined by 
\begin{equation}
\|P-Q\|_{\mathrm{TV}}:=\frac{1}{2}\sum_{x\in\calX}|P(x)-Q(x)|.
\end{equation}
By the definition of $\epsilon$-typical set, we have that for any
$x^{n}\in\mathcal{T}_{\epsilon}(Q)$, $\|T_{x^{n}}-Q\|_{\mathrm{TV}}\leq\frac{\epsilon}{2}$.

Fix distributions $P,Q\in\calP(\calX)$. Then the Rényi divergence
of order $q\in(-\infty,0)\cup(0,1)\cup(1,\infty)$ is defined in \eqref{eq:Renyidivergence}.
We adopt the convention $x/0=0$ for $x=0$, and $\infty$ for $x>0$
here and throughout this paper. The Rényi divergence of orders $0,1,\pm\infty$
is defined by continuous extension. In particular,\footnote{The constraint $P(x)>0$ in \eqref{eq:D-infty} cannot be removed
due to the convention $0/0=0$.} 
\begin{align}
D_{0}(Q\|P) & :=\lim_{q\downarrow0}D_{q}(Q\|P)=-\log P\{Q>0\};\nonumber \\
D_{1}(Q\|P) & :=\lim_{q\to1}D_{q}(Q\|P)=D(Q\|P);\nonumber \\
D_{\infty}(Q\|P) & :=\lim_{q\to\infty}D_{q}(Q\|P)=\log\sup_{x}\frac{Q(x)}{P(x)}\nonumber \\
D_{-\infty}(Q\|P) & :=\lim_{q\to-\infty}D_{q}(Q\|P)=\log\inf_{x:P(x)>0}\frac{Q(x)}{P(x)}.\label{eq:D-infty}
\end{align}
Hence a special case of the Rényi divergence is the usual relative
entropy. The conditional Rényi divergence is defined as 
\begin{align}
D_{q}(Q_{Y|X}\|P_{Y|X}|Q_{X}) & :=D_{q}(Q_{X}Q_{Y|X}\|Q_{X}P_{Y|X}),
\end{align}
We denote 
\[
H_{Q}(X):=-\sum_{x}Q_{X}(x)\log Q_{X}(x)
\]
as the Shannon entropy of $X\sim Q_{X}$ and 
\begin{equation}
I_{Q}(X;Y):=D(Q_{XY}\|Q_{X}Q_{Y})\label{eq:-76}
\end{equation}
as the mutual information between $(X,Y)\sim Q_{XY}$. When there
is no ambiguity, we omit the subscripts of $H_{Q}$ and $I_{Q}$.
The cross-entropy of the distribution $P$ relative to a distribution
$Q$ is defined as follows: 
\[
H(Q,P):=-\sum_{x}Q(x)\log P(x).
\]

The coupling set of $Q_{X}$ and $Q_{Y}$ is denoted as 
\begin{equation}
\Pi(Q_{X},Q_{Y}):=\big\{ P_{XY}\in\calP(\calX\times\calY):P_{X}=Q_{X},P_{Y}=Q_{Y}\big\}.
\end{equation}
The optimal transport divergence between $P_{X}$ and $P_{Y}$ with
respect to a distribution $P_{XY}$ is defined as 
\begin{equation}
\D(Q_{X},Q_{Y}\|P_{XY}):=\inf_{Q_{XY}\in\Pi(Q_{X},Q_{Y})}D(Q_{XY}\|P_{XY}).\label{eq:mathbbD}
\end{equation}

Finally, we use $o_{n}(1)$ to denote generic sequences tending to
zero as $n\rightarrow\infty$. We write $f(n)\apprge g(n)$ if $f(n)\ge g(n)+o_{n}(1)$.
We write $f(n)\approx g(n)$ if $f(n)\apprge g(n)$ and $g(n)\apprge f(n)$,
i.e., two sides coincide asymptotically. In addition, we write $f(n)\dotleq g(n)$
if $f(n)\ge g(n)e^{o(n)}$. We write $f(n)\doteq g(n)$ if $f(n)\dotleq g(n)$
and $g(n)\dotleq f(n)$, i.e., the exponents of two sides coincide
asymptotically.  For $a\in\bbR$, $[a]^{+}:=\max\{a,0\}$ denotes
positive clipping. We denote $q':=\frac{q}{q-1}$ as the Hölder conjugate
of $q$. For $q=1$, $q'=+\infty$. For $q=+\infty$ or $-\infty$,
$q'=1$. Denote $[m:n]:=\{m,m+1,\cdots,n\}$ and $[n]:=[1:n]$. By
convention, $\inf\emptyset=+\infty$ and $\sup\emptyset=-\infty$.
For brevity, sometimes we omit the argument $x$ in $P_{X}(x)$ when
the argument letter coincides with the subscript (and only differs
in letter case), e.g., we write $\sum_{x}Q_{X}(x)\log P_{X}(x)$ as
$\sum_{x}Q_{X}\log P_{X}$, or more briefly, $\sum Q_{X}\log P_{X}$.

\section{Rényi Resolvability }

\label{sec:Resolvability} We consider the channel resolvability problem
illustrated in Fig.~\ref{fig:Resolvability} with a channel $P_{Y|X}$
and a target distribution $P_{Y}$. Let $\mathcal{X},\mathcal{Y}$
be finite sets. We do not require that $P_{X}\circ P_{Y|X}=P_{Y}$
for some $P_{X}$, unless otherwise specified explicitly. We use Rényi
divergences to quantify the level of approximation. Rényi divergences
admit the following properties \cite{Erven}. 
\begin{enumerate}
\item (Skew Symmetry). For $q\in[-\infty,\infty]\backslash\{0,1\}$, $D_{q}(Q\|P)=\frac{q}{1-q}D_{1-q}(P\|Q)$
for probability measures $P,Q$. 
\item (Nonnegativity and Nonpositivity). For $q\in[0,\infty]$, $D_{q}(Q\|P)\ge0$
for probability measures $P,Q$. For $q\in[-\infty,0)$, $D_{q}(Q\|P)\le0$
for probability measures $P,Q$. Moreover, $D_{q}(Q\|P)=0$ for some
$q\in[-\infty,\infty]\backslash\{0\}$ if and only if $P=Q$; $D_{0}(Q\|P)=0$
if and only if $Q\ll P$. 
\end{enumerate}
By the skew symmetry, maximizing $D_{q}(Q_{Y^{n}}\|P_{Y}^{\otimes n})$
for $q<0$ is equivalent to minimizing $D_{1-q}(P_{Y}^{\otimes n}\|Q_{Y^{n}})$.
So, instead of using Rényi divergences of orders $q\in[-\infty,\infty]$,
we consider $D_{q}(Q_{Y^{n}}\|P_{Y}^{\otimes n})$ for $q\ge0$ and
$D_{q}(P_{Y}^{\otimes n}\|Q_{Y^{n}})$ for $q\ge1$ as the measures
of the level of approximation.

Observe that 
\begin{align*}
 & \e^{(q-1)D_{q}(Q_{Y^{n}}\|P_{Y}^{\otimes n})}\\
 & =\sum_{y^{n}}(\sum_{m}e^{-nR}P_{Y|X}^{\otimes n}(y^{n}|f(m)))^{q}P_{Y}^{\otimes n}(y^{n})^{1-q}.
\end{align*}
Hence to guarantee that $D_{q}(Q_{Y^{n}}\|P_{Y}^{\otimes n})$ is
finite for $q\geq1$, we assume $P_{Y|X=x}\ll P_{Y}$ for all $x\in\mathcal{X}$;
otherwise, we can remove all the values $x$ such that $P_{Y|X=x}\not\ll P_{Y}$
from $\mathcal{X}$. However, it is worth noting that we do not need
to do so for $0\leq q<1$, since $D_{q}(Q_{Y^{n}}\|P_{Y}^{\otimes n})$
is always finite regardless of whether $P_{Y|X=x}\ll P_{Y}$ for all
$x\in\mathcal{X}$. 

\subsection{\label{subsec:Direct-Part-for-1-1}Asymptotics }

Define 
\begin{equation}
R_{\min}:=\max_{Q_{Y}\ll P_{Y}}\min_{Q_{X|Y}:Q_{Y|X}\ll P_{Y|X}}I_{Q}(X;Y),\label{eq:-77}
\end{equation}
where $I_{Q}(X;Y)$ is the mutual information between $(X,Y)\sim Q_{XY}$.
The asymptotics of the normalized Rényi divergences is characterized
in the following theorem, whose proof is provided in Appendix \ref{sec:Proof-of-Theorem-Resolvability}. 
\begin{thm}[Rényi Resolvability]
\label{thm:Resolvability} The following hold. 
\begin{enumerate}
\item For $q\in[1,\infty]$ and $R\ge0$, 
\begin{align}
 & \lim_{n\to\infty}\frac{1}{n}\inf_{f:[e^{nR}]\to\mathcal{X}^{n}}D_{q}(Q_{Y^{n}}\|P_{Y}^{\otimes n})\nonumber \\
 & =\min_{Q_{X}}\max\{\mathbb{E}_{Q_{X}}[D_{q}(P_{Y|X}\|P_{Y})]-R,\nonumber \\
 & \qquad\qquad\max_{Q_{Y|X}}-q'D(Q_{Y|X}\|P_{Y|X}|Q_{X})+D(Q_{Y}\|P_{Y})\}.\label{eq:bound1}
\end{align}
\item For $q\in[0,1)$ and $R\ge0$, 
\begin{align}
 & \lim_{n\to\infty}\frac{1}{n}\inf_{f:[e^{nR}]\to\mathcal{X}^{n}}D_{q}(Q_{Y^{n}}\|P_{Y}^{\otimes n})\nonumber \\
 & =\min_{Q_{XY}:Q_{XY}\ll Q_{X}P_{Y|X}}\max\{-q'D(Q_{Y|X}\|P_{Y|X}|Q_{X})+D(Q_{Y|X}\|P_{Y}|Q_{X})-R,\nonumber \\
 & \qquad\qquad\qquad-q'D(Q_{Y|X}\|P_{Y|X}|Q_{X})+D(Q_{Y}\|P_{Y})\}.\label{eq:bound2}
\end{align}
\item For $q\in[1,\infty]$ and $R>R_{\min}$ (with $R_{\min}$ defined
in \eqref{eq:-77}), 
\begin{align}
 & \lim_{n\to\infty}\frac{1}{n}\inf_{f:[e^{nR}]\to\mathcal{X}^{n}}D_{q}(P_{Y}^{\otimes n}\|Q_{Y^{n}})\nonumber \\
 & =\max_{Q_{Y}}\min_{Q_{X|Y}:I_{Q}(X;Y)\le R}D(Q_{Y|X}\|P_{Y|X}|Q_{X})-q'D(Q_{Y}\|P_{Y}).\label{eq:-47}
\end{align}
For $q\in[1,\infty]$ and $R<R_{\min}$, 
\begin{align}
\lim_{n\to\infty}\frac{1}{n}\inf_{f:[e^{nR}]\to\mathcal{X}^{n}}D_{q}(P_{Y}^{\otimes n}\|Q_{Y^{n}}) & =+\infty.\label{eq:-47-1}
\end{align}
\item For $q=0$ and $R\ge0$, 
\begin{align*}
 & \lim_{n\to\infty}\frac{1}{n}\inf_{f:[e^{nR}]\to\mathcal{X}^{n}}D_{0}(P_{Y}^{\otimes n}\|Q_{Y^{n}})\\
 & =\min_{x}D_{0}(P_{Y}\|P_{Y|X=x}).
\end{align*}
\end{enumerate}
\end{thm}
\begin{rem}
For $q=1$ (i.e., $q'=+\infty$) and $q=+\infty$ (i.e., $q'=1$),
Statement 1 is understood as 
\begin{align}
 & \lim_{n\to\infty}\frac{1}{n}\inf_{f:[e^{nR}]\to\mathcal{X}^{n}}D_{q}(Q_{Y^{n}}\|P_{Y}^{\otimes n})\nonumber \\
 & =\begin{cases}
{\displaystyle \min_{Q_{X}}\max\{D(P_{Y|X}\|P_{Y}|Q_{X})-R,}\\
\qquad\qquad D(Q_{X}\circ P_{Y|X}\|P_{Y})\}, & q=1\\
{\displaystyle \min_{Q_{X}}\max\{\mathbb{E}_{Q_{X}}[D_{\infty}(P_{Y|X}\|P_{Y})]-R,}\\
\qquad\qquad{\displaystyle \max_{Q_{Y|X}}-D(Q_{Y|X}\|P_{Y|X}|Q_{X})+D(Q_{Y}\|P_{Y})\}}, & q=\infty
\end{cases}.\label{eq:bound1-1}
\end{align}
Equation \eqref{eq:bound1-1} with $q=1$ can be alternatively seen
as the special case of Statement 2 with $q'=-\infty$. 
\end{rem}
\begin{rem}
For $q\in(0,1)$, the constraint $Q_{XY}\ll Q_{X}P_{Y|X}$ in \eqref{eq:bound2}
can be removed. For $q=0$, Statement 2 is understood as
\begin{align*}
 & \lim_{n\to\infty}\frac{1}{n}\inf_{f:[e^{nR}]\to\mathcal{X}^{n}}D_{0}(Q_{Y^{n}}\|P_{Y}^{\otimes n})\\
 & =\min_{Q_{XY}:Q_{XY}\ll Q_{X}P_{Y|X}}\max\{D(Q_{Y|X}\|P_{Y}|Q_{X})-R,D(Q_{Y}\|P_{Y})\}.
\end{align*}
\end{rem}
\begin{rem}
For $q=1$ (i.e., $q'=+\infty$) and $q=+\infty$ (i.e., $q'=1$),
Statement 3 is understood as that for $R>R_{\min}$, 
\begin{align}
 & \lim_{n\to\infty}\frac{1}{n}\inf_{f:[e^{nR}]\to\mathcal{X}^{n}}D_{q}(P_{Y}^{\otimes n}\|Q_{Y^{n}})\nonumber \\
 & =\begin{cases}
{\displaystyle \min_{Q_{XY}:Q_{Y}=P_{Y},\,I_{Q}(X;Y)\le R}D(Q_{Y|X}\|P_{Y|X}|Q_{X})}, & q=1\\
{\displaystyle \max_{Q_{Y}}\min_{Q_{X|Y}:I_{Q}(X;Y)\le R}D(Q_{Y|X}\|P_{Y|X}|Q_{X})-D(Q_{Y}\|P_{Y})}, & q=\infty
\end{cases}.\label{eq:-47-3}
\end{align}
\end{rem}
\begin{rem}
Given $R$, the expressions in \eqref{eq:bound1} and \eqref{eq:bound2}
constitute a function of $q\in[0,\infty]$. Such a function is continuous
in $q\in[0,\infty]$. The expression in \eqref{eq:-47} is continuous
in $q\in[1,\infty]$. 
\end{rem}
We also consider the code restricted on a type class. The proof is
similar to that of Theorem \ref{thm:Resolvability}, and hence, omitted.
Define 
\begin{equation}
R_{\min,Q_{X}}:=\max_{Q_{Y}\ll P_{Y}}\min_{Q_{XY}\in\Pi(Q_{X},Q_{Y}):Q_{Y|X}\ll P_{Y|X}}I_{Q}(X;Y).\label{eq:-77-1}
\end{equation}

\begin{thm}[Rényi Resolvability]
\label{thm:Resolvability-3} Let $Q_{X}$ be an $n$-type. Then,
the following hold. 
\begin{enumerate}
\item For $q\in[1,\infty]$ and $R\ge0$, 
\begin{align}
 & \frac{1}{n}\inf_{f:[e^{nR}]\to\mathcal{T}_{Q_{X}}}D_{q}(Q_{Y^{n}}\|P_{Y}^{\otimes n})\nonumber \\
 & =\max\{\mathbb{E}_{Q_{X}}[D_{q}(P_{Y|X}\|P_{Y})]-R,\nonumber \\
 & \qquad\qquad\max_{Q_{Y|X}}-q'D(Q_{Y|X}\|P_{Y|X}|Q_{X})+D(Q_{Y}\|P_{Y})\}+o_{n}(1),\label{eq:-15}
\end{align}
where $o_{n}(1)$ is a term vanishing uniformly for all sequences
of types $Q_{X}$ as $n\to\infty$. 
\item For $q\in[0,1)$ and $R\ge0$, 
\begin{align}
 & \frac{1}{n}\inf_{f:[e^{nR}]\to\mathcal{T}_{Q_{X}}}D_{q}(Q_{Y^{n}}\|P_{Y}^{\otimes n})\nonumber \\
 & =\min_{Q_{Y|X}}\max\{-q'D(Q_{Y|X}\|P_{Y|X}|Q_{X})+D(Q_{Y|X}\|P_{Y}|Q_{X})-R,\nonumber \\
 & \qquad\qquad\qquad-q'D(Q_{Y|X}\|P_{Y|X}|Q_{X})+D(Q_{Y}\|P_{Y})\}+o_{n}(1),\label{eq:-21}
\end{align}
where $o_{n}(1)$ is a term vanishing uniformly for all sequences
of types $Q_{X}$ as $n\to\infty$. 
\item For $q\in[1,\infty]$ and $R>R_{\min,Q_{X}}$, 
\begin{align}
 & \lim_{n\to\infty}\frac{1}{n}\inf_{f:[e^{nR}]\to\mathcal{T}_{Q_{X}}}D_{q}(P_{Y}^{\otimes n}\|Q_{Y^{n}})\nonumber \\
 & =\max_{Q_{Y}}\min_{Q_{XY}\in\Pi(Q_{X},Q_{Y}):I_{Q}(X;Y)\le R}D(Q_{Y|X}\|P_{Y|X}|Q_{X})-q'D(Q_{Y}\|P_{Y})+o_{n}(1)\label{eq:-47-5}
\end{align}
where $o_{n}(1)$ is a term vanishing uniformly for all sequences
of types $Q_{X}$ as $n\to\infty$. For $q\in[1,\infty]$ and $R<R_{\min,Q_{X}}$,
\begin{align}
\lim_{n\to\infty}\frac{1}{n}\inf_{f:[e^{nR}]\to\mathcal{T}_{Q_{X}}}D_{q}(P_{Y}^{\otimes n}\|Q_{Y^{n}}) & =+\infty.\label{eq:-47-1-1}
\end{align}
\end{enumerate}
\end{thm}
We next derive dual formulas for the expressions in the two theorems
above, whose proof is given in Appendix \ref{sec:Proof-of-Proposition}. 
\begin{prop}[Dual Formulas]
 \label{prop:The-expressions-in} The following hold. 
\begin{enumerate}
\item The optimization expression in \eqref{eq:-15} (without the term $o_{n}(1)$)
is equal to
\begin{align}
 & \max\{\mathbb{E}_{Q_{X}}[D_{q}(P_{Y|X}\|P_{Y})]-R,\,\nonumber \\
 & \qquad\qquad\max_{S_{Y}}q'\mathbb{E}_{Q_{X}}\big[\log\mathbb{E}_{P_{Y|X}}[(\frac{S_{Y}}{P_{Y}})^{1/q'}]\big]\}.\label{eq:-89-1-1}
\end{align}
Consequently, the optimization expression in \eqref{eq:bound1} is
equal to the $\min_{Q_{X}}$ of \eqref{eq:-89-1-1}. 
\item The optimization expression in \eqref{eq:-21} (without the term $o_{n}(1)$)
is equal to 
\begin{align}
 & \max_{\lambda\in[0,1]}\max_{S_{Y}}-(\lambda-q')\mathbb{E}_{Q_{X}}\big[\log\sum_{y}P_{Y|X}^{\frac{-q'}{\lambda-q'}}P_{Y}^{\frac{1}{\lambda-q'}}S_{Y}^{\frac{\lambda-1}{\lambda-q'}}\big]-\lambda R,\label{eq:-90}
\end{align}
and the optimization expression in \eqref{eq:bound2} is equal to
\begin{equation}
\max_{\lambda\in[0,1]}\max_{S_{Y}}\min_{x}-(\lambda-q')\log(\sum_{y}P_{Y|X=x}^{\frac{-q'}{\lambda-q'}}P_{Y}^{\frac{1}{\lambda-q'}}S_{Y}^{\frac{\lambda-1}{\lambda-q'}})-\lambda R.\label{eq:-97}
\end{equation}
 
\item The optimization expression in \eqref{eq:-47-5} (without the term
$o_{n}(1)$) admits the following dual formula 
\begin{equation}
\sup_{\lambda\ge0,f}\min_{S_{X}}(q'-1)\log\mathbb{E}_{P_{Y}}\Big[\mathbb{E}_{S_{X}}[(\frac{P_{Y|X}}{P_{Y}}e^{f})^{\frac{1}{1+\lambda}}]^{\frac{1+\lambda}{1-q'}}\Big]-\mathbb{E}_{Q_{X}}[f]-\lambda R,\label{eq:-96}
\end{equation}
with $f$ denoting a real-valued function on $\mathcal{X}$, and the
optimization expression in \eqref{eq:-47} admits the following dual
formula 
\begin{equation}
\sup_{\lambda\ge0}\min_{S_{X}}(q'-1)\log\mathbb{E}_{P_{Y}}\Big[\mathbb{E}_{S_{X}}[(\frac{P_{Y|X}}{P_{Y}})^{\frac{1}{1+\lambda}}]^{\frac{1+\lambda}{1-q'}}\Big]-\lambda R.\label{eq:-98}
\end{equation}
\end{enumerate}
\end{prop}
\begin{rem}
The optimization expression in \eqref{eq:-15} (without the term $o_{n}(1)$)
can be also written as the one in \eqref{eq:-90}, in which the optimal
$\lambda$ is equal to $0$ or $1$.  
\end{rem}

\begin{rem}
\label{rem:iid}Using the proof techniques used in the proof of Theorem
\ref{thm:Resolvability}, one can identify the performance of i.i.d.
codes as follows. Consider a random codebook $\mathcal{C}_{n}=\{X^{n}(m)\}_{m\in\calM_{n}}$
with $X^{n}(m)\sim Q_{X}^{\otimes n},m\in\calM_{n}$, and set $f(m)=X^{n}(m)$.
Denote $Q_{Y^{n}|\mathcal{C}_{n}}$ as the (random) output distribution.
In this case, $\frac{1}{n}D_{q}(Q_{Y^{n}|\mathcal{C}_{n}}\|P_{Y}^{\otimes n})$
converges in probability to 
\[
\begin{cases}
\max\{D_{q}(P_{Y|X}\|P_{Y}|Q_{X})-R,D_{q}(Q_{Y}\|P_{Y})\}, & q\in[1,\infty]\\
\max_{\lambda\in[0,1]}\frac{1}{q-1}\log(\sum_{x,y}Q_{X}P_{Y|X}^{1-\lambda(1-q)}Q_{Y}^{(\lambda-1)(1-q)}P_{Y}^{1-q})-\lambda R, & q\in[0,1)
\end{cases}
\]
where $Q_{Y}=Q_{X}\circ P_{Y|X}$.  For $q\in[1,\infty]$ and $R>R_{\min}$
(with $R_{\min}$ defined in \eqref{eq:-77}),  $\frac{1}{n}D_{q}(P_{Y}^{\otimes n}\|Q_{Y^{n}|\mathcal{C}_{n}})$
converges in probability to
\[
\sup_{\lambda\ge0}(q'-1)\log\mathbb{E}_{P_{Y}}\Big[\mathbb{E}_{Q_{X}}[(\frac{P_{Y|X}}{P_{Y}})^{\frac{1}{1+\lambda}}]^{\frac{1+\lambda}{1-q'}}\Big]-\lambda R.
\]
\end{rem}

\subsection{Resolvability Rates}

We now compute the Rényi resolvability rate, which is defined as the
minimum rate $R$ of the input process $\{X^{n}(m):m\in\calM_{n}\}$
to ensure that the unnormalized Rényi divergence $D_{q}(Q_{Y^{n}}\|P_{Y}^{\otimes n})$
or the normalized Rényi divergence $\frac{1}{n}D_{q}(Q_{Y^{n}}\|P_{Y}^{\otimes n})$
vanishes. We assume that 
\begin{equation}
\mathcal{P}(P_{Y|X},P_{Y}):=\{P_{X}:P_{X}\circ P_{Y|X}=P_{Y}\}\ne\emptyset.\label{eq:-35-1}
\end{equation}
Otherwise, there does not exist a code such that $\frac{1}{n}D_{q}(Q_{Y^{n}}\|P_{Y}^{\otimes n})$
vanishes. By Theorem \ref{thm:Resolvability} we easily obtain the
following result. The proofs of the resolvability rates under normalized
Rényi divergences are provided in Appendix \ref{sec:Proof-of-Theorem-Resolvability-Rates}.
The resolvability rates under unnormalized Rényi divergences are implied
by the exponential convergence results in Section \ref{subsec:Exponential-Behavior}. 
\begin{thm}[Rényi Resolvability Rates]
\label{thm:ResolvabilityRates} We assume that $\mathcal{P}(P_{Y|X},P_{Y})\ne\emptyset.$
Then, the following hold. 
\begin{enumerate}
\item For $q\in[0,\infty]$, we have 
\begin{align}
 & \inf\{R:\inf_{f}D_{q}(Q_{Y^{n}}\|P_{Y}^{\otimes n})\rightarrow0\}\nonumber \\
 & =\inf\{R:\frac{1}{n}\inf_{f}D_{q}(Q_{Y^{n}}\|P_{Y}^{\otimes n})\rightarrow0\}\\
 & =R_{q}(P_{Y|X},P_{Y}),\label{eq:-117-1}
\end{align}
where 
\begin{align}
 & R_{q}(P_{Y|X},P_{Y})\nonumber \\
 & :=\begin{cases}
\underset{P_{X}\in\mathcal{P}(P_{Y|X},P_{Y})}{\min}\mathbb{E}_{P_{X}}[D_{q}(P_{Y|X}\|P_{Y})], & q\in(1,\infty]\\
\underset{P_{X}\in\mathcal{P}(P_{Y|X},P_{Y})}{\min}I_{P}(X;Y), & q\in(0,1]\\
\min_{Q_{XY}:Q_{XY}\ll Q_{X}P_{Y|X},\,Q_{Y}=P_{Y}}I_{Q}(X;Y), & q=0.
\end{cases}.\label{eq:R}
\end{align}
\item For $q\in[0,\infty]$, we have 
\begin{align}
 & \inf\{R:\inf_{f}D_{q}(P_{Y}^{\otimes n}\|Q_{Y^{n}})\rightarrow0\}\nonumber \\
 & =\inf\{R:\frac{1}{n}\inf_{f}D_{q}(P_{Y}^{\otimes n}\|Q_{Y^{n}})\rightarrow0\}\\
 & =\hat{R}_{q}(P_{Y|X},P_{Y}),\label{eq:-117-1-1}
\end{align}
where 
\begin{align}
 & \hat{R}_{q}(P_{Y|X},P_{Y})\nonumber \\
 & :=\begin{cases}
\underset{Q_{Y}}{\max}\underset{Q_{X|Y}:D(Q_{Y|X}\|P_{Y|X}|Q_{X})\le q'D(Q_{Y}\|P_{Y})}{\min}I_{Q}(X;Y), & q\in(1,\infty]\\
\max\big\{ R_{\min},\underset{P_{X}\in\mathcal{P}(P_{Y|X},P_{Y})}{\min}I_{P}(X;Y)\big\}, & q=1\\
\underset{P_{X}\in\mathcal{P}(P_{Y|X},P_{Y})}{\min}I_{P}(X;Y), & q\in(0,1)\\
0, & q=0.
\end{cases}.\label{eq:Rhat}
\end{align}
\end{enumerate}
\end{thm}
\begin{rem}
The first clause in \eqref{eq:Rhat} is no smaller than $R_{\min}$. 
\end{rem}
\begin{rem}
It has been already known that $R_{q}(P_{Y|X},P_{Y})$ is continuous
in $q\in(0,\infty]$ but not at $q=0$ \cite{yu2019renyi}. It is
easy to see that $\hat{R}_{q}(P_{Y|X},P_{Y})$ is continuous in $q\in(0,\infty]$
if $R_{\min}\le\underset{P_{X}\in\mathcal{P}(P_{Y|X},P_{Y})}{\min}I_{P}(X;Y)$;
otherwise, $\hat{R}_{q}(P_{Y|X},P_{Y})$ is continuous in $q\in(0,1)$
and $q\in[1,\infty]$ individually, but not left-continuous at $q=1$.
For example, if $P_{Y|X}$ is the identity channel, then 
\[
\hat{R}_{q}(P_{Y|X},P_{Y})=\begin{cases}
H_{0}(P_{Y}), & q\in[1,\infty]\\
H(P_{Y}), & q\in(0,1).\\
0, & q=0
\end{cases}
\]
On the other hand, if $P_{XY}(x,y)>0$ for all $(x,y)\in\mathcal{X}\times\mathcal{Y}$,
then $R_{\min}=0$, which implies that $\hat{R}_{q}(P_{Y|X},P_{Y})$
is continuous in $q\in(0,\infty]$.
\end{rem}
The already known part and new part of Theorem \ref{thm:ResolvabilityRates}
are summarized in Table \ref{tab:Summary-of-our}.

\begin{table*}
\centering{}\caption{\label{tab:Summary-of-our}Summary of results for  Rényi resolvability
on finite alphabets. }
\begin{tabular}{|c|>{\centering}p{8cm}|}
\hline 
\multicolumn{1}{|c|}{\textbf{Rényi Divergence $D_{q}(Q_{Y^{n}}\|P_{Y}^{\otimes n})$}} & \textbf{Results}\tabularnewline
\hline 
\hline 
\multicolumn{1}{|c|}{$q=1$} & Wyner \cite{WynerCI} and Han--Verdú \cite{Han93} for normalized
divergence; Hayashi~\cite{Hayashi06,Hayashi11} for unnormalized
divergence\tabularnewline
\hline 
$q\in(0,1)\cup(1,2]\cup\{\infty\}$ & Yu--Tan \cite{yu2019renyi}\tabularnewline
\hline 
$q\in[-\infty,0]\cup(2,\infty)$ & Theorem \ref{thm:ResolvabilityRates}\tabularnewline
\hline 
\end{tabular}
\end{table*}

\begin{prop}[Dual Formula]
 For $q\in(1,\infty)$, $\hat{R}_{q}(P_{Y|X},P_{Y})$ admits the
following dual formula: 
\begin{align*}
\hat{R}_{q}(P_{Y|X},P_{Y}) & =\sup_{\lambda\ge0}\min_{S_{X}}-\lambda(1-q')\log\mathbb{E}_{P_{Y}}\Big[\mathbb{E}_{S_{X}}[(\frac{P_{Y|X}}{P_{Y}})^{\frac{\lambda}{1+\lambda}}]^{\frac{1+\lambda}{\lambda(1-q')}}\Big].
\end{align*}
\end{prop}

\subsection{\label{subsec:Exponential-Behavior}Exponential Behavior }

We now consider the exponential convergence of the problem.  We first
consider a random map $f_{\mathcal{C}_{n}}:\calM_{n}:=[\e^{nR}]\rightarrow\calX^{n}$
induced by i.i.d. codes given below. Let $P_{X}\in\mathcal{P}(P_{Y|X},P_{Y})$.
We set $\mathcal{C}_{n}=\{X^{n}(m)\}_{m\in\calM_{n}}$ with $X^{n}(m)\sim P_{X}^{\otimes n},m\in\calM_{n}$
drawn independently, and set $f_{\mathcal{C}_{n}}(m)=X^{n}(m)$.
This forms a random i.i.d. code, and $\mathcal{C}_{n}$ is called
a random codebook. The joint distribution induced by the code is 
\[
Q_{\mathcal{C}_{n}M_{n}X^{n}Y^{n}}:=Q_{\mathcal{C}_{n}}Q_{M_{n}}Q_{X^{n}|M_{n},\mathcal{C}_{n}}P_{Y|X}^{\otimes n}(\cdot|X^{n}(M_{n})),
\]
where $Q_{M_{n}}=\Unif[\e^{nR}]$ and $Q_{X^{n}|M_{n},\mathcal{C}_{n}}(x^{n}|m,\mathcal{C}_{n})=\bone\{x^{n}=X^{n}(m)\}$.
When specialized to this ensemble of random codes,  the resolvability
problem reduces to the so-called soft-covering problem \cite{cuff13}.
We now characterize the convergence exponent of $D_{q}(Q_{Y^{n}|\mathcal{C}_{n}}\|P_{Y}^{\otimes n}|Q_{\mathcal{C}_{n}})$
for this case.  
\begin{thm}[Exponential Behavior of i.i.d.\ Codes]
\label{thm:exponentforiid} For the i.i.d.\ code described above
and $q\in[2,\infty)$, if the rate $R$ satisfies $R>D_{q}(P_{Y|X}\|P_{Y}|P_{X})$,
then we have 
\begin{align}
\lim_{n\to\infty}-\frac{1}{n}\log D_{q}(Q_{Y^{n}|\mathcal{C}_{n}}\|P_{Y}^{\otimes n}|P_{\mathcal{C}_{n}}) & =\min\{\gamma(1),\gamma(q-1)\},\label{eq:-26}
\end{align}
where 
\[
\gamma(s):=s(R-D_{1+s}(P_{Y|X}\|P_{Y}|P_{X})).
\]
Furthermore, for $q=\infty$, 
\begin{align}
\frac{1}{n}D_{\infty}(Q_{Y^{n}|\mathcal{C}_{n}}\|P_{Y}^{\otimes n}|P_{\mathcal{C}_{n}}) & =D_{\infty}(P_{Y|X}\|P_{Y}|P_{X}),\label{eq:-26-3}
\end{align}
which is zero if and only if $(X,Y)\sim P_{XY}$ are independent. 

\end{thm}
The proof of the above theorem is provided in Appendix~\ref{sec:Proof-of-Theorem-expiid},
which is based on the one-shot bounds in Lemma \ref{lem:oneshot}.
Furthermore, the upper bound in \eqref{eq:-26} can be also proven
by slightly modifying the  proof for converse part with $q\in(1,2]$
in \cite[Appendix F]{yu2019renyi}. Specifically, replace the minimization
operation in (353) therein with the maximization operation, and modify
subsequent operations correspondingly.

In contrast, for $q=1$ and for other $q\in(0,2]$, the optimal exponents
for i.i.d. codes were respectively completely characterized in \cite{Parizi}
and \cite{yu2019renyi}, i.e., 
\begin{align}
\lim_{n\to\infty}-\frac{1}{n}\log D_{q}(Q_{Y^{n}|\mathcal{C}_{n}}\|P_{Y}^{\otimes n}|Q_{\mathcal{C}_{n}}) & =\begin{cases}
\underset{s\in[q-1,1]}{\max}\gamma(s), & q\in(1,2]\\
\underset{s\in[0,1]}{\max}\gamma(s), & q\in(0,1]
\end{cases}.\label{eq:-28}
\end{align}
Combining \eqref{eq:-26} and \eqref{eq:-28} provides a complete
characterization of the optimal exponent for soft-covering with respect
to i.i.d. codes under Rényi divergences of positive orders. Furthermore,
for $q=0$, replacing the channel $P_{Y|X}$ with any $Q_{Y|X}$ such
that $P_{X}Q_{Y|X}\ll P_{X}P_{Y|X}$ will not decrease $D_{0}(Q_{Y^{n}|\mathcal{C}_{n}}\|P_{Y}^{\otimes n}|Q_{\mathcal{C}_{n}})$.
So, \eqref{eq:-28} implies the following achievability result for
$q=0$:
\begin{align}
 & \lim_{n\to\infty}-\frac{1}{n}\log D_{0}(Q_{Y^{n}|\mathcal{C}_{n}}\|P_{Y}^{\otimes n}|Q_{\mathcal{C}_{n}})\nonumber \\
 & \ge\underset{s\in[0,1]}{\max}s(R-\min_{Q_{XY}\in\Pi(P_{X},P_{Y}):Q_{XY}\ll P_{X}P_{Y|X}}D_{1+s}(Q_{Y|X}\|Q_{Y}|Q_{X})).
\end{align}
 In contrast, the optimal exponent for the same problem but under
the TV distance was completely characterized by Yagli and Cuff in
\cite{yagli2019exact}. The characterizations of the convergence exponents
for the resolvability problem for i.i.d. codes under various divergence
measures are summarized in Table \ref{tab:Summary-of-our-1}.

\begin{table*}
\centering{}\caption{\label{tab:Summary-of-our-1}Summary of results for the convergence
exponents for the resolvability problem for i.i.d. codes on finite
alphabets. }
\begin{tabular}{|c|c|c|}
\hline 
\multicolumn{2}{|c|}{\textbf{Cases}} & \textbf{Results}\tabularnewline
\hline 
\hline 
\multirow{3}{*}{Rényi Divergence \textbf{$D_{q}(Q_{Y^{n}}\|P_{Y}^{\otimes n})$}} & \multicolumn{1}{c|}{$q=1$} & Parizi--Telatar--Merhav \cite{Parizi} \tabularnewline
\cline{2-3} \cline{3-3} 
 & $q\in(0,1)\cup(1,2]$ & Yu--Tan \cite{yu2019renyi}\tabularnewline
\cline{2-3} \cline{3-3} 
 & $q\in(2,\infty]$ & Theorem \ref{thm:exponentforiid}\tabularnewline
\hline 
\multicolumn{2}{|c|}{TV Distance} & \multirow{1}{*}{Yagli--Cuff \cite{yagli2019exact}}\tabularnewline
\hline 
\end{tabular}
\end{table*}

We next consider typical codes which are specified by the random mapping
$f_{\mathcal{C}_{n}}:\calM_{n}\rightarrow\calX^{n}$ given below.
Let $P_{X}\in\mathcal{P}(P_{Y|X},P_{Y})$. We set $\mathcal{C}_{n}=\{X^{n}(m)\}_{m\in\calM_{n}}$
with $X^{n}(m),m\in\calM_{n}$ drawn independently for different $m$'s
and according to the same distribution $P_{X}^{\otimes n}(\cdot|\mathcal{T}_{\epsilon}^{(n)}(P_{X}))$
with $\epsilon\in(0,1)$, and set $f_{\mathcal{C}_{n}}(m)=X^{n}(m)$.
This forms a random code. For these codes, we show that the Rényi
divergence decays at least exponentially fast, as long as the code
rate is larger than the Rényi resolvability rate given in \eqref{eq:R}.
The proof is provided in Appendix \ref{sec:Proof-of-Theorem-exp},
which is almost same as the proof of Theorem 4 in \cite{yu2019renyi}
except for replacing the one-shot bound for $q\in(1,2]$ with the
one for $q\in[2,\infty)$ in Lemma \ref{lem:oneshot}.
\begin{thm}[Exponential Behavior of Typical Codes]
\label{thm:exponent} Let $q\in[2,\infty)$ and $\epsilon\in(0,1)$.
If $R>(1+\epsilon)\mathbb{E}_{P_{X}}[D_{q}(P_{Y|X}\|P_{Y})]$, then
the random code given above satisfies that 
\begin{align*}
 & \liminf_{n\to\infty}-\frac{1}{n}\log D_{q}(Q_{Y^{n}|\mathcal{C}_{n}}\|P_{Y}^{\otimes n}|Q_{\mathcal{C}_{n}})\\
 & \ge\min\{\frac{\epsilon^{2}P_{\min}}{3},\gamma(1,\epsilon),\gamma(q-1,\epsilon)\},
\end{align*}
where $P_{\min}:=\min_{x:P_{X}(x)>0}P_{X}(x)$ and 
\[
\gamma(s,\epsilon):=s(R-(1+\epsilon)\mathbb{E}_{P_{X}}[D_{1+s}(P_{Y|X}\|P_{Y})]).
\]
\end{thm}
In contrast, for $q\in(0,2]$, it was shown in \cite{yu2019renyi}
that if $R>(1+\epsilon)\mathbb{E}_{P_{X}}[D_{q}(P_{Y|X}\|P_{Y})]$,
then the random code given above satisfies that 
\begin{align}
 & \liminf_{n\to\infty}-\frac{1}{n}\log D_{q}(Q_{Y^{n}|\mathcal{C}_{n}}\|P_{Y}^{\otimes n}|Q_{\mathcal{C}_{n}})\nonumber \\
 & \ge\min\{\frac{\epsilon^{2}P_{\min}}{3},E_{q}(P_{XY},\epsilon)\},\label{eq:-74}
\end{align}
where 
\begin{align}
E_{q}(P_{XY},\epsilon) & :=\begin{cases}
\underset{s\in[q-1,1]}{\max}\gamma(s,\epsilon), & q\in(1,2]\\
\underset{s\in[0,1]}{\max}\gamma(s,\epsilon), & q\in(0,1]
\end{cases}.\label{eq:-28-1}
\end{align}

Theorem \ref{thm:exponent} and \eqref{eq:-74} imply the following
corollary. 
\begin{cor}[Exponential Behavior for $D_{q}(Q_{Y^{n}}\|P_{Y}^{\otimes n})$]
Given $q\in(0,\infty)$, if $R>R_{q}(P_{Y|X},P_{Y})$, then there
exists a sequence of typical codes such that the induced Rényi divergence
$D_{q}(Q_{Y^{n}}\|P_{Y}^{\otimes n})$ decays at least exponentially
fast, where $R_{q}(P_{Y|X},P_{Y})$ is the Rényi resolvability rate
given in \eqref{eq:R}. 
\end{cor}
We lastly consider the Rényi divergence $D_{q}(P_{Y}^{\otimes n}\|Q_{Y^{n}})$.
Let $\delta\in(0,1)$ and $R>\delta>0$. We consider typical codes
which are specified by the random mapping $f_{\tilde{\mathcal{C}}}:[e^{nR}-e^{n(R-\delta)}]\rightarrow\calX^{n}$
given below. Let $P_{X}\in\mathcal{P}(P_{Y|X},P_{Y})$. We set $\tilde{\mathcal{C}}=\{X^{n}(m)\}_{m\in[e^{nR}-e^{n(R-\delta)}]}$
with $X^{n}(m),m\in[e^{nR}-e^{n(R-\delta)}]$ drawn independently
for different $m$'s and according to the same distribution $P_{X}^{\otimes n}(\cdot|\mathcal{T}_{\delta}^{(n)}(P_{X}))$
with , and set $f_{\tilde{\mathcal{C}}}(m)=X^{n}(m)$.

Let $\mathcal{B}_{\delta/2}:=\mathcal{B}_{\delta/2}(P_{X}):=\{Q_{X}:\|Q_{X}-P_{X}\|_{\mathrm{TV}}\leq\frac{\delta}{2}\}.$
Denote $\e^{n\hat{R}}=e^{n(R-\delta)}/|\mathcal{P}_{n}(\mathcal{X})\backslash\mathcal{B}_{\delta/2}|$.
For each $T_{X}\in\mathcal{P}_{n}(\mathcal{X})\backslash\mathcal{B}_{\delta/2}$,
let $\hat{\mathcal{C}}_{T_{X}}:=\{X^{n}(m)\}_{m\in[\e^{n\hat{R}}]}$
be a set of random sequences such that $X^{n}(m),m\in[\e^{n\hat{R}}]$
are drawn independently for different $m$'s and according to the
same distribution $\Unif(\mathcal{T}_{T_{X}})$.

Let $\mathcal{C}_{n}:=\tilde{\mathcal{C}}\cup(\bigcup_{T_{X}\in\mathcal{P}_{n}(\mathcal{X})\backslash\mathcal{B}_{\delta/2}}\hat{\mathcal{C}}_{T_{X}})$.
Let $f_{\mathcal{C}_{n}}:[e^{nR}]\to\mathcal{X}^{n}$ be the random
resolvability code based on $\mathcal{C}_{n}$, i.e., the deterministic
map given by $f_{\mathcal{C}_{n}}(m)=X^{n}(m)$ with $x^{n}(m)$ being
the $m$-th codeword in $\mathcal{C}_{n}$. This forms a random code. 
\begin{thm}[Exponential Behavior for $D_{q}(P_{Y}^{\otimes n}\|Q_{Y^{n}})$]
\label{thm:exponent-1} Let $q\in[1,\infty)$ and $R>\hat{R}_{q}(P_{Y|X},P_{Y})$,
where $\hat{R}_{q}(P_{Y|X},P_{Y})$ is the Rényi resolvability rate
given in \eqref{eq:Rhat}. Let $P_{X}\in\mathcal{P}(P_{Y|X},P_{Y})$
attain $\hat{R}_{q}(P_{Y|X},P_{Y})$. For the code above with rate
$R$ and some parameter $\delta>0$, there is a sequence of realizations
$c_{n}$ of $\mathcal{C}_{n}$ such that the Rényi divergence $D_{q}(P_{Y}^{\otimes n}\|Q_{Y^{n}})$
induced by $f_{c_{n}}$ decays at least exponentially fast. 
\end{thm}
The proof of this theorem is provided in Appendix \ref{sec:Proof-of-Theorem-exp-2}.

\subsection{Binary Example}

In this subsection, the bases of logarithms are set to $2$. We next
focus on binary distributions and channels, and provide explicit expressions
for the results derived in previous subsections. We consider $P_{Y|X}=\mathrm{BSC}(\epsilon)$
(i.e., $Y=X\oplus V$ with $V\sim\Bern(\epsilon)$) and $P_{X}=P_{Y}=\Bern(1/2)$.
This joint distribution is known as the doubly symmetric binary source
with crossover probability $\epsilon$, denoted by $\DSBS(\epsilon)$.
For a $\mathrm{BSC}(\epsilon)$ with $\epsilon\in(1/2,1]$, we can
convert it to $\mathrm{BSC}(\epsilon)$ with $\epsilon\in[0,1/2)$
by replacing $X$ with $1-X$. Furthermore, excluding trivial cases
with $\epsilon\in\{0,1/2,1\}$, we assume that the crossover probability
satisfies $\epsilon\in(0,1/2)$. 

Define the binary Rényi entropy function as 
\[
H_{q}(a):=\begin{cases}
-\frac{1}{q-1}\log_{2}(a^{q}+\overline{a}^{q}), & q\in(0,\infty)\backslash\{1\}\\
-a\log_{2}a-\overline{a}\log_{2}\overline{a}, & q=1\\
-\log_{2}\max\{a,\overline{a}\}, & q=\infty\\
\bone\{0<a<1\}, & q=0
\end{cases},
\]
and the binary Rényi divergence as 
\[
D_{q}(a):=D_{q}(a\|1/2):=1-H_{q}(a),
\]
where $\overline{a}:=1-a$. In particular, for $q=1$, $H(a):=H_{1}(a)$
and $D(a):=D_{1}(a)$ are respectively known as the binary (Shannon)
entropy function and the binary relative entropy (or binary KL divergence). 

Using Theorem \ref{thm:Resolvability} and Proposition \ref{prop:The-expressions-in},
we obtain the following corollary, the proof of which is given in
Appendix \ref{sec:Proof-of-Corollary-asymptotics}. 
\begin{cor}[Binary Rényi Resolvability]
\label{cor:asymptotics} Consider $P_{Y|X}=\mathrm{BSC}(\epsilon)$
and $P_{Y}=\Bern(1/2)$. 
\begin{enumerate}
\item For any $q\in[1,\infty]$, 
\begin{align*}
\lim_{n\to\infty}\frac{1}{n}\inf_{f:[e^{nR}]\to\mathcal{X}^{n}}D_{q}(Q_{Y^{n}}\|P_{Y}^{\otimes n}) & =[D_{q}(\epsilon)-R]^{+}.
\end{align*}
\item For any $q\in[0,1)$, 
\begin{align*}
 & \lim_{n\to\infty}\frac{1}{n}\inf_{f:[e^{nR}]\to\mathcal{X}^{n}}D_{q}(Q_{Y^{n}}\|P_{Y}^{\otimes n})=\begin{cases}
\max_{\lambda\in[0,1]}\lambda\left(D_{\frac{-q'}{\lambda-q'}}(\epsilon)-R\right), & q\in(0,1)\\
0, & q=0
\end{cases}.
\end{align*}
where for $D(\frac{\epsilon^{q}}{\epsilon^{q}+\bar{\epsilon}^{q}})<R<D(\epsilon)$,
the optimal $\lambda$ is the unique $\lambda^{*}\in[0,1]$ such that
$D(\frac{\epsilon^{\frac{-q'}{\lambda^{*}-q'}}}{\epsilon^{\frac{-q'}{\lambda^{*}-q'}}+\bar{\epsilon}^{\frac{-q'}{\lambda^{*}-q'}}})=R$,
for $R\le D(\frac{\epsilon^{q}}{\epsilon^{q}+\bar{\epsilon}^{q}})$,
the optimal $\lambda$ is $1$, and for $R\ge D(\epsilon)$, the optimal
$\lambda$ is $0$.
\item For any $q\in[1,\infty]$,
\begin{align*}
 & \lim_{n\to\infty}\frac{1}{n}\inf_{f:[e^{nR}]\to\mathcal{X}^{n}}D_{q}(P_{Y}^{\otimes n}\|Q_{Y^{n}})=\sup_{\lambda\ge0}\lambda\left(D_{\frac{1}{1+\lambda}}(\epsilon)-R\right),
\end{align*}
where for $R<D(\epsilon)$, the optimal $\lambda$ is the unique $\lambda^{*}\ge0$
such that $D(\frac{\epsilon^{\frac{1}{1+\lambda^{*}}}}{\epsilon^{\frac{1}{1+\lambda^{*}}}+\bar{\epsilon}^{\frac{1}{1+\lambda^{*}}}})=R$,
and for $R\ge D(\epsilon)$, the optimal $\lambda$ is $0$.
\end{enumerate}
\end{cor}

Using Theorem \ref{thm:ResolvabilityRates} (or Corollary \ref{cor:asymptotics}),
we obtain the following corollary. 
\begin{cor}[Binary Rényi Resolvability Rates]
\label{thm:ResolvabilityRates-1} Consider $P_{Y|X}=\mathrm{BSC}(\epsilon)$
and $P_{Y}=\Bern(1/2)$. It holds that 
\begin{align}
R_{q}(P_{Y|X},P_{Y}) & =\begin{cases}
1-H_{q}(\epsilon), & q\in(1,\infty]\\
1-H(\epsilon), & q=(0,1]\\
0, & q=0
\end{cases},\label{eq:-35}
\end{align}
and 
\begin{align}
\hat{R}_{q}(P_{Y|X},P_{Y}) & =\begin{cases}
1-H(\epsilon), & q\in(0,\infty]\\
0, & q=0.
\end{cases}.\label{eq:-34}
\end{align}
\end{cor}
The result in \eqref{eq:-34} is new, but the result in \eqref{eq:-35}
is not. The result in \eqref{eq:-35} for $q\in[0,2]\cup\{\infty\}$
was proven by the present author and Tan \cite{yu2019renyi}, the
result for $q\in(2,\infty)\cap\mathbb{N}$ was proven by Samorodnitsky
\cite{samorodnitsky2022some}, and the result for $q\in(2,\infty)$
was independently proven by Pathegama and Barg \cite{pathegama2023smoothing}.

\section{$q$-Stability }

\label{sec:Stability}As mentioned in the introduction, the $q$-stability
problem concerns the following question: For a measurable event $A$,
if the probability $P_{X}^{\otimes n}(A)$ is given, then how large
and how small could the $q$-norm of the noisy version, $\Vert P_{X|Y}^{\otimes n}(A|\cdot)\Vert_{q}$,
be? The connection between this problem and the $q$-stability problem
is as follows. 

Consider a joint distribution $P_{XY}$. Denote $Q_{X^{n}}:=P_{X}^{\otimes n}(\cdot|A).$
Then, 
\begin{align}
-\log\Vert P_{X|Y}^{\otimes n}(A|\cdot)\Vert_{q} & =-\log P_{X}^{\otimes n}(A)-\frac{1}{q'}D_{q}(Q_{Y^{n}}\|P_{Y}^{\otimes n})\label{eq:-82}\\
 & =-\log P_{X}^{\otimes n}(A)+D_{1-q}(P_{Y}^{\otimes n}\|Q_{Y^{n}}).\label{eq:Stability-Resolvability}
\end{align}
Hence, the $q$-stability problem is equivalent to a variant of Rényi
resolvability problem, in which we minimize or maximize the Rényi
divergence of the output distributions under the condition that the
input distribution is set to $P_{X}^{\otimes n}(\cdot|A)$ for some
set $A$ of probability $P_{X}^{\otimes n}(A)=e^{-n\alpha}$. In particular,
if $P_{X}$ is the uniform distribution over $\mathcal{X}$, then
the $q$-stability problem on minimizing $\Vert P_{X|Y}^{\otimes n}(A|\cdot)\Vert_{q}$
for $q>1$ and maximizing $\Vert P_{X|Y}^{\otimes n}(A|\cdot)\Vert_{q}$
for $q<1$ is equivalent to the Rényi resolvability problem with the
coding function $f$ restricted to be a one-to-one map. 

It was known that maximizing $\Vert P_{X|Y}^{\otimes n}(A|\cdot)\Vert_{q}$
for $q>1$ and minimizing $\Vert P_{X|Y}^{\otimes n}(A|\cdot)\Vert_{q}$
for $q<1$ are in fact respectively equivalent to the BL inequality
and its reverse in some sense, and hence, sharp dimension-free bounds
were already obtained; see e.g., \cite{yu2021strong}. In this paper,
we focus on the other direction, i.e., minimizing $\Vert P_{X|Y}^{\otimes n}(A|\cdot)\Vert_{q}$
for $q>1$ and maximizing $\Vert P_{X|Y}^{\otimes n}(A|\cdot)\Vert_{q}$
for $q<1$, or equivalently, minimizing $D_{q}(Q_{Y^{n}}\|P_{Y}^{\otimes n})$
for $q>0$ and minimizing $D_{1-q}(P_{Y}^{\otimes n}\|Q_{Y^{n}})$
for $q<0$. By the connections between the $q$-stability problem
and the Rényi resolvability problem, we will derive sharp dimension-free
bounds for them. It will be shown in the next section that the part
of the $q$-stability problem considered here is equivalent to a new
kind of inequalities, called anti-contractivity inequalities.

\subsection{Asymptotics and Dimension-Free Bounds}

We now provide the sharp dimension-free lower bound on $\Vert P_{X|Y}^{\otimes n}(A|\cdot)\Vert_{q}$
for $q\ge1$ and the sharp dimension-free upper bound on $\Vert P_{X|Y}^{\otimes n}(A|\cdot)\Vert_{q}$
for $q<1$ The proof of Theorem \ref{thm:strongqs} is provided in
Appendix \ref{sec:Proof-of-Theorem-strongqs}. Define 
\begin{align}
\eta_{r}(Q_{XY}) & :=D(Q_{Y|X}\|P_{Y|X}|Q_{X})-\frac{1}{r}D(Q_{Y}\|P_{Y}),\label{eq:-62}\\
\hat{\eta}_{r}(Q_{XY},\alpha) & :=D(Q_{Y|X}\|P_{Y|X}|Q_{X})-\frac{1}{r}D(Q_{Y|X}\|P_{Y}|Q_{X})\nonumber \\
 & \qquad\qquad\qquad-\frac{1}{r}(\alpha-H(Q_{X},P_{X})).\label{eq:-63}
\end{align}

\begin{thm}[Strong $q$-Stability Theorem]
\label{thm:strongqs} Let $\mathcal{X},\mathcal{Y}$ be finite sets.
For any $n\ge1$ and any subsets $A\subseteq\mathcal{X}^{n}$ with
$\alpha:=-\frac{1}{n}\log P_{X}^{\otimes n}(A)$, the following hold. 
\begin{enumerate}
\item For $q\in[1,\infty]$, 
\begin{align}
 & -\frac{1}{n}\log\Vert P_{X|Y}^{\otimes n}(A|\cdot)\Vert_{q}\nonumber \\
 & \le\alpha+\max_{Q_{X}:D(Q_{X}\|P_{X})\le\alpha\le H(Q_{X},P_{X})}\min_{Q_{Y|X}}\min\{\hat{\eta}_{q'}(Q_{XY},\alpha),\eta_{q'}(Q_{XY})\}.\label{eq:-22}
\end{align}
\item For $q\in(0,1)$, 
\begin{align}
 & -\frac{1}{n}\log\Vert P_{X|Y}^{\otimes n}(A|\cdot)\Vert_{q}\nonumber \\
 & \ge\alpha+\min_{Q_{X}:D(Q_{X}\|P_{X})\le\alpha\le H(Q_{X},P_{X})}\min_{Q_{Y|X}}\max\{\hat{\eta}_{q'}(Q_{XY},\alpha),\eta_{q'}(Q_{XY})\}.\label{eq:-84}
\end{align}
\item For $q\in[-\infty,0)$, 
\begin{align}
 & -\frac{1}{n}\log\Vert P_{X|Y}^{\otimes n}(A|\cdot)\Vert_{q}\nonumber \\
 & \ge\alpha+\max_{Q_{Y}}\min_{Q_{X|Y}:I_{Q}(X;Y)\le H(Q_{X},P_{X})-\alpha}\eta_{q'}(Q_{XY})+[D(Q_{X}\|P_{X})-\alpha]^{+}.\label{eq:-23}
\end{align}
\item Moreover, the inequalities in Statements 1-3 are asymptotically sharp
in the sense that for each inequality in Statements 1-3, there exists
a sequence of sets $A_{n}$ such that $-\frac{1}{n}\log P_{X}^{\otimes n}(A_{n})\to\alpha$
and both sides of the inequality coincide asymptotically as $n\to\infty$. 
\end{enumerate}
\end{thm}
\begin{rem}
Dual formulas for the bounds in this theorem can be obtained by using
Proposition \ref{prop:The-expressions-in}.
\end{rem}
In the literature, existing works only focus on the other direction,
i.e., upper bounding on $\Vert P_{X|Y}^{\otimes n}(A|\cdot)\Vert_{q}$
for $q\ge1$ or lower bounding on $\Vert P_{X|Y}^{\otimes n}(A|\cdot)\Vert_{q}$
for $q<1$; see e.g., \cite{witsenhausen1975sequences,borell1985geometric,mossel2005coin,mossel2006non,ODonnell14analysisof,kamath2016non,ordentlich2020note,yu2021non,li2021boolean,kirshner2021moment,yu2021strong}.
In particular, exponentially sharp bounds for this case were provided
in \cite{kirshner2021moment,yu2021strong}. Theorem \ref{thm:strongqs}
complements the existing works (more specifically, the exponentially
sharp bounds in \cite{kirshner2021moment,yu2021strong}), since it
focus on lower bounding on $\Vert P_{X|Y}^{\otimes n}(A|\cdot)\Vert_{q}$
for $q\ge1$ and upper bounding on $\Vert P_{X|Y}^{\otimes n}(A|\cdot)\Vert_{q}$
for $q<1$ and provides exponentially sharp bounds for these cases. 

The noise stability result in Theorem \ref{thm:strongqs} is related
to the Rényi resolvability result in Theorem \ref{thm:Resolvability}
in the following way. If $P_{X}$ is the uniform distribution over
$\mathcal{X}$, then the inequality in \eqref{eq:-22} reduces to
that 
\begin{align}
 & -\frac{1}{n}\log\Vert P_{X|Y}^{\otimes n}(A|\cdot)\Vert_{q}\nonumber \\
 & \le\alpha+\max_{Q_{X}:\alpha_{0}-\alpha\le H_{Q}(X)}\min\{-\frac{1}{q'}\mathbb{E}_{Q_{X}}[D_{q}(P_{Y|X}\|P_{Y})]+\frac{1}{q'}(\alpha_{0}-\alpha),\nonumber \\
 & \qquad\min_{Q_{Y|X}}D(Q_{Y|X}\|P_{Y|X}|Q_{X})-\frac{1}{q'}D(Q_{Y}\|P_{Y})\},\label{eq:-83}
\end{align}
where $\alpha_{0}:=\log|\mathcal{X}|$. Note that for this case, if
we substitute the expression for the Rényi resolvability in \eqref{eq:bound1}
into \eqref{eq:-82}, we will obtain a formula similar to \eqref{eq:-83}
but without the constraint $\alpha_{0}-\alpha\le H_{Q}(X)$. This
difference comes from that in the Rényi resolvability problem, the
coding function $f$ is not necessarily one-to-one. Similar phenomena
can be observed for the inequalities in \eqref{eq:-84} and \eqref{eq:-23}. 

\subsection{Binary Example}

In this subsection, the bases of logarithms are set to $2$. We next
focus on binary distributions and channels, and provide explicit expressions
for the results derived in previous subsections. We consider $P_{XY}=\DSBS(\epsilon)$.
That is, $P_{X}=\Bern(1/2)$ and $P_{Y|X}=\mathrm{BSC}(\epsilon)$. 

Define $\kappa=(\frac{\overline{\epsilon}}{\epsilon})^{2}$. Denote
the binary versions of joint relative entropy and optimal-transport
divergence as 
\begin{align*}
D_{a,b}(\delta\|\epsilon) & :=D(\begin{bmatrix}\frac{\overline{a}+\overline{b}-\delta}{2} & \frac{b-a+\delta}{2}\\
\frac{a-b+\delta}{2} & \frac{a+b-\delta}{2}
\end{bmatrix}\|\DSBS(\epsilon)),\\
\DD(a,b\|\epsilon) & :=\min_{|a-b|\le\delta\le\min\{a+b,\overline{a}+\overline{b}\}}D_{a,b}(\delta\|\epsilon)=D_{a,b}(\delta^{*}\|\epsilon),
\end{align*}
where 
\begin{equation}
\delta^{*}=\frac{\sqrt{((\kappa-1)(a+b)+1)^{2}-4\kappa(\kappa-1)ab}-1}{\kappa-1}.\label{eq:opt_p-1-1}
\end{equation}
It is easy to verify that $|a-b|\le\delta^{*}\le a*b\le\min\{a+b,\overline{a}+\overline{b}\}$.
  Denote 
\[
Q^{*}:=\begin{bmatrix}\frac{\overline{a}+\overline{b}-\delta^{*}}{2} & \frac{b-a+\delta^{*}}{2}\\
\frac{a-b+\delta^{*}}{2} & \frac{a+b-\delta^{*}}{2}
\end{bmatrix},
\]
which attains $\DD(a,b)$. 

Define the information-constrained optimal-transport  divergence
as 
\[
\D_{R}(Q_{X},Q_{Y}\|P_{XY}):=\min_{Q_{XY}\in\Pi(Q_{X},Q_{Y}):I_{Q}(X;Y)\le R}D(Q_{XY}\|P_{XY}).
\]
For $\DSBS(\epsilon)$, 
\begin{equation}
\D_{R}(\Bern(a),\Bern(b)\|\DSBS(\epsilon))=\D_{R}(a,b\|\epsilon):=\begin{cases}
D_{a,b}(\delta^{*}\|\epsilon), & R\ge I_{Q^{*}}(X;Y)\\
D_{a,b}(\hat{\delta}\|\epsilon), & R<I_{Q^{*}}(X;Y)
\end{cases}\label{eq:-40-1}
\end{equation}
where $\hat{\delta}$ is the unique $\delta$ such that $\delta^{*}\le\delta\le a*b$
and 
\begin{align*}
H(\begin{bmatrix}\frac{\overline{a}+\overline{b}-\delta}{2} & \frac{b-a+\delta}{2}\\
\frac{a-b+\delta}{2} & \frac{a+b-\delta}{2}
\end{bmatrix}) & =H(a)+H(b)-R.
\end{align*}

We evaluate the bounds in Theorem \ref{thm:strongqs} for the binary
case as follows. 
\begin{cor}[Binary $q$-Stability]
\label{cor:asymptotics-1} Consider $P_{XY}=\DSBS(\epsilon)$. Let
$n\ge1$ and $A\subseteq\{0,1\}^{n}$ with $\alpha:=-\frac{1}{n}\log_{2}P_{X}^{\otimes n}(A)$. 
\begin{enumerate}
\item For $q\in[1,\infty]$, we have 
\[
-\frac{1}{n}\log_{2}\Vert P_{X|Y}^{\otimes n}(A|\cdot)\Vert_{q}\le\alpha-\frac{1}{q'}[\alpha-H_{q}(\epsilon)]^{+}.
\]
\item For $q\in(0,1)$, we have 
\begin{align*}
 & -\frac{1}{n}\log_{2}\Vert P_{X|Y}^{\otimes n}(A|\cdot)\Vert_{q}\\
 & \ge\alpha+\min_{a:D(a\|1/2)\le\alpha}\begin{cases}
-\frac{1}{q'}(D_{q}(\epsilon)-(1-\alpha)), & 1-\alpha<I_{Q}(X;Y)\\
{\displaystyle \min_{b}\{\D_{1-\alpha}(a,b\|\epsilon)-\frac{1}{q'}D(b)-D(a)\}}, & 1-\alpha>I_{Q}(X;Y)
\end{cases},
\end{align*}
where $\D_{1-\alpha}(a,b\|\epsilon)$ was defined in \eqref{eq:-40-1},
and $Q_{XY}$ is a joint distribution induced by $Q_{X}=\Bern(a)$
and $Q_{Y|X}=\BSC(\frac{\epsilon^{q}}{\epsilon^{q}+\bar{\epsilon}^{q}})$
with $\overline{\epsilon}:=1-\epsilon$. 
\item For $q\in[-\infty,0)$, we have 
\begin{align*}
 & -\frac{1}{n}\log_{2}\Vert P_{X|Y}^{\otimes n}(A|\cdot)\Vert_{q}\\
 & \ge\alpha+\max_{b}\min_{a}\D_{1-\alpha}(a,b\|\epsilon)-D(a)-\frac{1}{q'}D(b)+[D(a)-\alpha]^{+}.
\end{align*}
\item Moreover, the inequalities in Statements 1-3 are asymptotically sharp
as $n\to\infty$. 
\end{enumerate}
\end{cor}

\section{Anti-contractivity }

\label{subsec:Single-Function-Version}

Recall the forward and reverse contractivity-type inequalities in
\eqref{eq:FIFBL-2} and \eqref{eq:FIRBL-3} and the forward and reverse
contractivity-type inequalities in \eqref{eq:FAC} and \eqref{eq:RAC}.
It is well known that the forward and reverse contractivity-type
inequalities are in fact equivalent to the forward and reverse BL-type
inequalities for $q\ge1$ and $q<1$ respectively.   Here, for
any two real numbers $p,q$, the forward and reverse BL-type inequalities
are respectively

\begin{align}
\langle f,g\rangle & \le e^{-\underline{C}}\Vert f\Vert_{p}\Vert g\Vert_{q},\forall f\ge0,g\ge0,\label{eq:FBL}\\
\langle f,g\rangle & \geq e^{-\overline{C}}\Vert f\Vert_{p}\Vert g\Vert_{q},\forall f\ge0,g\ge0.\label{eq:RBL}
\end{align}
These equivalences are consequences of the following observations.
By Hölder's inequality and it reverse version, for any $f:\mathcal{X}\to[0,\infty)$,
\begin{equation}
\Vert P_{X|Y}(f)\Vert_{q}=\begin{cases}
{\displaystyle \sup_{g\ge0}\;\frac{\langle f,g\rangle}{\Vert g\Vert_{q'}}}, & q\ge1\vspace{.03in}\\
{\displaystyle \inf_{g\ge0}\;\frac{\langle f,g\rangle}{\Vert g\Vert_{q'}},} & q\le1
\end{cases}.\label{eq:-24}
\end{equation}
Hence, the inequality in \eqref{eq:FIFBL-2} for $q\ge1$ holds if
and only if~\eqref{eq:FBL} holds but with $q$ in the latter replaced
by its Hölder conjugate $q'=\frac{q}{q-1}$. Similarly, the inequality
in \eqref{eq:FIRBL-3} for $q\le1$ holds if and only if~\eqref{eq:RBL}
holds but with $q$ in the latter replaced by its Hölder conjugate
$q'$. In other words, if we denote the optimal exponents  in \eqref{eq:FIFBL-2}
and \eqref{eq:FIRBL-3} respectively as $\underline{\Gamma}_{p,q}$
and $\overline{\Gamma}_{p,q}$, then for $q\ge1$, 
\begin{align*}
\underline{\Gamma}_{p,q} & =-\log\sup_{f\ge0,g\ge0}\;\frac{\langle f,g\rangle}{\Vert f\Vert_{p}\Vert g\Vert_{q'}},
\end{align*}
and for $q\ge1$, 
\[
\overline{\Gamma}_{p,q}=-\log\inf_{f\ge0,g\ge0}\;\frac{\langle f,g\rangle}{\Vert f\Vert_{p}\Vert g\Vert_{q'}}.
\]
Here, we run over all $f\ge0,g\ge0$ such that the denominators in
the objective functions are nonzero. 

The optimal exponents in the anti-contractivity inequalities in \eqref{eq:FAC}
and \eqref{eq:RAC} are still respectively denoted by $\underline{\Gamma}_{p,q}$
and $\overline{\Gamma}_{p,q}$ (but for values of $q$ different from
those in \eqref{eq:FIFBL-2} and \eqref{eq:FIRBL-3}). These two anti-contractivity
inequalities are not equivalent to BL-type inequalities anymore. In
fact, we can write for $q\ge1$, 
\[
\overline{\Gamma}_{p,q}=-\log\inf_{f\ge0}\;\frac{\Vert P_{X|Y}(f)\Vert_{q}}{\Vert f\Vert_{p}}=-\log\inf_{f\ge0}\sup_{g\ge0}\;\frac{\langle f,g\rangle}{\Vert f\Vert_{p}\Vert g\Vert_{q'}},
\]
and for $q\le1$, 
\begin{align*}
\underline{\Gamma}_{p,q} & =-\log\sup_{f\ge0}\;\frac{\Vert P_{X|Y}(f)\Vert_{q}}{\Vert f\Vert_{p}}=-\log\sup_{f\ge0}\inf_{g\ge0}\;\frac{\langle f,g\rangle}{\Vert f\Vert_{p}\Vert g\Vert_{q'}}.
\end{align*}

As mentioned in the introduction, if $f$ is set to a Boolean-valued
function, i.e., $f=\bone_{A}$, then the contractivity-type and anti-contractivity
inequalities reduce to the $q$-stability problem. So, any contractivity-type
and anti-contractivity inequalities can yield bounds on the $q$-stability
problem. In fact, the implication in the opposite direction is also
true. This is the intuition behind the main results in this section. 

\subsection{Information-Theoretic Characterizations}

Define the following linear combination of OT divergence and relative
entropies 
\begin{equation}
\theta_{p,q}(Q_{X},Q_{Y}):=\rvD(Q_{X},Q_{Y}\|P_{XY})-\frac{1}{p}\,D(Q_{X}\|P_{X})-\frac{1}{q}\,D(Q_{Y}\|P_{Y}).
\end{equation}
By minimax theorems, the anti-contractivity exponents also can be
written in the following alternative information-theoretic forms. 

\textbf{Condition 1:} For $p<0,q<0$, we assume that there is some
set $A$ such that $0<P_{X}(A)<1$ and $P_{X|Y}(A|y)<1$ for $P_{Y}$-almost
all $y$. 
\begin{thm}[Information-Theoretic Characterizations]
\label{thm:ITcharacterization} Let $\calX,\calY$ be Polish spaces.
Let $p,q\in\mathbb{R}\backslash\{0\}$. Then, it holds that 
\begin{align}
\underline{\Gamma}_{p,q} & =\begin{cases}
{\displaystyle \inf_{Q_{X},Q_{Y}}\theta_{p,q'}(Q_{X},Q_{Y})}, & p,q>0\\
{\displaystyle \sup_{Q_{Y}}\;\inf_{Q_{X}}\;\theta_{p,q'}(Q_{X},Q_{Y})}, & q<0<p\\
{\displaystyle -\infty}, & p<0,q>0,\\
 & \textrm{or }p<0,q<0,\textrm{Condition }1
\end{cases},\label{eq:FIInfChLambdaUnderline-1}
\end{align}
and 
\begin{align}
\overline{\Gamma}_{p,q} & =\begin{cases}
{\displaystyle \sup_{Q_{X},Q_{Y}}\;\theta_{p,q'}(Q_{X},Q_{Y})}, & p>0>q\vspace{.03in}\\
{\displaystyle \sup_{Q_{X}}\;\inf_{Q_{Y}}\;\theta_{p,q'}(Q_{X},Q_{Y})}, & p,q>0\vspace{.03in}\\
{\displaystyle \sup_{Q_{Y}}\;\inf_{Q_{X}}\;\theta_{p,q'}(Q_{X},Q_{Y})}, & p,q<0\vspace{.03in}\\
{\displaystyle 0}, & p\!<\!0<q
\end{cases}.\label{eq:FIInfChLambdaOverline-1}
\end{align}
In fact, for $p>1$ and $p>q$, $\underline{\Gamma}_{p,q}=0$, and
for $p<1$ and $p<q$, $\overline{\Gamma}_{p,q}=0.$ 
\end{thm}
The expressions for contractivity-type exponents (i.e., the BL exponents)
in the theorem above (i.e., \eqref{eq:FIInfChLambdaUnderline-1} for
$q\ge1$ and \eqref{eq:FIInfChLambdaOverline-1} for $q\le1$) are
existing results \cite{carlen2009subadditivity,nair2014equivalent,kamath2015reverse,beigi2016equivalent,liu2016brascamp,liu2018forward},
while the expressions for anti-contractivity exponents (i.e., \eqref{eq:FIInfChLambdaUnderline-1}
for $q<1$ and \eqref{eq:FIInfChLambdaOverline-1} for $q>1$) are
new. The proof for anti-contractivity exponents is provided in Appendix
\ref{sec:Proof-of-Theorem-IT}. 

Another information-theoretic characterization of contractivity-type
and anti-contractivity exponents is expressed in terms of Rényi divergences.
We may assume, by homogeneity, that $\Vert f\Vert_{1}=1$. Then, without
loss of generality, we can write $f=\frac{\mathrm{d}Q_{X}}{\mathrm{d}P_{X}}$
for some probability measures $Q_{X}\ll P_{X}$. Moreover, we require
$f<\infty$. We then rewrite 
\begin{align*}
\log\|f\|_{p} & =\frac{1}{p'}D_{p}(Q_{X}\|P_{X})=-D_{1-p}(P_{X}\|Q_{X}),\\
\log\|P_{X|Y}(f)\|_{q} & =\frac{1}{q'}D_{q}(Q_{Y}\|P_{Y})=-D_{1-q}(P_{Y}\|Q_{Y}),
\end{align*}
where $Q_{Y}:=Q_{X}\circ P_{Y|X}$. So, the following characterizations
hold, which extend the one for hypercontractivity given in \cite{raginsky2013logarithmic}. 
\begin{thm}[Information-Theoretic Characterizations based on Rényi Divergences]
\label{thm:ITcharacterization-1} Let $\calX,\calY$ be Polish spaces.
For $p,q\in\mathbb{R}\backslash\{0\}$, it holds that $\underline{\Gamma}_{p,q}=\inf_{Q_{X}}a-b$
and $\overline{\Gamma}_{p,q}=\sup_{Q_{X}}a-b$, where $a=\frac{1}{p'}D_{p}(Q_{X}\|P_{X})=-D_{1-p}(P_{X}\|Q_{X})$,
and $b=\frac{1}{q'}D_{q}(Q_{Y}\|P_{Y})=-D_{1-q}(P_{Y}\|Q_{Y})$. 
\end{thm}
This theorem illustrates the relation between the contractivity-type/anti-contractivity
inequalities and the data-processing inequalities. The data-processing
inequalities here refer to the inequalities quantify the best tradeoff
between the $p$-Rényi divergence $D_{p}(Q_{X}\|P_{X})$ of the input
distributions and the $q$-Rényi divergence $D_{q}(Q_{Y}\|P_{Y})$
of the output distributions. Detailed information on this relation
could be found in \cite{raginsky2013logarithmic,yu2021strong}.

\subsection{Asymptotics and Dimension-Free Bounds}

We define the exponents for the $n$-product distribution $P_{XY}^{\otimes n}$
as 
\begin{align*}
\underline{\Gamma}_{p,q}^{(n)} & :=\frac{1}{n}\underline{\Gamma}_{p,q}(P_{XY}^{\otimes n})\textrm{ and }\overline{\Gamma}_{p,q}^{(n)}:=\frac{1}{n}\overline{\Gamma}_{p,q}(P_{XY}^{\otimes n}),
\end{align*}
and their limits as $\underline{\Gamma}_{p,q}^{(\infty)}$ and $\overline{\Gamma}_{p,q}^{(\infty)}$.
Observe that by the product construction, the optimal exponents $n\overline{\Gamma}_{p,q}^{(n)}$
is superadditive in $n$. So, by Fekete's lemma, $\overline{\Gamma}_{p,q}^{(\infty)}=\sup_{n\ge1}\overline{\Gamma}_{p,q}^{(n)}$.
Similarly, $\underline{\Gamma}_{p,q}^{(\infty)}=\inf_{n\ge1}\underline{\Gamma}_{p,q}^{(n)}$.
These imply that if we have expressions of $\overline{\Gamma}_{p,q}^{(\infty)}$
and $\underline{\Gamma}_{p,q}^{(\infty)}$, then we immediately obtain
sharp dimension-free bounds 
\[
\overline{\Gamma}_{p,q}^{(n)}\le\overline{\Gamma}_{p,q}^{(\infty)}\textrm{ and }\underline{\Gamma}_{p,q}^{(n)}\ge\underline{\Gamma}_{p,q}^{(\infty)}
\]
for all $n$. To this end, we only need to focus on the asymptotic
case.

It is well known that contractivity-type inequalities satisfy the
tensorization property. That is, $\underline{\Gamma}_{p,q}^{(n)}=\underline{\Gamma}_{p,q}$
for $q\ge1$, and $\overline{\Gamma}_{p,q}^{(n)}=\overline{\Gamma}_{p,q}$
for $q<1$. However, anti-contractivity inequalities do not satisfy
tensorization anymore (i.e., their exponents are not additive, but
subadditive or supperadditive). The asymptotics of the anti-contractivity
exponents are characterized in the following theorem. The proof is
provided in Appendix \ref{sec:Proof-of-Theorem}. 
\begin{thm}[Anti-contractivity Exponents]
\label{thm:anticontractivity} Let $\calX,\calY$ be finite spaces.
For $p,q\in[-\infty,\infty]\backslash\{0\}$, it holds that 
\begin{align}
\underline{\Gamma}_{p,q}^{(n)}\ge\underline{\Gamma}_{p,q}^{(\infty)} & =\begin{cases}
{\displaystyle \min_{x}\frac{1}{p'}\log\frac{1}{P_{X}(x)}-\frac{1}{q'}D_{q}(P_{Y|X=x}\|P_{Y})}, & 0<p\le q<1\\
{\displaystyle \min_{Q_{XY}}\eta_{q'}(Q_{XY})+\frac{1}{p'}(H(Q_{X},P_{X})-I_{Q}(X;Y))}, & 0<q<p<1\\
{\displaystyle \max_{Q_{Y}}\min_{Q_{X|Y}}\eta_{q'}(Q_{XY})+\frac{1}{p'}(H(Q_{X},P_{X})-I_{Q}(X;Y)),} & q<0<p<1\\
0, & q<1\le p\\
{\displaystyle -\infty,} & p<0,\,q<1
\end{cases}\label{eq:RAH-1}
\end{align}
and 
\begin{align}
\overline{\Gamma}_{p,q}^{(n)}\le\overline{\Gamma}_{p,q}^{(\infty)} & =\begin{cases}
{\displaystyle \max_{x}\frac{1}{p'}\log\frac{1}{P_{X}(x)}-\frac{1}{q'}D_{q}(P_{Y|X=x}\|P_{Y})}, & 1\le q\le p\\
\max_{Q_{X}}\frac{1}{p'}H(Q_{X},P_{X})-\frac{1}{p'}\mathbb{E}_{Q_{X}}[D_{q}(P_{Y|X}\|P_{Y})]\\
\qquad+(1-\frac{q'}{p'})\min_{Q_{Y|X}}\eta_{q'}(Q_{XY}), & 1\le p<q\\
{\displaystyle 0}, & p\!<\!1\le q
\end{cases},\label{eq:FAH-1}
\end{align}
where $\eta_{q'}$ is defined in \eqref{eq:-62}. 
\end{thm}

We next derive dual formulas for the expressions in the theorem above.
The proof is similar to that of Proposition \ref{prop:The-expressions-in},
and omitted here. 
\begin{prop}[Dual Formulas]
 \label{prop:For-both-the} The following hold. 
\begin{enumerate}
\item For both the second and third clauses in \eqref{eq:RAH-1} (i.e.,
for $0<p<1,q<p$),  
\begin{equation}
\underline{\Gamma}_{p,q}^{(\infty)}={\displaystyle \min_{S_{X}}-\frac{1}{q}\log\mathbb{E}_{P_{Y}}\big[\mathbb{E}_{S_{X}}[\frac{P_{X|Y}^{p}}{P_{X}}]^{q/p}\big]}.\label{eq:q0p1}
\end{equation}
\item For the second clause in \eqref{eq:FAH-1} (i.e., for $1\le p<q$),
 
\begin{align}
\overline{\Gamma}_{p,q}^{(\infty)} & =\min_{S_{Y}}\max_{x}\frac{1}{p'}\log\frac{1}{P_{X}(x)}-\frac{1}{p'}D_{q}(P_{Y|X=x}\|P_{Y})\nonumber \\
 & \qquad-(1-\frac{q'}{p'})\log\mathbb{E}_{P_{Y|X=x}}[(\frac{S_{Y}}{P_{Y}})^{1/q'}].\label{eq:-85}
\end{align}
\end{enumerate}
\end{prop}

The connection between the anti-contractivity inequalities in Theorem
\ref{thm:anticontractivity} and the bounds for the noise stability
in Theorem \ref{thm:strongqs} is as follows. By checking our proofs,
it is seen that Theorem \ref{thm:anticontractivity} is proven by
invoking Theorem \ref{thm:strongqs} since indicator functions are
asymptotic extremers in anti-contractivity inequalities. On the other
hand, setting $f$ to indicator functions $\bone_{A}$, we can obtain
asymptotically sharp bounds for the noise stability with certain value
of $P_{X}^{\otimes n}(A)$. 

\subsection{Binary Example and Others }

Applying Theorem \ref{thm:anticontractivity} to the DSBS, we obtain
sharp dimension-free anti-contractivity inequalities for the DSBS.
The proof is provided in Appendix \ref{sec:Proof-of-Corollary}. 
\begin{cor}[Binary Anti-contractivity Inequalities]
\label{cor:binary} Let $P_{XY}$ be the DSBS with crossover probability
$\epsilon\in(0,1/2)$. For $p,q\in[-\infty,\infty]\backslash\{0\}$,
it holds that 
\begin{align}
\underline{\Gamma}_{p,q}^{(n)}\ge\underline{\Gamma}_{p,q}^{(\infty)} & =\begin{cases}
\frac{1}{q'}H_{q}(\epsilon)-\frac{1}{q'}+\frac{1}{p'}, & 0<p\le q<\!1\\
\frac{1}{p'}H_{p}(\epsilon), & 0<p<\!1,\,q<p\\
-\infty, & p<0,\,q\!<\!1\\
0, & q<1<p
\end{cases}\label{eq:RAH}
\end{align}
and 
\begin{align}
\overline{\Gamma}_{p,q}^{(n)}\le\overline{\Gamma}_{p,q}^{(\infty)} & =\begin{cases}
\frac{1}{q'}H_{q}(\epsilon)-\frac{1}{q'}+\frac{1}{p'}, & 1\le q<p\\
\frac{1}{p'}H_{q}(\epsilon), & 1\le p\le q\\
0, & p\!<\!1\le q
\end{cases}.\label{eq:FAH}
\end{align}
Moreover, the second clause in \eqref{eq:RAH} (even for $q<0$)
is with high probability asymptotically attained by the indicator
of a random subset of $\{0,1\}^{n}$ of size $2^{n(1-H(\hat{\epsilon}))}$
with $\hat{\epsilon}=\frac{\epsilon^{p}}{\epsilon^{p}+(1-\epsilon)^{p}}$;
 the second clause in \eqref{eq:FAH} is with high probability asymptotically
attained by the indicator of a random subset of $\{0,1\}^{n}$ of
size $2^{n(1-H_{q}(\epsilon))}$; the first and third clauses in \eqref{eq:RAH}
and the first clause in \eqref{eq:FAH} are attained by the indicator
of a single point; and the last clause in \eqref{eq:RAH} and the
last clause in \eqref{eq:FAH} are attained by positive constant functions.
%Moreover, the   indicator of a random subset of $\mathcal{T}_{\Bern(1/2+o_{n}(1))}^{(n)}$ of size $ 2^{n(1-H_{q}(\epsilon))}$ asymptotically attains the expression in the first  clause  in  \eqref{eq:FAH};   the   indicator of a random subset of $\mathcal{T}_{\Bern(1/2+o_{n}(1))}^{(n)}$ of size $2^{n( 1-H_{q}( \hat{\epsilon}))}$ with $\hat{\epsilon}=\frac{\epsilon^p}{\epsilon^p+(1-\epsilon)^p}$  asymptotically attains the expression in the first  clause  in  \eqref{eq:RAH} (even for $q<0$); and  the indicator of a single point  asymptotically attains the expressions in the second clauses in \eqref{eq:RAH} and \eqref{eq:FAH}. 
\end{cor}
 The special case of \eqref{eq:FAH} with $p=q\ge0$ was first shown
by Samorodnitsky \cite{samorodnitsky2022some}. The corollary above
is the opposite part of the classic hyercontractivity inequalities
(or more generally, the single function version of BL inequalities)
\cite{bonami1968ensembles,kiener1969uber,schreiber1969fermeture,bonami1970etude,beckner1975inequalities,gross1975logarithmic,borell1982positivity}.
See the comparison in Section \ref{subsec:Main-Contributions}.

We now discuss anti-contractivity inequalities for other distributions.
Theorem \ref{thm:ITcharacterization} implies that for any joint distribution
$P_{XY}$, it holds that $\underline{\Gamma}_{p,q}^{(n)}=0$ for $p\ge1\ge q$,
and $\overline{\Gamma}_{p,q}^{(n)}=0$ for $p\!\le\!1\le q$. This
point can be also observed from the monotonicity of the ``norm''.
 We next consider other cases, i.e., $\underline{\Gamma}_{p,q}^{(n)}$
for $p<1,q\!\le\!1$ and $\overline{\Gamma}_{p,q}^{(n)}$ for $p>1,q\ge1$.
For these cases, we consider the case that $P_{X}$ is a continuous
distribution and $D_{q}(P_{Y|X=x}\|P_{Y})$ is finite for some $x$
(e.g., $P_{XY}$ is jointly Gaussian). Then, in this setting, $\underline{\Gamma}_{p,q}^{(n)}=-\infty$
for $p<1,q\!\le\!1$ and $\overline{\Gamma}_{p,q}^{(n)}={\displaystyle \infty}$
for $p>1,q\ge1$.  This result seems uninteresting to us. So, the
interesting case for anti-contractivity inequalities is that $P_{X}$
is discrete.

\section{Conclusion and Future Work}

\label{sec:concl}In this paper, we provided a complete solution to
the Rényi resolvability problem of Rényi orders in the entire range
$\mathbb{R}\cup\{\pm\infty\}$, in which the (normalized or unnormalized)
Rényi divergence is used to measure the level of approximation in
the channel resolvability problem. Our results generalize several
classical and recent results. Our resolvability results extend those
by Han and Verdú \cite{Han93} with the normalized KL divergence measure,
by Hayashi \cite{Hayashi06,Hayashi11} with unnormalized KL divergence
measure, by the present author and Tan \cite{yu2019renyi} with the
Rényi divergence of orders in $(0,1)\cup(1,2]\cup\{\infty\}$, and
by Samorodnitsky \cite{samorodnitsky2022some} as well as by Pathegama
and Barg \cite{pathegama2023smoothing} with the binary setting and
the Rényi parameter taking values in the range $(2,\infty)$.

We then connect the Rényi resolvability problem to the noise stability
problem, by observing that the $q$-stability of a set can be expressed
in terms of the Rényi divergence between the true output distribution
and the target distribution in a variant of the Rényi resolvability
problem. By such a connection, we provide sharp dimension-free bounds
on the $q$-stability.

We lastly relate the noise stability problem to the anti-contractivity
of a Markov operator (i.e., conditional expectation operator), where
anti-contractivity refers to as an opposite property of the well-known
contractivity (or hyercontractivity). We derive sharp dimension-free
anti-contractivity inequalities. 

All of the results in this paper are evaluated for binary distributions,
which recovers an existing result of Samorodnitsky \cite{samorodnitsky2022some}.
Our proofs in this paper are mainly based on the method of types,
especially strengthened versions of packing-covering lemmas. We expect
more applications of our results in this paper, especially applications
in analyzing Fourier spectra of Boolean functions.

As mentioned below Theorem \ref{thm:exponentforiid}, the convergence
exponent for i.i.d. codes in the channel resolvability problem with
$q>0$ is already known. It is interesting to investigate the convergence
exponent for i.i.d. codes with $q<0$. Furthermore, another but currently
seemingly hopeless open problem is to  find the optimal convergence
exponent over all codes for the channel resolvability problem for
all $q$. 

\appendices{}

\section{Useful Lemmas }

\label{sec:usefullemmas} 

\subsection{Basic Lemmas }
\begin{lem}
\cite[Lem.~2.1.2]{Dembo} \cite{yu2019renyi} \label{lem:typecovering} 
\begin{enumerate}
\item Assume $\mathcal{X}$ is a finite set. Then for any $P_{X}\in\mathcal{P}(\mathcal{X})$,
one can find a type $T_{X}^{(n)}\in\mathcal{P}_{n}(\mathcal{X})$
such that $\big|P_{X}-T_{X}^{(n)}\big|\le\frac{\left|\mathcal{X}\right|}{2n}$. 
\item Assume $\mathcal{X},\mathcal{Y}$ are finite sets. Then for any type
$T_{X}^{(n)}\in\mathcal{P}_{n}(\mathcal{X})$ and any $P_{Y|X}\in\mathcal{P}(\mathcal{Y}|\mathcal{X})$,
one can find a conditional type $T_{Y|X}^{(n)}\in\mathcal{P}_{n}\big(\mathcal{Y}|T_{X}^{(n)}\big)$
such that $\big|T_{X}^{(n)}P_{Y|X}-T_{X}^{(n)}T_{Y|X}^{(n)}\big|\le\frac{\left|\mathcal{X}\right|\left|\mathcal{Y}\right|}{2n}$. 
\end{enumerate}
\end{lem}
We also have the following property concerning the optimization over
the set of types and conditional types. 
\begin{lem}
\cite{yu2019renyi} \label{lem:minequality} 
\begin{enumerate}
\item Assume $\mathcal{X}$ is a finite set. Then for any continuous (under
TV distance) function $f:\mathcal{P}(\mathcal{X})\rightarrow\mathbb{R}$,
we have\footnote{Since $\mathcal{P}(\mathcal{X})$ and $\mathcal{P}_{n}(\mathcal{X})$
are compact (closed and bounded) and $f$ is continuous on $\mathcal{P}(\mathcal{X})$,
the infima of $\inf_{P_{X}\in\mathcal{P}(\mathcal{X})}f(P_{X})$ and
$\inf_{P_{X}\in\mathcal{P}_{n}(\mathcal{X})}f(P_{X})$ are actually
minima. } 
\begin{equation}
\lim_{n\rightarrow\infty}\min_{P_{X}\in\mathcal{P}_{n}(\mathcal{X})}f(P_{X})=\min_{P_{X}\in\mathcal{P}(\mathcal{X})}f(P_{X}).
\end{equation}
\item Assume $\mathcal{X},\mathcal{Y}$ are finite sets. Then for any continuous
function $f:\mathcal{P}(\mathcal{X}\times\mathcal{Y})\rightarrow\mathbb{R}$
and any sequence of types $T_{X}^{(n)}\in\mathcal{P}_{n}(\mathcal{X}),n\in\bbN$,
we have 
\begin{align}
 & \min_{P_{Y|X}\in\mathcal{P}_{n}(\mathcal{Y}|T_{X}^{(n)})}f(T_{X}^{(n)}P_{Y|X})\nonumber \\
 & =\min_{P_{Y|X}\in\mathcal{P}(\mathcal{Y}|\mathcal{X})}f(T_{X}^{(n)}P_{Y|X})+o_{n}(1).\label{eqn:state2}
\end{align}
\end{enumerate}
\end{lem}
\begin{lem}
\label{lem:typeequality}\cite{yu2019renyi} For any joint type $T_{XY}\in\mathcal{P}_{n}(\mathcal{X\times Y})$
and any distribution $P_{X^{n}}\in\mathcal{P}(\mathcal{X}^{n})$ (not
restricted to be i.i.d.), we have 
\begin{equation}
\sum_{y^{n}\in\mathcal{T}_{T_{Y}}}P_{X^{n}}(\mathcal{T}_{T_{X|Y}}(y^{n}))=\e^{nH_{T}(Y|X)+o(n)}P_{X^{n}}(\mathcal{T}_{T_{X}}).
\end{equation}
\end{lem}
\begin{lem}
\label{lem:norm} \cite[Problem 4.15(f)]{gallagerIT} Assume $\{a_{i}\}$
are non-negative real numbers. Then for $p\geq1$, we have 
\begin{equation}
\sum_{i}a_{i}^{p}\leq(\sum_{i}a_{i})^{p},
\end{equation}
and for $0<p\leq1$, we have 
\begin{equation}
\sum_{i}a_{i}^{p}\geq(\sum_{i}a_{i})^{p}.\label{eqn:p_ineq}
\end{equation}
\end{lem}
%\section{\label{sec:min}Proof of Lemma \ref{lem:minequality}}

\begin{rem}
Note that $(\sum_{i}a_{i}^{p})^{1/p}$ is a norm for $p\geq1$, but
not for $0<p<1$. 
\end{rem}
\begin{lem}[{{{{{\cite[Lemma 14]{Tan11_IT}}}}}}]
\label{lemma:continuity} Let $\mathcal{X}$ and $\mathcal{Y}$ be
two metric spaces and let $\calK\subseteq\mathcal{X}$ be a compact
set. Let $f:\mathcal{X}\times\mathcal{Y}\rightarrow\mathbb{R}$ be
a (jointly) continuous real-valued function. Then the function $g:\mathcal{Y}\rightarrow\mathbb{R}$,
defined as 
\begin{equation}
g(y):=\min_{x\in\calK}\,f(x,y),\quad\forall\,y\in\mathcal{Y},\label{eqn:gy}
\end{equation}
is continuous on $\mathcal{Y}$. 
\end{lem}

\subsection{Strong Packing-Covering Lemma for Constant Composition Codes }

We consider constant composition codes. Let $\mathcal{C}:=\{X^{n}(m)\}_{m\in[e^{nR}]}$
be a set (precisely, a multiset) of random sequences such that $X^{n}(m),m\in[e^{nR}]$
are drawn independently for different $m$'s and according to the
same distribution $\Unif(\mathcal{T}_{T_{X}})$. For $\epsilon>0$,
define two events on $\mathcal{C}$ as 
\begin{align}
\mathcal{B}_{1}(\epsilon|T_{XY}) & :=\Big\{\Big|\frac{\phi_{\mathcal{C}}(y^{n})}{\mathbb{E}[\phi_{\mathcal{C}}(y^{n})]}-1\Big|\leq e^{-n\epsilon},\forall y^{n}\in\mathcal{T}_{T_{Y}}\Big\},\\
\mathcal{B}_{2}(\epsilon|T_{XY}) & :=\Big\{0\le\phi_{\mathcal{C}}(y^{n})\leq e^{3n\epsilon},\forall y^{n}\in\mathcal{T}_{T_{Y}}\Big\},
\end{align}
where 
\begin{align*}
\phi_{\mathcal{C}}(y^{n}):=\phi_{\mathcal{C}}(y^{n}|T_{X|Y}) & :=|\mathcal{T}_{T_{X|Y}}(y^{n})\cap\mathcal{C}|\\
 & =\sum_{m\in[e^{nR}]}\bone\{X^{n}(m)\in\mathcal{T}_{T_{X|Y}}(y^{n})\}
\end{align*}
is the number of codewords belonging to the conditional type class
$\mathcal{T}_{T_{X|Y}}(y^{n})$. It is easy to estimate $\mathbb{E}[\phi_{\mathcal{C}}(y^{n})]$
as 
\begin{align}
\mathbb{E}[\phi_{\mathcal{C}}(y^{n})] & =e^{nR}|\mathcal{T}_{T_{X|Y}}(y^{n})|/|\mathcal{T}_{T_{X}}|\nonumber \\
 & =e^{n(R-I_{T}(X;Y)-o_{n}(1))}.\label{eq:thetaMean-2}
\end{align}
If the event $\mathcal{B}_{1}(\epsilon|T_{XY})$ occurs, then each
$y^{n}\in\mathcal{T}_{T_{Y}}$ is covered by $\mathbb{E}[\phi_{\mathcal{C}}(y^{n})](1\pm e^{-n\epsilon})$
conditional type classes $\mathcal{T}_{T_{Y|X}}(x^{n})$ with $x^{n}\in\mathcal{C}$.
In other words, for the channel $Y^{n}|X^{n}\sim\Unif(\mathcal{T}_{T_{Y|X}}(X^{n}))$,
the output is closed to the distribution $\Unif(\mathcal{T}_{T_{Y}})$
within a factor $1\pm e^{-n\epsilon}$ when the input $X^{n}\sim\Unif(\mathcal{C})$.
The following lemma provides a sufficient condition for this phenomenon
on multiple packing-covering.
\begin{lem}
\label{lem:B} Let $\epsilon>0$ and $R>3\epsilon$ be fixed. Then,
the following hold. 
\begin{enumerate}
\item It holds that 
\begin{equation}
\mathbb{P}[\mathcal{B}_{1}(\epsilon|T_{XY})]\ge1-e^{-\exp(n(\epsilon-o_{n}(1)))}.\label{eq:-4}
\end{equation}
for all joint $n$-types $T_{XY}$ such that $I_{T}(X;Y)\le R-3\epsilon$,
where $o_{n}(1)$ is a term independent of $(T_{XY},R)$ and vanishes
as $n\to\infty$. 
\item It holds that 
\begin{equation}
\mathbb{P}[\mathcal{B}_{2}(\epsilon|T_{XY})]\ge1-e^{-\exp(n(\epsilon-o_{n}(1)))}\label{eq:-5}
\end{equation}
for all joint $n$-types $T_{XY}$ such that $I_{T}(X;Y)\ge R-3\epsilon$,
where $o_{n}(1)$ is a term independent of $(T_{XY},R)$ and vanishes
as $n\to\infty$. 
\end{enumerate}
\end{lem}
\begin{rem}
This lemma is essentially a soft-covering lemma under the (forward
and reverse) $\infty$-Rényi divergence in which the channel input
is a uniform distribution on a type class and the channel corresponds
to a conditional uniform distribution on a family of conditional type
classes. A similar result but for a uniform distribution on a typical
set and a conditional uniform distribution on a family of conditional
typical sets is given by the present author and Tan in \cite{YuTan2020b}. 
\end{rem}
\begin{IEEEproof}[Proof of Lemma \ref{lem:B}]
Proof of Statement 1: The proof follows an idea from \cite{schieler2016henchman}.
By using a union bound, 
\begin{align}
 & \mathbb{P}_{{\mathcal{C}}}[\mathcal{B}_{1}(\epsilon|T_{XY})^{c}]\nonumber \\
 & =\mathbb{P}_{{\mathcal{C}}}\Big\{\Big|\frac{\phi_{\mathcal{C}}(y^{n})}{\mathbb{E}[\phi_{\mathcal{C}}(y^{n})]}-1\Big|>e^{-n\epsilon},\exists y^{n}\in\mathcal{T}_{T_{Y}}\Big\}\nonumber \\
 & \le|\mathcal{T}_{T_{Y}}|\mathop{\max}\limits _{y^{n}\in\mathcal{T}_{T_{Y}}}\mathbb{P}_{\mathcal{C}}\Big\{\Big|\frac{\phi_{\mathcal{C}}(y^{n})}{\mathbb{E}[\phi_{\mathcal{C}}(y^{n})]}-1\Big|>e^{-n\epsilon}\Big\}.\label{eq:-9-1-2-1}
\end{align}
Given $y^{n}\in\mathcal{T}_{T_{Y}}$, $\theta_{m}(y^{n}):=\bone\{X^{n}(m)\in\mathcal{T}_{T_{X|Y}}(y^{n})\},m\in[e^{nR}]$
are i.i.d. random variables, with mean

\noindent 
\begin{align}
p_{T_{XY}}:=\mathbb{E}_{\mathcal{C}}[\theta_{m}(y^{n})] & =|\mathcal{T}_{T_{X|Y}}(y^{n})|/|\mathcal{T}_{T_{X}}|\nonumber \\
 & =e^{-n(I_{T}(X;Y)-o_{n}(1))},\label{eq:thetaMean}
\end{align}
where \eqref{eq:thetaMean} follows from the basic estimation of the
size of a (conditional) type class \cite{elgamal}.

\noindent On the other hand, $|\mathcal{T}_{T_{Y}}|\leq e^{nH_{T}(Y)}$,
which is exponential in $n$. Hence if we can show that the probability
in \eqref{eq:-9-1-2-1} decays doubly exponentially fast with $n$,
then the proof will be complete. To that end, we first introduce the
following lemma on Chernoff bounds. 
\begin{lem}
\label{lem:chernoff} \cite{Mitzenmacher} If $X^{k}$ is a sequence
of i.i.d. $\text{Bern}(p)$ random variables with $0\le p\le1$, then
for $0<\delta<1$, 
\begin{equation}
\mathbb{P}\Big[|\sum_{i=1}^{k}X_{i}-kp|\ge\delta kp\Big]\leq2e^{-\frac{\delta^{2}kp}{3}}.
\end{equation}
\end{lem}
By identifying that 
\begin{align}
k & =e^{nR},\label{eq:-60-2}\\
p & =p_{T_{XY}}=e^{-n(I_{T}(X;Y)-o_{n}(1))},\\
\delta & =e^{-n\epsilon},
\end{align}
and applying Lemma \ref{lem:chernoff}, we have 
\begin{equation}
\mathbb{P}_{\mathcal{C}}\Big\{\Big|\frac{\phi_{\mathcal{C}}(y^{n})}{\mathbb{E}[\phi_{\mathcal{C}}(y^{n})]}-1\Big|>e^{-n\epsilon}\Big\}\le2e^{-\frac{1}{3}\cdot\exp(n(\beta-o_{n}(1)))}=e^{-\exp(n(\beta-o_{n}(1)))},\label{eq:DoubleExp-1}
\end{equation}

\noindent where 
\begin{align}
 & \beta=R-I_{T}(X;Y)-2\epsilon.\label{eq:-28-4}
\end{align}

For fixed $\epsilon$ and $R>I_{T}(X;Y)+3\epsilon$, it holds that
$\beta>\epsilon$. Hence \eqref{eq:DoubleExp-1} vanishes doubly exponentially
fast. This completes the proof of Statement 1.

Proof of Statement 2:\textbf{ }Statement 2 follows from Statement
1. This is because, on one hand, it is trivial that 
\[
\mathbb{P}\{\phi_{{\mathcal{C}}}(y^{n})\ge0,\forall y^{n}\in\mathcal{T}_{T_{Y}}\}=1.
\]
On the other hand, $\phi_{{\mathcal{C}}}(y^{n})$ is increasing in
$R$, and by setting $R=I_{T}(X;Y)+3\epsilon$, it follows that 
\[
\mathbb{P}[\mathcal{B}_{1}(\epsilon|T_{XY})]\to1
\]
doubly exponentially fast. That is, 
\[
\mathbb{P}\{\phi_{{\mathcal{C}}}(y^{n})\le e^{3n\epsilon},\forall y^{n}\in\mathcal{T}_{T_{Y}}\}\to1
\]
doubly exponentially fast. 
\end{IEEEproof}
The lemma above implies that with high probability, $\phi_{\mathcal{C}}(y^{n})$
is around $e^{n(R-I_{T}(X;Y))}$ if $R\ge I_{T}(X;Y)$, and around
$0$ if $R\le I_{T}(X;Y)$. Moreover, both the probabilities in \eqref{eq:-4}
and \eqref{eq:-5} converge to one doubly exponentially fast and uniformly
for all sequences of types $T_{XY}$ as $n\to\infty$.

Define two events on $\mathcal{C}$ as 
\begin{align*}
\mathcal{B}_{1}(\epsilon|T_{X}) & :=\bigcap_{T_{Y|X}:I_{T}(X;Y)\le R-3\epsilon}\mathcal{B}_{1}(\epsilon|T_{XY}),\\
\mathcal{B}_{2}(\epsilon|T_{X}) & :=\bigcap_{T_{Y|X}:I_{T}(X;Y)\ge R-3\epsilon}\mathcal{B}_{2}(\epsilon|T_{XY}).
\end{align*}
We next simultaneously estimate for each conditional type $T_{Y|X}$,
how many conditional type classes $\mathcal{T}_{T_{Y|X}}(x^{n})$
with $x^{n}\in\mathcal{C}$ each point in $\mathcal{Y}^{n}$ is covered
by. 
\begin{lem}[Strong Packing-Covering Lemma for Constant Composition Codes]
\label{lem:B-1} Let $\epsilon>0$. It holds that 
\[
\mathbb{P}[\mathcal{B}_{1}(\epsilon|T_{X})\cap\mathcal{B}_{2}(\epsilon|T_{X})]\ge1-e^{-\exp(n(\epsilon-o_{n}(1)))}
\]
for all joint $n$-type $T_{X}$, where $o_{n}(1)$ is a term independent
of $(T_{X},R)$ and vanishes as $n\to\infty$. That is, the probability
above converges to one doubly exponentially fast and uniformly for
all types $T_{X}$ as $n\to\infty$. 
\end{lem}
\begin{IEEEproof}
Using a union bound, 
\begin{align*}
 & \mathbb{P}[(\mathcal{B}_{1}(\epsilon|T_{X})\cap\mathcal{B}_{2}(\epsilon|T_{X}))^{c}]\\
 & \le\mathbb{P}[\mathcal{B}_{1}(\epsilon|T_{X})^{c}]+\mathbb{P}[\mathcal{B}_{2}(\epsilon|T_{X})^{c}]\\
 & \le\sum_{T_{Y|X}:I_{T}(X;Y)\le R-3\epsilon}\mathbb{P}[\mathcal{B}_{1}(\epsilon|T_{XY})^{c}]+\sum_{T_{Y|X}:I_{T}(X;Y)\ge R-3\epsilon}\mathbb{P}[\mathcal{B}_{2}(\epsilon|T_{XY})^{c}]\\
 & \to0\textrm{ doubly exponentially fast},
\end{align*}
where the last line follows since the number of types is polynomial
in $n$. 
\end{IEEEproof}

\subsection{\label{subsec:Strong-Packing-Covering-Lemma}Strong Packing-Covering
Lemma for Typical Codes }

Let $P_{XY}$ be a joint distribution such that $(X,Y)\sim P_{XY}$
are not independent. Let $\mathcal{C}:=\{X^{n}(m)\}_{m\in[e^{nR}]}$
be a set (precisely, a multiset) of random sequences such that $X^{n}(m),m\in[e^{nR}]$
are drawn independently for different $m$'s and according to the
same distribution $Q_{X^{n}}:=P_{X}^{\otimes n}(\cdot|\mathcal{T}_{\delta}^{(n)}(P_{X}))$
with $\delta>0$.

For $y^{n}\in\mathcal{T}_{\delta'}^{(n)}(P_{Y})$ and $m\in[e^{nR}]$,
define 
\[
\theta_{m}(y^{n}):=P_{Y|X}^{\otimes n}(y^{n}|X^{n}(m))\bone\{(X^{n}(m),y^{n})\in\mathcal{T}_{\delta}^{(n)}(P_{XY})\}.
\]

Given $y^{n}\in\mathcal{T}_{\delta'}^{(n)}(P_{Y})$, $\theta_{m}(y^{n})$
with $m\in[e^{nR}]$ are i.i.d. random variables, with mean 
\begin{align*}
\mu & :=\mathbb{E}_{\mathcal{C}}[\theta_{m}(y^{n})]\\
 & =\mathbb{E}_{Q_{X^{n}}}[P_{Y|X}^{\otimes n}(y^{n}|X^{n})\bone\{(X^{n},y^{n})\in\mathcal{T}_{\delta}\}]\\
 & =\frac{\sum_{x^{n}\in\mathcal{T}_{\delta}^{(n)}(P_{XY}|y^{n})}P_{X}^{\otimes n}(x^{n})P_{Y|X}^{\otimes n}(y^{n}|x^{n})}{P_{X}^{\otimes n}(\mathcal{T}_{\delta}^{(n)}(P_{X}))}\\
 & =\frac{P_{Y}^{\otimes n}(y^{n})P_{X|Y}^{\otimes n}(\mathcal{T}_{\delta}^{(n)}(P_{XY}|y^{n})|y^{n})}{P_{X}^{\otimes n}(\mathcal{T}_{\delta}^{(n)}(P_{X}))}\\
 & \in P_{Y}^{\otimes n}(y^{n})(1\pm e^{-n\epsilon'})
\end{align*}
for some $\epsilon'>0$ independent of $y^{n}$. Here, we apply the
typicality lemma and the conditional typicality lemma in \cite{elgamal}.

For $\epsilon>0$ and $\delta>\delta'>0$, define an event on $\mathcal{C}$
as 
\begin{align}
\mathcal{B}(\epsilon,\delta') & :=\Big\{\Big|\frac{e^{-nR}\sum_{m}\theta_{m}(y^{n})}{\mu}-1\Big|\leq e^{-n\epsilon},\forall y^{n}\in\mathcal{T}_{\delta'}^{(n)}(P_{Y})\Big\}.
\end{align}

\begin{lem}[Strong Packing-Covering Lemma for Typical Codes]
\label{lem:B-3} Let $\epsilon>0$ and $R>I(X;Y)+\epsilon$. Then,
there exist $\delta>\delta'>0$ such that 
\begin{equation}
\mathbb{P}[\mathcal{B}(\epsilon,\delta')]\ge1-e^{-\exp(n(\beta+o_{n}(1)))},\label{eq:-4-3-1}
\end{equation}
where $\beta>0$ and $o_{n}(1)$ is a term that vanishes as $n\to\infty$
for given $(\delta,\delta')$. 
\end{lem}
\begin{IEEEproof}[Proof of Lemma \ref{lem:B-3}]
Proof of Statement 1: By using a union bound, 
\begin{align}
 & \mathbb{P}_{{\mathcal{C}}}[\mathcal{B}(\epsilon,\delta')^{c}]\nonumber \\
 & =\mathbb{P}_{{\mathcal{C}}}\Big\{\Big|\frac{e^{-nR}\sum_{m}\theta_{m}(y^{n})}{\mu}-1\Big|>e^{-n\epsilon},\exists y^{n}\in\mathcal{T}_{\delta'}^{(n)}(P_{Y})\Big\}\nonumber \\
 & \le|\mathcal{T}_{\delta'}^{(n)}(P_{Y})|\mathop{\max}\limits _{y^{n}\in\mathcal{T}_{\delta'}^{(n)}(P_{Y})}\mathbb{P}_{\mathcal{C}}\Big\{\Big|\frac{e^{-nR}\sum_{m}\theta_{m}(y^{n})}{\mu}-1\Big|>e^{-n\epsilon}\Big\}.\label{eq:-42}
\end{align}

\noindent On the other hand, $|\mathcal{T}_{\delta'}^{(n)}(P_{Y})|$
is exponential in $n$. Hence if we can show that the probability
in \eqref{eq:-42} decays doubly exponentially fast with $n$, then
the proof will be complete. To that end, we first introduce the Bernstein
inequality, which is a generalization of Chernoff bounds. 
\begin{lem}[{{Bernstein Inequality \cite[Corollary 2.11]{boucheron2013concentration}}}]
\label{lem:Bernstein} If $X^{k}$ is a sequence of i.i.d. zero-mean
random variables taking values on the interval $[-a,a]$ with variance
$\sigma^{2}$, then for all $t>0$, 
\[
{\displaystyle \mathbb{P}\Big[|\sum_{i=1}^{k}X_{i}|\geq t\Big]\leq2\exp\left(-\frac{\tfrac{1}{2}t^{2}}{k\sigma^{2}+\tfrac{1}{3}at}\right).}
\]
\end{lem}
To apply the lemma above, we first observe that 
\[
\theta_{m}(y^{n})\le e^{-n(H(Y|X)+o_{\delta}(1))}\textrm{ a.s.}
\]
where $o_{\delta}(1)$ is a term vanishing as $\delta\downarrow0$.
We then estimate 
\begin{align*}
\mathbb{E}_{\mathcal{C}}[\theta_{m}(y^{n})^{2}] & =\mathbb{E}_{\mathcal{C}}[P(y^{n}|X^{n}(m))^{2}\bone\{(X^{n}(m),y^{n})\in\mathcal{T}_{\delta}\}]\\
 & =\sum_{x^{n}}Q_{X^{n}}(x^{n})P_{Y|X}^{\otimes n}(y^{n}|x^{n})^{2}\bone\{(x^{n},y^{n})\in\mathcal{T}_{\delta}\}\\
 & =\frac{\sum_{x^{n}\in\mathcal{T}_{\delta}^{(n)}(P_{XY}|y^{n})}P_{X}^{\otimes n}(x^{n})P_{Y|X}^{\otimes n}(y^{n}|x^{n})^{2}}{P_{X}^{\otimes n}(\mathcal{T}_{\delta}^{(n)}(P_{X}))}\\
 & \le\frac{\sum_{x^{n}\in\mathcal{T}_{\delta}^{(n)}(P_{XY}|y^{n})}P_{X|Y}^{\otimes n}(x^{n}|y^{n})e^{-n(H(Y|X)+o_{\delta}(1))}e^{-n(H(Y)+o_{\delta}(1))}}{P_{X}^{\otimes n}(\mathcal{T}_{\delta}^{(n)}(P_{X}))}\\
 & =\frac{P_{X|Y}^{\otimes n}(\mathcal{T}_{\delta}^{(n)}(P_{XY}|y^{n})|y^{n})}{P_{X}^{\otimes n}(\mathcal{T}_{\delta}^{(n)}(P_{X}))}e^{-n(H(Y|X)+H(Y)+o_{\delta}(1))}\\
 & \doteq e^{-n(H(Y|X)+H(Y)+o_{\delta}(1))}.
\end{align*}
Hence, 
\begin{align*}
\Var(\theta_{m}(y^{n})) & \dotleq e^{-n(H(Y|X)+H(Y)+o_{\delta}(1))}-e^{-2n(H(Y)+o_{\delta}(1))}\\
 & \doteq e^{-n(H(Y|X)+H(Y)+o_{\delta}(1))},
\end{align*}
where $H(Y|X)<H(Y)$ since $(X,Y)$ are not independent.

By identifying that 
\begin{align}
a & =e^{-n(H(Y|X)+o_{\delta}(1))},\label{eq:-60-2-1}\\
k & =e^{nR},\\
t & =\mu e^{n(R-\epsilon)}=e^{n(R-H(Y)+o_{\delta}(1)-\epsilon)},\\
\sigma^{2} & =\Var(\theta_{m}(y^{n}))\dotleq e^{-n(H(Y|X)+H(Y)+o_{\delta}(1))},
\end{align}
and applying Lemma \ref{lem:Bernstein}, we have 
\begin{align}
 & -\log\mathbb{P}_{\mathcal{C}}\Big\{\Big|\frac{e^{-nR}\sum_{m}\theta_{m}(y^{n})}{\mu}-1\Big|>e^{-n\epsilon}\Big\}\nonumber \\
 & \dotgeq\frac{\tfrac{1}{2}e^{2n(R-H(Y)+o_{\delta}(1)-\epsilon)}}{e^{nR}e^{-n(H(Y|X)+H(Y)+o_{\delta}(1))}+\tfrac{1}{3}e^{-n(H(Y|X)+o_{\delta}(1))}e^{n(R-H(Y)+o_{\delta}(1)-\epsilon)}}-\log2\nonumber \\
 & \doteq e^{n(R-I(X;Y)+o_{\delta}(1)-\epsilon)}-\log2\nonumber \\
 & \doteq e^{n(R-I(X;Y)+o_{\delta}(1)-\epsilon)},\label{eq:-41}
\end{align}

\noindent where $R-I(X;Y)+o_{\delta}(1)-\epsilon>0$ for some $\epsilon,\delta>0$,
since $R>I(X;Y)$. Hence \eqref{eq:-41} goes to infinity exponentially
fast, i.e., the probability in \eqref{eq:-42} vanishes doubly exponentially
fast. This means that $\mathbb{P}_{{\mathcal{C}}}[\mathcal{B}(\epsilon,\delta')^{c}]$
vanishes doubly exponentially fast as $n\rightarrow\infty$.
\end{IEEEproof}

\subsection{One-Shot Bounds for Rényi Resolvability }

\label{sec:one-shot} We now consider the one-shot (i.e., blocklength
$n$ equal to 1) version of the Rényi resolvability problem and provide
bounds for this setting. Consider a random map $f_{\mathcal{C}}:\calM=[\e^{R}]\rightarrow\calX$
given below. We set $\mathcal{C}=\{X(m)\}_{m\in\calM}$ with $X(m),m\in\calM$
drawn independently for different $m$'s and according to the same
distribution $Q_{X}$, and set $f_{\mathcal{C}}(m)=X(m)$. This forms
a random code, and $\mathcal{C}$ is called a random codebook. The
joint distribution induced by the code is 
\[
Q_{\mathcal{C}MXY}:=Q_{\mathcal{C}}Q_{M}Q_{X|M,\mathcal{C}}P_{Y|X=X(M)},
\]
where $Q_{M}=\Unif[\e^{R}]$ and $Q_{X|M,\mathcal{C}}(x|m,\mathcal{C})=\bone\{x=X(m)\}$.
We provide several bounds for this code in the following lemma. 
\begin{lem}[One-Shot Bounds]
\label{lem:oneshot} For the random code described above, the following
hold. 
\begin{enumerate}
\item For any $q\in[2,\infty)$ and any distribution $P_{Y}$, 
\begin{equation}
\e^{(q-1)D_{q}(Q_{Y|\mathcal{C}}\|P_{Y}|Q_{\mathcal{C}})}\le\sum_{t=1}^{\tilde{q}-1}S(\tilde{q},t)\left(t\beta(t)+\beta(t+\hat{q})\right)+\tilde{q}\beta(q-1)+\beta(q),\label{eq:-123}
\end{equation}
where $\tilde{q}:=\left\lceil q\right\rceil -1$, $\hat{q}:=q-\tilde{q}\in(0,1]$,
\begin{align}
\beta(t) & :=\inf_{s\ge q-t}\e^{(q-t)(D_{1+s}(P_{Y|X}\|P_{Y}|Q_{X})-R)+(t-1)D_{1+\frac{s(t-1)}{s+t-q}}(Q_{Y}\|P_{Y})},\label{eq:-106}
\end{align}
and 
\[
S(m,k):=\frac{1}{k!}\sum_{i=0}^{k}(-1)^{k-i}\binom{k}{i}i^{m}=\sum_{i=0}^{k}\frac{(-1)^{k-i}i^{m}}{(k-i)!i!}
\]
is the Stirling partition number (i.e., the number of ways to partition
a set of $m$ elements into $k$ non-empty subsets). 
\item For any $q\in[2,\infty)$ and any distribution $P_{Y}$,
\begin{align}
 & \e^{(q-1)D_{q}(Q_{Y|\mathcal{C}}\|P_{Y}|Q_{\mathcal{C}})}\nonumber \\
 & \ge(1-\e^{-R})^{q-1}\e^{(q-1)D_{q}(Q_{Y}\|P_{Y})}+\e^{(q-1)(D_{q}(P_{Y|X}\|P_{Y}|Q_{X})-R)}.\label{eq:-123-4}
\end{align}
\end{enumerate}
\end{lem}
\begin{rem}
\label{rem:simplebound}By definition, it holds that 
\begin{align*}
\beta(1) & =\e^{(q-1)(D_{q}(P_{Y|X}\|P_{Y}|Q_{X})-R)},\\
\beta(q) & =\e^{(q-1)D_{q}(Q_{Y}\|P_{Y})}.
\end{align*}
Moreover, for any $t\in[1,q]$, setting $s=q-1$ and $s=q-t$ respectively,
one can obtain that 
\begin{equation}
\beta(t)\le\e^{(q-t)(D_{q}(P_{Y|X}\|P_{Y}|Q_{X})-R)+(t-1)D_{q}(Q_{Y}\|P_{Y})},\label{eq:-18}
\end{equation}
and 
\begin{equation}
\beta(t)\le\e^{(q-t)(D_{q-t+1}(P_{Y|X}\|P_{Y}|Q_{X})-R)+(t-1)D_{\infty}(Q_{Y}\|P_{Y})}.\label{eq:-16}
\end{equation}
In fact, the bound in \eqref{eq:-18} is good for $R<D_{q}(P_{Y|X}\|P_{Y}|Q_{X})$
in the sense that it implies the achievability part  in \eqref{eq:bound1}
in Theorem \ref{thm:Resolvability} and the achievability part in
\eqref{eq:-15} in Theorem \ref{thm:Resolvability-3} by setting $Q_{X}$
to the uniform distribution over a type class (similarly to the proof
of Theorem 1 in \cite{yu2019renyi}). The bound in \eqref{eq:-16}
is good for $R>D_{q}(P_{Y|X}\|P_{Y}|Q_{X})$ in the sense that it
yields desirable bounds on the exponential convergence of the Rényi
divergence; see Section \ref{subsec:Exponential-Behavior}. 
\end{rem}
\begin{rem}
 In contrast, for $q\in(1,2]$, the present author and Tan \cite{yu2019renyi}
shows the following inequality for the random code: 
\begin{align}
 & \e^{(q-1)D_{q}(Q_{Y|\mathcal{C}}\|P_{Y}|Q_{\mathcal{C}})}\nonumber \\
 & \le\e^{(q-1)(D_{q}(P_{Y|X}\|P_{Y}|Q_{X})-R)}+\e^{(q-1)D_{q}(Q_{Y}\|P_{Y})}.\label{eq:-123-8}
\end{align}
This coincides with \eqref{eq:-123} for $q=2$. Furthermore, a
lower bound for general codes (not only for the random codes given
above) was provided in \cite{yu2019renyi}. 
\end{rem}
\begin{rem}
If $M\sim Q_{M}$ is an arbitrary (not necessarily uniform) random
variable defined on an arbitrary alphabet $\calM$ and $X(m)\sim Q_{X|M=m}$
for each $m\in\calM$, then  by checking our proofs, \eqref{eq:-123}
 and \eqref{eq:-123-8} still hold if all terms like $D_{1+s}(P_{Y|X}\|P_{Y}|Q_{X})-R$
therein replaced by $D_{1+s}(P_{Y|X}\|P_{Y}|Q_{M}^{(1+s)}Q_{X|M})-H_{1+s}(Q_{M})$,
where $Q_{M}^{(1+s)}:=\frac{Q_{M}^{1+s}}{\sum Q_{M}^{1+s}}$ is the
$(1+s)$-tilted version of $Q_{M}$, and $H_{1+s}(Q_{M})$ is the
$(1+s)$-Rényi entropy of $Q_{M}$. 
\end{rem}
\begin{IEEEproof}[Proof of Lemma \ref{lem:oneshot}]
 The main idea used here is similar to \cite{kavian2023statistics},
which can be seen as a combination of the ideas from \cite{yu2019renyi}
and \cite{pathegama2023smoothing}. Throughout the proof, we use the
notation ${\rmM}=\e^{R}$.

\emph{Statement 1: }For brevity, denote $\theta_{m,y}:=\frac{P_{Y|X}(y|f_{\mathcal{C}}(m))}{P_{Y}(y)}$.
The output distribution is then given by 
\[
\frac{Q_{Y|\mathcal{C}}(y|\mathcal{C})}{P_{Y}(y)}=\frac{1}{\rmM}\sum_{m=1}^{\rmM}\theta_{m,y}.
\]
In the following, we denote $\mathcal{A}:=\{A_{1},\cdots,A_{t}\}$
as a $t$-partition of $[\tilde{q}]$.

Observe that 
\begin{align}
 & \e^{(q-1)D_{q}(Q_{Y|\mathcal{C}}\|P_{Y}|Q_{\mathcal{C}})}\nonumber \\
 & =\mathbb{E}_{\mathcal{C}}\sum_{y}Q^{q}(y|\mathcal{C})P^{1-q}(y)\nonumber \\
 & =\sum_{y}P(y)\mathbb{E}_{\mathcal{C}}[(\frac{1}{\rmM}\sum_{m=1}^{\rmM}\theta_{m,y})^{q}]\nonumber \\
 & =\sum_{y}P(y){\rmM}^{-q}\mathbb{E}_{\mathcal{C}}[(\sum_{m=1}^{\rmM}\theta_{m,y})^{\tilde{q}}(\sum_{m=1}^{\rmM}\theta_{m,y})^{\hat{q}}]\nonumber \\
 & =\sum_{y}P(y){\rmM}^{-q}\mathbb{E}_{\mathcal{C}}[(\sum_{t=1}^{\tilde{q}}\sum_{\mathcal{A}:\,t\textrm{-partition of }[\tilde{q}]}\sum_{m_{1}\in[{\rmM}]}\sum_{m_{2}\in[{\rmM}]\backslash\{m_{1}\}}\nonumber \\
 & \qquad\qquad\cdots\sum_{m_{t}\in[{\rmM}]\backslash\{m_{1},\cdots,m_{t-1}\}}\theta_{m_{1},y}^{|A_{1}|}\theta_{m_{2},y}^{|A_{2}|}\cdots\theta_{m_{t},y}^{|A_{t}|})(\sum_{m=1}^{\rmM}\theta_{m,y})^{\hat{q}}]\label{eq:-39-3}\\
 & =\sum_{y}P(y){\rmM}^{-q}\sum_{t=1}^{\tilde{q}}\sum_{\mathcal{A}:\,t\textrm{-partition of }[\tilde{q}]}\sum_{m_{1}\in[{\rmM}]}\sum_{m_{2}\in[{\rmM}]\backslash\{m_{1}\}}\nonumber \\
 & \qquad\qquad\cdots\sum_{m_{t}\in[{\rmM}]\backslash\{m_{1},\cdots,m_{t-1}\}}\mathbb{E}_{\mathcal{C}}[\theta_{m_{1},y}^{|A_{1}|}\theta_{m_{2},y}^{|A_{2}|}\cdots\theta_{m_{t},y}^{|A_{t}|}(\sum_{m=1}^{\rmM}\theta_{m,y})^{\hat{q}}]\nonumber \\
 & =\sum_{y}P(y){\rmM}^{-q}\sum_{t=1}^{\tilde{q}}\sum_{\mathcal{A}:\,t\textrm{-partition of }[\tilde{q}]}{\rmM}({\rmM}-1)\cdots({\rmM}+1-t)\mathbb{E}_{\mathcal{C}}[\theta_{1,y}^{|A_{1}|}\theta_{2,y}^{|A_{2}|}\cdots\theta_{t,y}^{|A_{t}|}(\sum_{m=1}^{\rmM}\theta_{m,y})^{\hat{q}}]\label{eq:-14-3}\\
 & \le\sum_{t=1}^{\tilde{q}}{\rmM}^{t-q}\sum_{\mathcal{A}:\,t\textrm{-partition of }[\tilde{q}]}\mathbb{E}_{P_{Y}}\mathbb{E}_{\mathcal{C}}[\theta_{1,Y}^{|A_{1}|}\theta_{2,Y}^{|A_{2}|}\cdots\theta_{t,Y}^{|A_{t}|}(\sum_{m=1}^{\rmM}\theta_{m,Y})^{\hat{q}}],\label{eq:-13-3}
\end{align}
where in \eqref{eq:-39-3}, $(\sum_{m=1}^{\rmM}\theta_{m,y})^{\tilde{q}}$
is expanded, and \eqref{eq:-14-3} follows since $\theta_{m,y},m\in[{\rmM}]$
obey the same distribution.  

On one hand, for $t\le\tilde{q}-1$, we apply Lemma \ref{lem:norm}
to split $(\sum_{m=1}^{\rmM}\theta_{m,Y})^{\hat{q}}$ in the following
way: 
\[
(\sum_{m=1}^{\rmM}\theta_{m,Y})^{\hat{q}}\le\sum_{m=1}^{t}\theta_{m,Y}^{\hat{q}}+(\sum_{m=t+1}^{\rmM}\theta_{m,Y})^{\hat{q}}.
\]
On the other hand, for $t=\tilde{q}$, note that $\mathcal{A}$ is
the unique partition consisting of singletons, i.e., $|A_{i}|=1$
for all $i$. For this case, by Jensen's inequality and using the
independence of different codewords, we obtain that 
\begin{align*}
\mathbb{E}_{\mathcal{C}}[\theta_{1,Y}\theta_{2,Y}\cdots\theta_{\tilde{q},Y}(\sum_{m=1}^{\rmM}\theta_{m,Y})^{\hat{q}}] & \le\mathbb{E}_{\mathcal{C}}\big[\theta_{1,Y}\theta_{2,Y}\cdots\theta_{\tilde{q},Y}(\sum_{m=1}^{\tilde{q}}\theta_{m,Y}+\mathbb{E}_{\mathcal{C}}[\sum_{m=\tilde{q}+1}^{\rmM}\theta_{m,Y}])^{\hat{q}}\big]\\
 & \le(\rmM\frac{Q_{Y}}{P_{Y}})^{\hat{q}}\mathbb{E}_{\mathcal{C}}\big[\theta_{1,Y}\theta_{2,Y}\cdots\theta_{\tilde{q},Y}((\rmM\frac{Q_{Y}}{P_{Y}})^{-1}\sum_{m=1}^{\tilde{q}}\theta_{m,Y}+1)^{\hat{q}}\big]\\
 & \le(\rmM\frac{Q_{Y}}{P_{Y}})^{\hat{q}}\mathbb{E}_{\mathcal{C}}\big[\theta_{1,Y}\theta_{2,Y}\cdots\theta_{\tilde{q},Y}((\rmM\frac{Q_{Y}}{P_{Y}})^{-1}\sum_{m=1}^{\tilde{q}}\theta_{m,Y}+1)\big]\\
 & =\mathbb{E}_{\mathcal{C}}\big[\theta_{1,Y}\theta_{2,Y}\cdots\theta_{\tilde{q},Y}\big((\rmM\frac{Q_{Y}}{P_{Y}})^{\hat{q}-1}\sum_{m=1}^{\tilde{q}}\theta_{m,Y}+(\rmM\frac{Q_{Y}}{P_{Y}})^{\hat{q}}\big)\big].
\end{align*}

We hence obtain that 

\begin{align}
 & \e^{(q-1)D_{q}(Q_{Y|\mathcal{C}}\|P_{Y}|Q_{\mathcal{C}})}\nonumber \\
 & \le\sum_{t=1}^{\tilde{q}-1}{\rmM}^{t-q}\sum_{\mathcal{A}:\,t\textrm{-partition of }[\tilde{q}]}\mathbb{E}_{P_{Y}}\mathbb{E}_{\mathcal{C}}\Bigl[\theta_{1,Y}^{|A_{1}|+\hat{q}}\theta_{2,Y}^{|A_{2}|}\cdots\theta_{t,Y}^{|A_{t}|}+\theta_{1,Y}^{|A_{1}|}\theta_{2,Y}^{|A_{2}|+\hat{q}}\cdots\theta_{t,Y}^{|A_{t}|}+\nonumber \\
 & \qquad\qquad\cdots+\theta_{1,Y}^{|A_{1}|}\theta_{2,Y}^{|A_{2}|}\cdots\theta_{t,Y}^{|A_{t}|+\hat{q}}+\theta_{1,Y}^{|A_{1}|}\theta_{2,Y}^{|A_{2}|}\cdots\theta_{t,Y}^{|A_{t}|}(\sum_{m=t+1}^{\rmM}\theta_{m,Y})^{\hat{q}}\Bigr]\nonumber \\
 & \qquad+{\rmM}^{-\hat{q}}\mathbb{E}_{P_{Y}}\mathbb{E}_{\mathcal{C}}\Bigl[(\rmM\frac{Q_{Y}}{P_{Y}})^{\hat{q}-1}(\theta_{1,Y}^{2}\theta_{2,Y}\cdots\theta_{\tilde{q},Y}+\theta_{1,Y}\theta_{2,Y}^{2}\cdots\theta_{\tilde{q},Y}+\nonumber \\
 & \qquad\qquad\cdots+\theta_{1,Y}\theta_{2,Y}\cdots\theta_{\tilde{q},Y}^{2})+(\rmM\frac{Q_{Y}}{P_{Y}})^{\hat{q}}\theta_{1,Y}\theta_{2,Y}\cdots\theta_{\tilde{q},Y}\Bigr].\label{eq:-48-3}
\end{align}

We first bound the $k$-th term in the first summation in \eqref{eq:-48-3}.
For  $k\in[t]$, 
\begin{align}
 & {\rmM}^{t-q}\mathbb{E}_{P_{Y}}\mathbb{E}_{\mathcal{C}}[\theta_{1,Y}^{|A_{1}|}\cdots\theta_{k-1,Y}^{|A_{k-1}|}\theta_{k,Y}^{|A_{k}|+\hat{q}}\theta_{k+1,Y}^{|A_{k+1}|}\cdots\theta_{t,Y}^{|A_{t}|}]\nonumber \\
 & ={\rmM}^{t-q}\mathbb{E}_{P_{Y}}\Big[\mathbb{E}_{Q_{X}}[(\frac{P_{Y|X}}{P_{Y}})^{|A_{k}|+\hat{q}}]\prod_{i\in[t]\backslash\{k\}}\mathbb{E}_{Q_{X}}[(\frac{P_{Y|X}}{P_{Y}})^{|A_{i}|}]\Big]\nonumber \\
 & ={\rmM}^{t-q}\mathbb{E}_{Q_{Y}}\Big[(\frac{Q_{Y}}{P_{Y}})^{t-1}\mathbb{E}_{Q_{X|Y}}[(\frac{P_{Y|X}}{P_{Y}})^{|A_{k}|+\hat{q}-1}]\prod_{i\in[t]\backslash\{k\}}\mathbb{E}_{Q_{X|Y}}[(\frac{P_{Y|X}}{P_{Y}})^{|A_{i}|-1}]\Big]\label{eq:-72-3}\\
 & \le{\rmM}^{t-q}\mathbb{E}_{Q_{Y}}\Big[(\frac{Q_{Y}}{P_{Y}})^{t-1}\mathbb{E}_{Q_{X|Y}}[(\frac{P_{Y|X}}{P_{Y}})^{q-t}]^{\frac{|A_{k}|+\hat{q}-1}{q-t}}\prod_{i\in[t]\backslash\{k\}}\mathbb{E}_{Q_{X|Y}}[(\frac{P_{Y|X}}{P_{Y}})^{q-t}]^{\frac{|A_{i}|-1}{q-t}}\Big]\nonumber \\
 & ={\rmM}^{t-q}\mathbb{E}_{Q_{XY}}[(\frac{Q_{Y}}{P_{Y}})^{t-1}(\frac{P_{Y|X}}{P_{Y}})^{q-t}]\nonumber \\
 & \le\inf_{s\ge q-t}{\rmM}^{t-q}\mathbb{E}_{Q_{Y}}[(\frac{Q_{Y}}{P_{Y}})^{\frac{s(t-1)}{s+t-q}}]^{\frac{s+t-q}{s}}\mathbb{E}_{Q_{XY}}[(\frac{P_{Y|X}}{P_{Y}})^{s}]^{\frac{q-t}{s}}\nonumber \\
 & =\inf_{s\ge q-t}\e^{(q-t)(D_{1+s}(P_{Y|X}\|P_{Y}|Q_{X})-R)+(t-1)D_{1+\frac{s(t-1)}{s+t-q}}(Q_{Y}\|P_{Y})},\nonumber 
\end{align}
where $Q_{XY}:=Q_{X}P_{Y|X}$, the first inequality follows by Jensen's
inequality, and the second inequality follows by Hölder's inequality.

We now bound the last term in the first summation in \eqref{eq:-48-3}
as follows: 
\begin{align*}
 & {\rmM}^{t-q}\mathbb{E}_{P_{Y}}\mathbb{E}_{\mathcal{C}}[\theta_{1,Y}^{|A_{1}|}\cdots\theta_{t,Y}^{|A_{t}|}(\sum_{m=t+1}^{\rmM}\theta_{m,y})^{\hat{q}}]\\
 & ={\rmM}^{t-q}\mathbb{E}_{P_{Y}}\Big[\mathbb{E}_{\mathcal{C}}[\theta_{1,Y}^{|A_{1}|}\cdots\theta_{t,Y}^{|A_{t}|}]\mathbb{E}_{\mathcal{C}}[(\sum_{m=t+1}^{\rmM}\theta_{m,y})^{\hat{q}}]\Big]\\
 & \le{\rmM}^{t-q}\mathbb{E}_{P_{Y}}\Big[\mathbb{E}_{\mathcal{C}}[\theta_{1,Y}^{|A_{1}|}\cdots\theta_{t,Y}^{|A_{t}|}]\mathbb{E}_{\mathcal{C}}[\sum_{m=t+1}^{\rmM}\theta_{m,y}]^{\hat{q}}\Big]\\
 & ={\rmM}^{t-q}\mathbb{E}_{P_{Y}}\Big[\prod_{i\in[t]}\mathbb{E}_{Q_{X}}[(\frac{P_{Y|X}}{P_{Y}})^{|A_{i}|}](({\rmM}-t)\frac{Q_{Y}}{P_{Y}})^{\hat{q}}\Big]\\
 & \le{\rmM}^{t-\tilde{q}}\mathbb{E}_{Q_{Y}}\Big[\prod_{i\in[t]}\mathbb{E}_{Q_{X|Y}}[(\frac{P_{Y|X}}{P_{Y}})^{|A_{i}|-1}](\frac{Q_{Y}}{P_{Y}})^{\hat{q}+t-1}\Big]\\
 & \le{\rmM}^{t-\tilde{q}}\mathbb{E}_{Q_{Y}}\Big[\prod_{i\in[t]}\mathbb{E}_{Q_{X|Y}}[(\frac{P_{Y|X}}{P_{Y}})^{\tilde{q}-t}]^{\frac{|A_{i}|-1}{\tilde{q}-t}}(\frac{Q_{Y}}{P_{Y}})^{\hat{q}+t-1}\Big]\\
 & ={\rmM}^{t-\tilde{q}}\mathbb{E}_{Q_{Y}}\Big[\mathbb{E}_{Q_{X|Y}}[(\frac{P_{Y|X}}{P_{Y}})^{\tilde{q}-t}](\frac{Q_{Y}}{P_{Y}})^{\hat{q}+t-1}\Big]\\
 & \le\inf_{s\ge\tilde{q}-t}\e^{(\tilde{q}-t)(D_{1+s}(P_{Y|X}\|P_{Y}|Q_{X})-R)+(\hat{q}+t-1)D_{1+\frac{s(\hat{q}+t-1)}{s+t-\tilde{q}}}(Q_{Y}\|P_{Y})},
\end{align*}
where the first and third inequalities follow by Jensen's inequality,
and the last inequality follows by Hölder's inequality.

We next bound the $k$-th term  in the second summation in \eqref{eq:-48-3}.
 Observe that 
\begin{align*}
 & {\rmM}^{-1}\mathbb{E}_{P_{Y}}\mathbb{E}_{\mathcal{C}}\Bigl[(\frac{Q_{Y}}{P_{Y}})^{\hat{q}-1}\theta_{1,Y}\cdots\theta_{k-1,Y}\theta_{k,Y}^{2}\theta_{k+1,Y}\cdots\theta_{\tilde{q},Y}\Big]\\
 & ={\rmM}^{-1}\mathbb{E}_{P_{Y}}\Bigl[(\frac{Q_{Y}}{P_{Y}})^{q-2}\mathbb{E}_{Q_{X}}[(\frac{P_{Y|X}}{P_{Y}})^{2}]\Big]\\
 & ={\rmM}^{-1}\mathbb{E}_{Q_{Y}}\Bigl[(\frac{Q_{Y}}{P_{Y}})^{q-2}\mathbb{E}_{Q_{X|Y}}[\frac{P_{Y|X}}{P_{Y}}]\Big]\\
 & \le\inf_{s\ge1}{\rmM}^{-1}\mathbb{E}_{Q_{Y}}[(\frac{Q_{Y}}{P_{Y}})^{\frac{s(q-2)}{s-1}}]^{\frac{s-1}{s}}\mathbb{E}_{Q_{XY}}[(\frac{P_{Y|X}}{P_{Y}})^{s}]^{\frac{1}{s}}\\
 & =\inf_{s\ge1}\e^{D_{1+s}(P_{Y|X}\|P_{Y}|Q_{X})-R+(q-2)D_{1+\frac{s(q-2)}{s-1}}(Q_{Y}\|P_{Y})},
\end{align*}
where the inequality follows by Hölder's inequality. 

We lastly simplify the last term in the second summation in \eqref{eq:-48-3}
as follows: 
\begin{align*}
 & \mathbb{E}_{P_{Y}}\mathbb{E}_{\mathcal{C}}\Bigl[(\frac{Q_{Y}}{P_{Y}})^{\hat{q}}\theta_{1,Y}\theta_{2,Y}\cdots\theta_{\tilde{q},Y}\Big]\\
 & =\mathbb{E}_{P_{Y}}\Bigl[(\frac{Q_{Y}}{P_{Y}})^{q}\Big]\\
 & =\e^{(q-1)D_{q}(Q_{Y}\|P_{Y})}.
\end{align*}

Therefore, the desired upper bound follows.

\emph{Statement 2: }We next prove the lower bound. Observe that for
$q\ge2$, 
\begin{align}
 & \e^{(q-1)D_{q}(Q_{Y|\mathcal{C}}\|P_{Y}|Q_{\mathcal{C}})}\nonumber \\
 & =\mathbb{E}_{P_{Y}}\Big[{\rmM}^{-q}\mathbb{E}_{\mathcal{C}}[(\sum_{m=1}^{\rmM}\theta_{m,Y})^{q}]\Big]\nonumber \\
 & =\mathbb{E}_{P_{Y}}\Big[{\rmM}^{-q}\mathbb{E}_{\mathcal{C}}[\sum_{m=1}^{\rmM}\theta_{m,Y}(\theta_{m,Y}+\sum_{m'\neq m}\theta_{m',Y})^{q-1}]\Big]\nonumber \\
 & \ge\mathbb{E}_{P_{Y}}\Big[{\rmM}^{-q}\mathbb{E}_{\mathcal{C}}[\sum_{m=1}^{\rmM}\theta_{m,Y}^{q}+\sum_{m=1}^{\rmM}\theta_{m,Y}(\sum_{m'\neq m}\theta_{m',Y})^{q-1}]\Big]\label{eq:-75-2}\\
 & =\e^{(q-1)(D_{q}(P_{Y|X}\|P_{Y}|Q_{X})-R)}+{\rmM}^{-q}\mathbb{E}_{P_{Y}}\Big[\mathbb{E}_{\mathcal{C}}[\sum_{m=1}^{\rmM}\theta_{m,Y}(\sum_{m'\neq m}\theta_{m',Y})^{q-1}]\Big],\nonumber 
\end{align}
where \eqref{eq:-75-2} follows by Lemma \ref{lem:norm}. We now lower
bound the second term in the last line. 
\begin{align*}
 & \mathbb{E}_{P_{Y}}\Big[\mathbb{E}_{\mathcal{C}}[\sum_{m=1}^{\rmM}\theta_{m,Y}(\sum_{m'\neq m}\theta_{m',Y})^{q-1}]\Big]\\
 & =\mathbb{E}_{P_{Y}}\Big[\sum_{m=1}^{\rmM}\mathbb{E}_{X(m)}[\theta_{m,Y}]\mathbb{E}_{\mathcal{C}\backslash\{X(m)\}}[(\sum_{m'\neq m}\theta_{m',Y})^{q-1}]\Big]\\
 & \ge\mathbb{E}_{P_{Y}}\Big[\sum_{m=1}^{\rmM}\frac{Q_{Y}}{P_{Y}}(\mathbb{E}_{\mathcal{C}}[\sum_{m'\neq m}\theta_{m',Y}])^{q-1}\Big]\\
 & =\mathbb{E}_{P_{Y}}\Big[\sum_{m=1}^{\rmM}\frac{Q_{Y}}{P_{Y}}({\rmM}-1)^{q-1}(\frac{Q_{Y}}{P_{Y}})^{q-1}\Big]\\
 & ={\rmM}({\rmM}-1)^{q-1}\e^{(q-1)D_{q}(Q_{Y}\|P_{Y})},
\end{align*}
where the inequality follows by Jensen's inequality. 

So, we obtain that 
\begin{align*}
 & \e^{(q-1)D_{q}(Q_{Y|\mathcal{C}}\|P_{Y}|Q_{\mathcal{C}})}\\
 & \ge(1-\e^{-R})^{q-1}\e^{(q-1)D_{q}(Q_{Y}\|P_{Y})}+\e^{(q-1)(D_{q}(P_{Y|X}\|P_{Y}|Q_{X})-R)}.
\end{align*}
\end{IEEEproof}

\section{\label{sec:Proof-of-Theorem-Resolvability}Proof of Theorem \ref{thm:Resolvability}}

Statements 1 and 3 for $q=1$ follow from the corresponding statements
for $q\in(1,\infty]$. Statement 2 for $q=0$ and Statement 4 follow
from Statement 2 for $q\in(0,1)$. The arguments for these two points
are as follows.

By the monotonicity of the Rényi divergence in its order, for Statement
1 with $q=1$, it holds that 
\begin{align*}
 & \lim_{n\to\infty}\frac{1}{n}\inf_{f:[e^{nR}]\to\mathcal{X}^{n}}D(Q_{Y^{n}}\|P_{Y}^{\otimes n})\\
 & =\inf_{n\ge1}\frac{1}{n}\inf_{f:[e^{nR}]\to\mathcal{X}^{n}}\inf_{q>1}D_{q}(Q_{Y^{n}}\|P_{Y}^{\otimes n})\\
 & =\inf_{q>1}\inf_{n\ge1}\frac{1}{n}\inf_{f:[e^{nR}]\to\mathcal{X}^{n}}D_{q}(Q_{Y^{n}}\|P_{Y}^{\otimes n})\\
 & =\inf_{q>1}\min_{Q_{X}}\max\{\mathbb{E}_{Q_{X}}[D_{q}(P_{Y|X}\|P_{Y})]-R,\\
 & \qquad\qquad\qquad\max_{Q_{Y|X}}-q'D(Q_{Y|X}\|P_{Y|X}|Q_{X})+D(Q_{Y}\|P_{Y})\}\\
 & =\min_{Q_{X}}\lim_{q\downarrow1}\max\{\mathbb{E}_{Q_{X}}[D_{q}(P_{Y|X}\|P_{Y})]-R,\\
 & \qquad\qquad\qquad\max_{Q_{Y|X}}-q'D(Q_{Y|X}\|P_{Y|X}|Q_{X})+D(Q_{Y}\|P_{Y})\}\\
 & =\min_{Q_{X}}\max\{D(P_{Y|X}\|P_{Y}|Q_{X})-R,\,D(Q_{X}\circ P_{Y|X}\|P_{Y})\}.
\end{align*}

Statement 3 with $q=1$ follows in a similar way. By steps similar
to the above, for $q=1$ and $R>R_{\min}$, we have 
\begin{align}
 & \lim_{n\to\infty}\frac{1}{n}\inf_{f:[e^{nR}]\to\mathcal{X}^{n}}D(P_{Y}^{\otimes n}\|Q_{Y^{n}})\nonumber \\
 & =\lim_{k\to\infty}\max_{Q_{Y}}\varpi(R,Q_{Y})-kD(Q_{Y}\|P_{Y}),\label{eq:-47-2}
\end{align}
where 
\[
\varpi(R,Q_{Y}):=\min_{Q_{X|Y}:I_{Q}(X;Y)\le R}D(Q_{Y|X}\|P_{Y|X}|Q_{X}).
\]
What we want to prove is that the expression in \eqref{eq:-47-2}
is exactly $\varpi(R,P_{Y})$.

On one hand, by setting $Q_{Y}=P_{Y}$, $\eqref{eq:-47-2}\ge\varpi(R,P_{Y}).$
On the other hand, let $Q_{Y}^{(k)}$ attain $\max_{Q_{Y}}\varpi(R,Q_{Y})-kD(Q_{Y}\|P_{Y})$.
Since the probability simplex is compact, by passing to a convergent
subsequence, we assume that $Q_{Y}^{(k)}\to Q_{Y}^{*}$ as $k\to\infty$
for some $Q_{Y}^{*}$. So, 
\begin{align*}
\eqref{eq:-47-2} & =\lim_{k\to\infty}\varpi(R,Q_{Y}^{(k)})-kD(Q_{Y}^{(k)}\|P_{Y})\\
 & =\varpi(R,Q_{Y}^{*})-\lim_{k\to\infty}kD(Q_{Y}^{(k)}\|P_{Y}),
\end{align*}
where the last line follows by the continuity of $\varpi(R,Q_{Y})$
in $Q_{Y}$. If $Q_{Y}^{*}\neq P_{Y}$, then $\lim_{k\to\infty}kD(Q_{Y}^{(k)}\|P_{Y})=+\infty$,
which implies $\eqref{eq:-47-2}\le-\infty,$ contradicting with the
nonnegativity of $D(P_{Y}^{\otimes n}\|Q_{Y^{n}})$. So, we must
have $Q_{Y}^{*}=P_{Y}$. Hence, $\eqref{eq:-47-2}\le\varpi(R,P_{Y}).$
Combining two points above yields Statement 3 with $q=1$.

In a similar way, we can show that Statement 2 for $q=0$ and Statement
4 follow from Statement 2 for $q\in(0,1)$. So, in the following,
we only prove Statements 1 and 3 for $q\in(1,\infty]$ and Statement
2 for $q\in(0,1)$.

\subsection{Upper Bound in Statement 1 }

We first consider the case $q\in(1,\infty)$. Let $R>3\epsilon>0$.
Let $\mathcal{C}:=\{X^{n}(m)\}_{m\in[e^{nR}]}$ be a set of random
sequences such that $X^{n}(m),m\in[e^{nR}]$ are drawn independently
for different $m$'s and according to the same distribution $\Unif(\mathcal{T}_{T_{X}})$.
By Lemma \ref{lem:B-1}, there is a realization $c$ of $\mathcal{C}$
satisfying $\mathcal{B}_{1}(\epsilon|T_{X})\cap\mathcal{B}_{2}(\epsilon|T_{X})$.
Let $f:[e^{nR}]\to\mathcal{X}^{n}$ be the resolvability code based
on $c$, i.e., the deterministic map given by $f(m)=x^{n}(m)$ with
$x^{n}(m)$ being the $m$-th codeword in $c$. Denote $q=1+s$. Observe
that for $s\in(-1,\infty)\backslash\{0\}$, 
\begin{align}
 & \e^{sD_{1+s}(Q_{Y^{n}}\|P_{Y}^{\otimes n})}\nonumber \\
 & =\sum_{y^{n}}(\sum_{m}e^{-nR}P(y^{n}|x^{n}(m)))^{1+s}P^{-s}(y^{n})\\
 & =\sum_{T_{Y}}\sum_{y^{n}\in\mathcal{T}_{T_{Y}}}e^{-(1+s)nR-sn\sum T_{Y}\log P_{Y}}(\sum_{T_{XY}\in\Pi(T_{X},T_{Y})}e^{n\sum T_{XY}\log P_{Y|X}}|\mathcal{T}_{T_{X|Y}}(y^{n})\cap c|)^{1+s}\\
 & \doteq\max_{T_{Y|X}}\sum_{y^{n}\in\mathcal{T}_{T_{Y}}}e^{-(1+s)nR-sn\sum T_{Y}\log P_{Y}+(1+s)n\sum T_{XY}\log P_{Y|X}}|\mathcal{T}_{T_{X|Y}}(y^{n})\cap c|^{1+s},\label{eq:}
\end{align}
where \eqref{eq:} follows since the number of types is polynomial
in $n$. In order to estimate the expression at the last line above,
we partition the set of conditional types $T_{Y|X}$ into two parts:
$\mathcal{T}_{1}:=\{T_{Y|X}:I_{T}(X;Y)\le R-3\epsilon\}$ and $\mathcal{T}_{2}:=\{T_{Y|X}:I_{T}(X;Y)\ge R-3\epsilon\}$.
By Lemma \ref{lem:B-1}, for all $T_{Y|X}\in\mathcal{T}_{1}$ and
$y^{n}\in\mathcal{T}_{T_{Y}}$, 
\[
|\mathcal{T}_{T_{X|Y}}(y^{n})\cap c|\leq e^{n(R-I_{T}(X;Y)+3\epsilon)};
\]
and for all $T_{Y|X}\in\mathcal{T}_{2}$ and $y^{n}\in\mathcal{T}_{T_{Y}}$,
\[
|\mathcal{T}_{T_{X|Y}}(y^{n})\cap c|\le e^{3n\epsilon}\cdot\bone\{y^{n}\in\bigcup_{x^{n}\in c}\mathcal{T}_{T_{Y|X}}(x^{n})\}.
\]

So, the expression in \eqref{eq:} is upper bounded by the maximum
of $\gamma_{1}$ and $\gamma_{2}$ defined below: 
\begin{align*}
\gamma_{1} & :=\max_{T_{Y|X}:I_{T}(X;Y)\le R-3\epsilon}\sum_{y^{n}\in\mathcal{T}_{T_{Y}}}e^{(1+s)n(\sum T_{XY}\log P_{Y|X}-I_{T}(X;Y)+3\epsilon)-sn\sum T_{Y}\log P_{Y}}\\
 & \doteq\max_{T_{Y|X}:I_{T}(X;Y)\le R-3\epsilon}e^{-(1+s)n(D(T_{Y|X}\|P_{Y|X}|T_{X})-3\epsilon)+snD(T_{Y}\|P_{Y})},
\end{align*}
and 
\begin{align}
\gamma_{2} & :=\max_{T_{Y|X}:I_{T}(X;Y)\ge R-3\epsilon}\sum_{y^{n}\in\mathcal{T}_{T_{Y}}}e^{-sn\sum T_{Y}\log P_{Y}+(1+s)n(\sum T_{XY}\log P_{Y|X}-R+3\epsilon)}\bone\{y^{n}\in\bigcup_{x^{n}\in c}\mathcal{T}_{T_{Y|X}}(x^{n})\}\nonumber \\
 & \le\max_{T_{Y|X}:I_{T}(X;Y)\ge R-3\epsilon}e^{n(R+H_{T}(Y|X))-(1+s)nR+(1+s)n(\sum T_{XY}\log P_{Y|X}+3\epsilon)-sn\sum T_{Y}\log P_{Y}}\label{eq:-2}\\
 & =\max_{T_{Y|X}:I_{T}(X;Y)\ge R-3\epsilon}e^{-snR-(1+s)n(D(T_{Y|X}\|P_{Y|X}|T_{X})-3\epsilon)+snD(T_{Y|X}\|P_{Y}|T_{X})},\nonumber 
\end{align}
where \eqref{eq:-2} follows since $|\bigcup_{x^{n}\in c}\mathcal{T}_{T_{Y|X}}(x^{n})|\le e^{n(R+H_{T}(Y|X))}$.

Therefore, for $s>0$, 
\begin{align*}
\e^{sD_{1+s}(Q_{Y^{n}}\|P_{Y}^{\otimes n})} & \dotleq\max_{T_{Y|X}}e^{-(1+s)n(D(T_{Y|X}\|P_{Y|X}|T_{X})-3\epsilon)+snD(T_{Y}\|P_{Y})}\max\{1,e^{-snR+snI_{T}(X;Y)}\}.
\end{align*}
That is, 
\begin{align}
\frac{1}{n}D_{1+s}(Q_{Y^{n}}\|P_{Y}^{\otimes n}) & \apprle\max_{T_{Y|X}}-\frac{1+s}{s}D(T_{Y|X}\|P_{Y|X}|T_{X})+D(T_{Y}\|P_{Y})+[I_{T}(X;Y)-R]^{+}+3\epsilon\frac{1+s}{s}\nonumber \\
 & =\max_{T_{Y|X}}\max\{-q'D(T_{Y|X}\|P_{Y|X}|T_{X})+D(T_{Y|X}\|P_{Y}|T_{X})-R,\nonumber \\
 & \qquad\qquad-q'D(T_{Y|X}\|P_{Y|X}|T_{X})+D(T_{Y}\|P_{Y})\}+3q'\epsilon\\
 & \le\max\{\mathbb{E}_{T_{X}}[D_{q}(P_{Y|X}\|P_{Y})]-R,\nonumber \\
 & \qquad\qquad\qquad\max_{Q_{Y|X}}-q'D(Q_{Y|X}\|P_{Y|X}|T_{X})+D(T_{X}\circ Q_{Y|X}\|P_{Y})\}+3q'\epsilon.\label{eq:-3}
\end{align}

We now show the continuity of the objective function in $T_{X}$.
To prove this, we only need to show 
\begin{equation}
\max_{Q_{Y|X}}-q'D(Q_{Y|X}\|P_{Y|X}|Q_{X})+D(Q_{X}\circ Q_{Y|X}\|P_{Y})\label{eq:-105-2}
\end{equation}
is uniformly continuous in $Q_{X}$. Observe that $\mathcal{P}(\mathcal{Y}|\mathcal{X})$
is compact, and $-q'D(Q_{Y|X}\|P_{Y|X}|Q_{X})+D(Q_{X}\circ Q_{Y|X}\|P_{Y})$
is (jointly) continuous in $(Q_{X},Q_{Y|X})$. By Lemma \ref{lemma:continuity},
we have that \eqref{eq:-105-2} is continuous in $Q_{X}$ on the compact
set $\mathcal{P}(\mathcal{X})$, and hence, it is also uniformly continuous
in $Q_{X}$ on $\mathcal{P}(\mathcal{X})$. 

By the continuity of the objective function in \eqref{eq:-3} in $T_{X}$
and by the fact that the set $\mathcal{P}_{n}(\mathcal{X})$ is dense
in $\mathcal{P}(\mathcal{X})$, the expression in \eqref{eq:-3} is
upper bounded by
\begin{align*}
 & \min_{Q_{X}}\max\{\mathbb{E}_{Q_{X}}[D_{q}(P_{Y|X}\|P_{Y})]-R,\\
 & \qquad\qquad\qquad\max_{Q_{Y|X}}-q'D(Q_{Y|X}\|P_{Y|X}|Q_{X})+D(Q_{Y}\|P_{Y})\}+3q'\epsilon.
\end{align*}
Letting $\epsilon\downarrow0$ yields the desired upper bound for
$q\in(1,\infty)$.

The desired upper bound for $q=\infty$ follows similarly.

\begin{rem}
The proof above is based on the strong packing-covering lemma (i.e.,
Lemma \eqref{lem:B-1}). However, the upper bound in Statement 2 can
be also proven by using the one-shot bound in Lemma \ref{lem:oneshot};
see remark \ref{rem:simplebound}.
\end{rem}

\subsection{Upper Bound in Statement 2 }

Similarly to the upper bound in Statement 1, for $-1<s<0$, we can
prove that 
\begin{align*}
\e^{sD_{1+s}(Q_{Y^{n}}\|P_{Y}^{\otimes n})} & \dotgeq\max_{T_{Y|X}}e^{-(1+s)nD(T_{Y|X}\|P_{Y|X}|T_{X})+snD(T_{Y}\|P_{Y})-3n\epsilon}\min\{1,e^{-sn(R-3\epsilon-I_{T}(X;Y))}\}.
\end{align*}
That is, 
\begin{align*}
\frac{1}{n}D_{1+s}(Q_{Y^{n}}\|P_{Y}^{\otimes n}) & \apprle\min_{T_{Y|X}}-q'D(T_{Y|X}\|P_{Y|X}|T_{X})+D(T_{Y}\|P_{Y})+[I_{T}(X;Y)-R+3\epsilon]^{+}-\frac{\epsilon}{q-1}\\
 & =\min_{T_{Y|X}}\max\{-q'D(T_{Y|X}\|P_{Y|X}|T_{X})+D(T_{Y}\|P_{Y}),\\
 & \qquad\qquad-q'D(T_{Y|X}\|P_{Y|X}|T_{X})+D(T_{Y|X}\|P_{Y}|T_{X})-R+3\epsilon\}-\frac{\epsilon}{q-1}.
\end{align*}
By Lemma \ref{lemma:continuity} again and letting $n\to\infty$ and
$\epsilon\downarrow0$ yields the desired upper bound.

\subsection{\label{subsec:Upper-Bound-in}Upper Bound in Statement 3 }

We first consider $q\in(1,\infty)$. Let $R>3\epsilon>0$. Let $\mathcal{C}_{T_{X}}:=\{X^{n}(m)\}_{m\in[e^{nR}/|\mathcal{P}_{n}(\mathcal{X})|]}$
be a set of random sequences such that $X^{n}(m),m\in[e^{nR}/|\mathcal{P}_{n}(\mathcal{X})|]$
are drawn independently for different $m$'s and according to the
same distribution $\Unif(\mathcal{T}_{T_{X}})$. The rate of $\mathcal{C}_{T_{X}}$
is hence $R'=R-o_{n}(1)$. By Lemma \ref{lem:B-1}, there is a realization
$c_{T_{X}}$ of $\mathcal{C}_{T_{X}}$ satisfying $\mathcal{B}_{1}(\epsilon|T_{X})\cap\mathcal{B}_{2}(\epsilon|T_{X})$.
That is, 
\begin{equation}
e^{n(R'-I_{T}(X;Y)-\epsilon)}\le|\mathcal{T}_{T_{X|Y}}(y^{n})\cap c_{T_{X}}|\leq e^{n(R'-I_{T}(X;Y)+\epsilon)},\qquad\forall y^{n}\in\mathcal{T}_{T_{Y}}\label{eq:-73}
\end{equation}
for all $T_{Y|X}$ such that $I_{T}(X;Y)\le R'-3\epsilon$. Here,
$T_{Y}T_{X|Y}=T_{X}T_{Y|X}.$

Let $c:=\bigcup_{T_{X}}c_{T_{X}}$. Let $f:[e^{nR}]\to\mathcal{X}^{n}$
be the resolvability code based on $c$, i.e., the deterministic map
given by $f(m)=x^{n}(m)$ with $x^{n}(m)$ being the $m$-th codeword
in $c$.

Denote $q=1+s$. Observe that 
\begin{align}
 & \e^{sD_{1+s}(P_{Y}^{\otimes n}\|Q_{Y^{n}})}\nonumber \\
 & =\sum_{y^{n}}P^{1+s}(y^{n})(\sum_{m}e^{-nR}P(y^{n}|x^{n}(m)))^{-s}\label{eq:-44}\\
 & =\sum_{T_{Y}}\sum_{y^{n}\in\mathcal{T}_{T_{Y}}}e^{snR+(1+s)n\sum T_{Y}\log P_{Y}}(\sum_{T_{X|Y}}e^{n\sum T_{XY}\log P_{Y|X}}\cdot|\mathcal{T}_{T_{X|Y}}(y^{n})\cap c|)^{-s}\\
 & \le\sum_{T_{Y}}\sum_{y^{n}\in\mathcal{T}_{T_{Y}}}e^{snR+(1+s)n\sum T_{Y}\log P_{Y}}(\sum_{T_{X|Y}:I_{T}(X;Y)\le R'-3\epsilon}e^{n(R'-I_{T}(X;Y)+\sum T_{XY}\log P_{Y|X}-\epsilon)})^{-s}\label{eq:-101}\\
 & \dotleq\max_{T_{Y}}\min_{T_{X|Y}:I_{T}(X;Y)\le R'-3\epsilon}e^{sn(D(T_{Y|X}\|P_{Y|X}|T_{X})+\epsilon)-(1+s)nD(T_{Y}\|P_{Y})},\label{eq:-45}
\end{align}
where \eqref{eq:-101} follows since we impose the constraint $I_{T}(X;Y)\le R'-3\epsilon$
to the sum over $T_{X|Y}$, and under this constraint, \eqref{eq:-73}
holds. 

Therefore, 
\begin{align}
\frac{1}{n}D_{1+s}(P_{Y}^{\otimes n}\|Q_{Y^{n}}) & \apprle\max_{T_{Y}}\min_{T_{X|Y}:I_{T}(X;Y)\le R-4\epsilon}D(T_{Y|X}\|P_{Y|X}|T_{X})-\frac{1+s}{s}D(T_{Y}\|P_{Y})+\epsilon.\label{eq:-46}
\end{align}
By Lemma \ref{lemma:continuity} again and letting $\epsilon\downarrow0$
yields the desired upper bound for $q\in(1,\infty)$.

The desired upper bound for $q=\infty$ follows similarly.

\subsection{Lower Bounds in Statements 1 and 2 }

The lower bound in Statements 1 for $q\in(1,\infty]$ and the lower
bound in Statement 2 were already shown by the present author and
Tan in \cite[Theorem 1 \& Remark 10]{yu2019renyi}.

\subsection{\label{subsec:Lower-Bound-in}Lower Bound in Statement 3}

It is easy to see that if $R<R_{\min}$, then the support of $P_{Y}^{\otimes n}$
cannot be covered by the support of $Q_{Y^{n}}$. So, in this case,
$D_{q}(P_{Y}^{\otimes n}\|Q_{Y^{n}})=+\infty$ for any resolvability
code with rate $R$. We next focus on the case $R>R_{\min}$. For
this case, we first consider $q\in(1,\infty)$.

Denote $A=\{f(m):m\in[e^{nR}]\}$ and $A_{T_{X}}:=A\cap\mathcal{T}_{T_{X}}$.
Denote 
\begin{align}
B_{T_{Y}} & :=\bigcup_{T_{X|Y}:I_{T}(X;Y)>R+\epsilon}\bigcup_{x^{n}\in A_{T_{X}}}\mathcal{T}_{T_{Y|X}}(x^{n})\label{eq:-6}\\
 & =\{y^{n}:\exists T_{X|Y},\exists x^{n}\in A_{T_{X}}\textrm{ s.t. }I_{T}(X;Y)>R+\epsilon,(x^{n},y^{n})\in\mathcal{T}_{T_{XY}}\}\nonumber \\
 & =\{y^{n}:\exists T_{X|Y},\exists x^{n}\in A\textrm{ s.t. }I_{T}(X;Y)>R+\epsilon,(x^{n},y^{n})\in\mathcal{T}_{T_{XY}}\}\nonumber \\
 & =\{y^{n}:\exists x^{n}\in B_{y^{n}}\}\nonumber \\
 & =\{y^{n}:B_{y^{n}}\neq\emptyset\},\nonumber 
\end{align}
where 
\[
B_{y^{n}}:=\bigcup_{T_{X|Y}:I_{T}(X;Y)>R+\epsilon}\mathcal{T}_{T_{X|Y}}(y^{n})\cap A.
\]
By definition in \eqref{eq:-6}, it is obvious that 
\begin{align*}
|B_{T_{Y}}| & \le\max_{T_{X|Y}:I_{T}(X;Y)>R+\epsilon}e^{nR+nH_{T}(Y|X)+o(n)}\\
 & \le\max_{T_{X|Y}:I_{T}(X;Y)>R+\epsilon}e^{nI_{T}(X;Y)+nH_{T}(Y|X)-n\epsilon+o(n)}\\
 & \le e^{nH_{T}(Y)-2n\epsilon}
\end{align*}
for sufficiently large $n$. So, the set $\mathcal{T}_{T_{Y}}$ is
exponentially larger than $B_{T_{Y}}$. The set $\mathcal{T}_{T_{Y}}\backslash B_{T_{Y}}$
contains only the sequences $y^{n}$ such that $B_{y^{n}}$ is empty.
That is, for each $y^{n}\in\mathcal{T}_{T_{Y}}\backslash B_{T_{Y}}$,
there is no sequence $x^{n}$ in $c$ together with $y^{n}$ having
joint type $T_{XY}$ such that $I_{T}(X;Y)>R+\epsilon$. In other
words, for each $y^{n}\in\mathcal{T}_{T_{Y}}\backslash B_{T_{Y}}$,
all sequences $x^{n}$ in $c$ together with $y^{n}$ have a joint
type $T_{XY}$ such that $I_{T}(X;Y)\le R+\epsilon$.

Denote $q=1+s$. Observe that 
\begin{align}
 & \e^{sD_{1+s}(P_{Y}^{\otimes n}\|Q_{Y^{n}})}\nonumber \\
 & =\sum_{y^{n}}P^{1+s}(y^{n})(\sum_{m}e^{-nR}P(y^{n}|x^{n}(m)))^{-s}\label{eq:-104}\\
 & =\sum_{T_{Y}}\sum_{y^{n}\in\mathcal{T}_{T_{Y}}}e^{snR+(1+s)n\sum T_{Y}\log P_{Y}}(\sum_{T_{X|Y}}e^{n\sum T_{XY}\log P_{Y|X}}\cdot|\mathcal{T}_{T_{X|Y}}(y^{n})\cap A|)^{-s}\\
 & \ge\sum_{T_{Y}}\sum_{y^{n}\in\mathcal{T}_{T_{Y}}\backslash B_{T_{Y}}}e^{snR+(1+s)n\sum T_{Y}\log P_{Y}}(\sum_{T_{X|Y}}e^{n\sum T_{XY}\log P_{Y|X}}\cdot|\mathcal{T}_{T_{X|Y}}(y^{n})\cap A|)^{-s}\\
 & =\sum_{T_{Y}}\sum_{y^{n}\in\mathcal{T}_{T_{Y}}\backslash B_{T_{Y}}}e^{snR+(1+s)n\sum T_{Y}\log P_{Y}}(\sum_{T_{X|Y}:I_{T}(X;Y)\le R+\epsilon}e^{n\sum T_{XY}\log P_{Y|X}}\cdot|\mathcal{T}_{T_{X|Y}}(y^{n})\cap A|)^{-s}\label{eq:-7}\\
 & \ge\sum_{T_{Y}}e^{snR+(1+s)n\sum T_{Y}\log P_{Y}}|\mathcal{T}_{T_{Y}}\backslash B_{T_{Y}}|\nonumber \\
 & \qquad\times(\frac{1}{|\mathcal{T}_{T_{Y}}\backslash B_{T_{Y}}|}\sum_{y^{n}\in\mathcal{T}_{T_{Y}}\backslash B_{T_{Y}}}\sum_{T_{X|Y}:I_{T}(X;Y)\le R+\epsilon}e^{n\sum T_{XY}\log P_{Y|X}}\cdot|\mathcal{T}_{T_{X|Y}}(y^{n})\cap A|)^{-s}\label{eq:-8}\\
 & \doteq\max_{T_{Y}}\min_{T_{X|Y}:I_{T}(X;Y)\le R+\epsilon}e^{snR+(1+s)n\sum T_{Y}\log P_{Y}}|\mathcal{T}_{T_{Y}}\backslash B_{T_{Y}}|^{1+s}\nonumber \\
 & \qquad\times(\sum_{y^{n}\in\mathcal{T}_{T_{Y}}\backslash B_{T_{Y}}}e^{n\sum T_{XY}\log P_{Y|X}}\cdot|\mathcal{T}_{T_{X|Y}}(y^{n})\cap A|)^{-s}\\
 & \dotgeq\max_{T_{Y}}\min_{T_{X|Y}:I_{T}(X;Y)\le R+\epsilon}e^{snR+(1+s)n\sum T_{Y}\log P_{Y}+n(1+s)H_{T}(Y)-sn\sum T_{XY}\log P_{Y|X}}\nonumber \\
 & \qquad\times(\sum_{y^{n}\in\mathcal{T}_{T_{Y}}}|\mathcal{T}_{T_{X|Y}}(y^{n})\cap A|)^{-s}\\
 & =\max_{T_{Y}}\min_{T_{X|Y}:I_{T}(X;Y)\le R+\epsilon}e^{-(1+s)nD(T_{Y}\|P_{Y})-sn\sum T_{XY}\log P_{Y|X}}(\sum_{y^{n}\in\mathcal{T}_{T_{Y}}}Q_{X^{n}}(\mathcal{T}_{T_{X|Y}}(y^{n})))^{-s}\label{eq:-102}\\
 & \doteq\max_{T_{Y}}\min_{T_{X|Y}:I_{T}(X;Y)\le R+\epsilon}e^{-(1+s)nD(T_{Y}\|P_{Y})-sn\sum T_{XY}\log P_{Y|X}}(e^{nH_{T}(Y|X)}Q_{X^{n}}(\mathcal{T}_{T_{X}}))^{-s}\label{eq:-12}\\
 & \ge\max_{T_{Y}}\min_{T_{X|Y}:I_{T}(X;Y)\le R+\epsilon}e^{-(1+s)nD(T_{Y}\|P_{Y})+snD(T_{Y|X}\|P_{Y|X}|T_{X})},\label{eq:-105}
\end{align}
where \eqref{eq:-7} follows by the property given above the equation
chain, \eqref{eq:-8} follows by Jensen's inequality, in \eqref{eq:-102},
$Q_{X^{n}}$ is the uniform distribution over $A$, \eqref{eq:-12}
follows by Lemma \ref{lem:typeequality}, and the last line follows
since $Q_{X^{n}}(\mathcal{T}_{T_{X}})\le1$.

Therefore, 
\begin{align*}
\lim_{n\to\infty}\frac{1}{n}D_{1+s}(P_{Y}^{\otimes n}\|Q_{Y^{n}}) & \ge\max_{Q_{Y}}\min_{Q_{X|Y}:I_{Q}(X;Y)\le R+\epsilon}D(Q_{Y|X}\|P_{Y|X}|Q_{X})-\frac{1+s}{s}D(Q_{Y}\|P_{Y}).
\end{align*}
By Lemma \ref{lemma:continuity} again and letting $\epsilon\downarrow0$
yields the desired lower bound for $q\in(1,\infty)$.

By the monotonicity of the Rényi divergence in its order, for $q=\infty$,
it holds that 
\begin{align*}
 & \lim_{n\to\infty}\frac{1}{n}\inf_{f:[e^{nR}]\to\mathcal{X}^{n}}D_{\infty}(P_{Y}^{\otimes n}\|Q_{Y^{n}})\\
 & =\inf_{n\ge1}\frac{1}{n}\inf_{f:[e^{nR}]\to\mathcal{X}^{n}}\lim_{q\to\infty}D_{q}(P_{Y}^{\otimes n}\|Q_{Y^{n}})\\
 & \ge\lim_{q\to\infty}\inf_{n\ge1}\frac{1}{n}\inf_{f:[e^{nR}]\to\mathcal{X}^{n}}D_{q}(P_{Y}^{\otimes n}\|Q_{Y^{n}})\\
 & =\lim_{q\to\infty}\max_{Q_{Y}}\min_{Q_{X|Y}:I_{Q}(X;Y)\le R}D(Q_{Y|X}\|P_{Y|X}|Q_{X})-q'D(Q_{Y}\|P_{Y})\\
 & =\max_{Q_{Y}}\min_{Q_{X|Y}:I_{Q}(X;Y)\le R}D(Q_{Y|X}\|P_{Y|X}|Q_{X})-D(Q_{Y}\|P_{Y}),
\end{align*}
where the last line follows by Lemma \ref{lemma:continuity}.

\section{\label{sec:Proof-of-Proposition}Proof of Proposition \ref{prop:The-expressions-in}}

Statement 1: For the last term in \eqref{eq:-15},
\begin{align}
 & \max_{Q_{Y|X}}-q'D(Q_{Y|X}\|P_{Y|X}|Q_{X})+D(Q_{Y}\|P_{Y})\nonumber \\
 & =\max_{S_{Y}}\max_{Q_{Y|X}}-q'D(Q_{Y|X}\|P_{Y|X}|Q_{X})+D(Q_{Y}\|P_{Y})-D(Q_{Y}\|S_{Y})\label{eq:-80}\\
 & =\max_{S_{Y}}\max_{Q_{Y|X}}-q'D(Q_{Y|X}\|Q_{Y|X}^{*}|Q_{X})+q'\mathbb{E}_{Q_{X}}\big[\log\mathbb{E}_{P_{Y|X}}[(\frac{S_{Y}}{P_{Y}})^{1/q'}]\big]\nonumber \\
 & =\max_{S_{Y}}q'\mathbb{E}_{Q_{X}}\big[\log\mathbb{E}_{P_{Y|X}}[(\frac{S_{Y}}{P_{Y}})^{1/q'}]\big],\label{eq:-81}
\end{align}
where $Q_{Y|X}^{*}=\frac{P_{Y|X}\cdot(\frac{S_{Y}}{P_{Y}})^{1/q'}}{\mathbb{E}_{P_{Y|X}}[(\frac{S_{Y}}{P_{Y}})^{1/q'}]}$. 

Statement 2: Denoting
\begin{align*}
\theta(Q_{Y|X},\lambda) & :=-q'D(Q_{Y|X}\|P_{Y|X}|Q_{X})+\lambda\left(D(Q_{Y|X}\|P_{Y}|Q_{X})-R\right)\\
 & \qquad\qquad\qquad+(1-\lambda)D(Q_{Y}\|P_{Y}),
\end{align*}
the expression in \eqref{eq:-21} (without the term $o_{n}(1)$) is
equal to 
\begin{align}
 & \min_{Q_{Y|X}}\max_{\lambda\in[0,1]}\theta(Q_{Y|X},\lambda)\nonumber \\
 & =\max_{\lambda\in[0,1]}\min_{Q_{Y|X}}\theta(Q_{Y|X},\lambda)\label{eq:-78}\\
 & =\max_{\lambda\in[0,1]}\min_{Q_{Y|X}}\max_{S_{Y}}\theta(Q_{Y|X},\lambda)-(1-\lambda)D(Q_{Y}\|S_{Y})\nonumber \\
 & =\max_{\lambda\in[0,1]}\max_{S_{Y}}\min_{Q_{Y|X}}\theta(Q_{Y|X},\lambda)-(1-\lambda)D(Q_{Y}\|S_{Y})\label{eq:-79}\\
 & =\max_{\lambda\in[0,1]}\max_{S_{Y}}-(\lambda-q')\mathbb{E}_{Q_{X}}[\log\sum_{y}P_{Y|X}^{\frac{-q'}{\lambda-q'}}P_{Y}^{\frac{1}{\lambda-q'}}S_{Y}^{\frac{\lambda-1}{\lambda-q'}}]-\lambda R,\nonumber 
\end{align}
where \eqref{eq:-78} and \eqref{eq:-79} follow by the minimax theorem,
and the last line follows by similar steps from \eqref{eq:-80} to
\eqref{eq:-81}.  

Similarly, one can show that the expression in \eqref{eq:bound2}
is equal to the one in \eqref{eq:-97}. 

Statement 3: The expression in \eqref{eq:-47-5} (without the term
$o_{n}(1)$) is equal to
\begin{align*}
 & \max_{\hat{Q}_{Y}}\min_{\hat{Q}_{X|Y}}\sup_{\lambda\ge0,f}D(\hat{Q}_{X|Y}\|P_{X|Y}|\hat{Q}_{Y})-D(\hat{Q}_{X}\|P_{X})-(q'-1)D(\hat{Q}_{Y}\|P_{Y})\\
 & \qquad+\lambda(I_{\hat{Q}}(X;Y)-R)+\mathbb{E}_{\hat{Q}}f-\mathbb{E}_{Q}f.
\end{align*}
By the minimax theorem, we swap $\min_{\hat{Q}_{X|Y}}$ and $\sup_{\lambda\ge0,f}$,
and then add the term $(1+\lambda)D(\hat{Q}_{X}\|S_{X})$ and insert
$\min_{S_{X}}$ as follows. 
\begin{align*}
 & \max_{\hat{Q}_{Y}}\sup_{\lambda\ge0,f}\min_{\hat{Q}_{X|Y}}\min_{S_{X}}D(\hat{Q}_{X|Y}\|P_{X|Y}|\hat{Q}_{Y})-D(\hat{Q}_{X}\|P_{X})-(q'-1)D(\hat{Q}_{Y}\|P_{Y})\\
 & \qquad+\lambda(I_{\hat{Q}}(X;Y)-R)+\mathbb{E}_{\hat{Q}}f-\mathbb{E}_{Q}f+(1+\lambda)D(\hat{Q}_{X}\|S_{X}).
\end{align*}
Swapping the two minimizations, we obtain 
\begin{align*}
 & \max_{\hat{Q}_{Y}}\sup_{\lambda\ge0,f}\min_{S_{X}}-(1+\lambda)\mathbb{E}_{\hat{Q}_{Y}}\log\mathbb{E}_{R_{X}}[(\frac{P_{Y|X}}{P_{Y}}e^{f})^{\frac{1}{1+\lambda}}]-(q'-1)D(\hat{Q}_{Y}\|P_{Y})-\lambda R-\mathbb{E}_{Q}f.
\end{align*}
By the minimax theorem again, we swap $\max_{\hat{Q}_{Y}}$ and $\min_{S_{X}}$,
and then obtain the desired expression in \eqref{eq:-96}. 

Similarly, one can show that the expression in \eqref{eq:-47} is
equal to the one in \eqref{eq:-98}.

\section{\label{sec:Proof-of-Theorem-Resolvability-Rates}Proof of Theorem
\ref{thm:ResolvabilityRates}}

Statement 1 for $q\in(1,\infty]$: It was already shown by the present
author and Tan in \cite{yu2019renyi} that 
\[
\inf\{R:\frac{1}{n}\inf_{f}D_{q}(Q_{Y^{n}}\|P_{Y}^{\otimes n})\rightarrow0\}\ge R_{q}(P_{Y|X},P_{Y}).
\]
So, it only remains to show that 
\begin{equation}
\inf\{R:\frac{1}{n}\inf_{f}D_{q}(Q_{Y^{n}}\|P_{Y}^{\otimes n})\rightarrow0\}\le R_{q}(P_{Y|X},P_{Y}).\label{eq:-114}
\end{equation}
If $R>R_{q}(P_{Y|X},P_{Y})$, i.e., $R>\mathbb{E}_{P_{X}}[D_{q}(P_{Y|X}\|P_{Y})]$
for some $P_{X}\in\mathcal{P}(P_{Y|X},P_{Y})$, then 
\begin{align*}
 & \lim_{n\to\infty}\frac{1}{n}\inf_{f:[e^{nR}]\to\mathcal{X}^{n}}D_{q}(Q_{Y^{n}}\|P_{Y}^{\otimes n})\\
 & =\min_{Q_{X}}\max\{\mathbb{E}_{Q_{X}}[D_{q}(P_{Y|X}\|P_{Y})]-R,\\
 & \qquad\qquad\qquad\max_{Q_{Y|X}}-q'D(Q_{Y|X}\|P_{Y|X}|Q_{X})+D(Q_{Y}\|P_{Y})\}\\
 & \le\max\{\mathbb{E}_{P_{X}}[D_{q}(P_{Y|X}\|P_{Y})]-R,\\
 & \qquad\qquad\qquad\max_{Q_{Y|X}}-q'D(Q_{Y|X}\|P_{Y|X}|P_{X})+D(Q_{Y}\|P_{Y})\}\\
 & \le0,
\end{align*}
where the last inequality follow since for $Q_{X}=P_{X}$, 
\begin{align*}
 & -q'D(Q_{Y|X}\|P_{Y|X}|P_{X})+D(Q_{Y}\|P_{Y})\\
 & =-q'D(Q_{XY}\|P_{XY})+D(Q_{Y}\|P_{Y})\\
 & \le0.
\end{align*}
Hence, \eqref{eq:-114} holds.

Statement 1 for $q\in[0,1]$: Theorem \ref{thm:ResolvabilityRates}
for $q\in(0,1]$ was already shown by the present author and Tan in
\cite{yu2019renyi}. Theorem \ref{thm:ResolvabilityRates} for $q=0$
follows by Theorem \ref{thm:Resolvability}. 

Statement 2 for $q\in(1,\infty]$: It suffices to prove that $\hat{R}_{q}(P_{Y|X},P_{Y})$
is exactly the infimum of $R$, denoted by $R^{*}$, such that 
\begin{equation}
\max_{Q_{Y}}\min_{Q_{X|Y}:I_{Q}(X;Y)\le R}D(Q_{Y|X}\|P_{Y|X}|Q_{X})-q'D(Q_{Y}\|P_{Y})\le0.\label{eq:-115}
\end{equation}
The inequality in \eqref{eq:-115} is equivalent to 
\begin{align*}
 & \min_{Q_{X|Y}:I_{Q}(X;Y)\le R}D(Q_{Y|X}\|P_{Y|X}|Q_{X})\le q'D(Q_{Y}\|P_{Y}),\forall Q_{Y}\\
\Longleftrightarrow & \forall Q_{Y},\exists Q_{X|Y}\textrm{ s.t. }I_{Q}(X;Y)\le R,\,D(Q_{Y|X}\|P_{Y|X}|Q_{X})\le q'D(Q_{Y}\|P_{Y})\\
\Longleftrightarrow & R^{*}=\max_{Q_{Y}}\min_{Q_{X|Y}:D(Q_{Y|X}\|P_{Y|X}|Q_{X})\le q'D(Q_{Y}\|P_{Y})}I_{Q}(X;Y)\\
\Longleftrightarrow & R^{*}=\hat{R}_{q}(P_{Y|X},P_{Y}).
\end{align*}

Statement 2 for $q=1$: For $q=1$ and $R>R_{\min}$, 
\begin{align*}
 & \lim_{n\to\infty}\frac{1}{n}\inf_{f:[e^{nR}]\to\mathcal{X}^{n}}D_{q}(P_{Y}^{\otimes n}\|Q_{Y^{n}})\\
 & =\min_{Q_{XY}:I_{Q}(X;Y)\le R,\,Q_{Y}=P_{Y}}D(Q_{Y|X}\|P_{Y|X}|Q_{X}),
\end{align*}
which is zero if and only if $Q_{Y|X}=P_{Y|X}$, $Q_{Y}=P_{Y}$, and
$I_{Q}(X;Y)\le R$ for some $Q_{XY}$. So, $R^{*}=\hat{R}_{1}(P_{Y|X},P_{Y})$.

Statement 2 for $q\in[0,1)$: Statement 2 for $q\in(0,1)$ was implied
by Statement 1 for $q\in(0,1)$ by the skew symmetry $D_{q}(Q\|P)=-q'D_{1-q}(P\|Q)$
\cite{Erven}.

\section{\label{sec:Proof-of-Theorem-expiid}Proof of Theorem \ref{thm:exponentforiid}}

To evaluate the performance of i.i.d. codes, we substitute $Q_{X}\leftarrow P_{X}^{\otimes n},P_{Y|X}\leftarrow P_{Y|X}^{\otimes n},P_{Y}\leftarrow P_{Y}^{\otimes n}$
into Lemma \ref{lem:oneshot}, and by Remark \ref{rem:simplebound},
obtain that for $q\in[2,\infty)$, 
\begin{align}
 & \e^{(q-1)D_{q}(Q_{Y^{n}|\mathcal{C}_{n}}\|P_{Y}^{\otimes n}|Q_{\mathcal{C}_{n}})}\nonumber \\
 & \le\sum_{t=1}^{\tilde{q}-1}\big\{ tS(\tilde{q},t)\Gamma(q-t)+S(\tilde{q},t)\Gamma(\tilde{q}-t)\big\}+\tilde{q}\Gamma(1)+\Gamma(0),\label{eq:-20}
\end{align}
and 
\begin{align}
 & \e^{(q-1)D_{q}(Q_{Y^{n}|\mathcal{C}_{n}}\|P_{Y}^{\otimes n}|Q_{\mathcal{C}_{n}})}\nonumber \\
 & \ge(1-\e^{-nR})^{q-1}+\Gamma(q-1),\label{eq:-19}
\end{align}
where $\Gamma(s):=e^{-n\gamma(s)}.$ 

Using \eqref{eq:-20} and noting that $\Gamma(0)=1$, we have that
\begin{align*}
 & (q-1)D_{q}(Q_{Y^{n}|\mathcal{C}_{n}}\|P_{Y}^{\otimes n}|Q_{\mathcal{C}_{n}})\\
 & \le\log[1+\tilde{q}\Gamma(1)+\sum_{t=1}^{\tilde{q}-1}tS(\tilde{q},t)\Gamma(q-t)+S(\tilde{q},t)\Gamma(\tilde{q}-t)]\\
 & \le\tilde{q}\Gamma(1)+\sum_{t=1}^{\tilde{q}-1}tS(\tilde{q},t)\Gamma(q-t)+S(\tilde{q},t)\Gamma(\tilde{q}-t)\\
 & \dotleq\max_{s\in[\tilde{q}-1]\cup([1:\tilde{q}-1]+\hat{q})}\Gamma(s).
\end{align*}
Given $(Q,P)$, $sD_{1+s}(Q\|P)$ is convex in $s\in\mathbb{R}$.
So, the maximum at the last line is attained at $s=1$ or $s=q-1$,
yielding the upper bound $\e^{-n\min\{\gamma(1),\gamma(q-1)\}}$.

We next focus on the other direction. Using \eqref{eq:-19}, we obtain
that 
\begin{align}
 & D_{q}(Q_{Y^{n}|\mathcal{C}_{n}}\|P_{Y}^{\otimes n}|Q_{\mathcal{C}_{n}})\nonumber \\
 & \ge\frac{1}{q-1}\log[(1-\e^{-nR})^{q-1}+\Gamma(q-1)]\nonumber \\
 & \doteq(1-\e^{-nR})^{q-1}-1+\e^{n(q-1)(D_{q}(P_{Y|X}\|P_{Y}|P_{X})-R)}\label{eq:-1-1}\\
 & \doteq\e^{n(q-1)(D_{q}(P_{Y|X}\|P_{Y}|P_{X})-R)}\label{eq:-9-1}\\
 & =\e^{-n\gamma(q-1)}.
\end{align}
On the other hand, by using the monotonicity of the Rényi divergence,
this lower bound further implies that 
\[
D_{q}(Q_{Y^{n}|\mathcal{C}_{n}}\|P_{Y}^{\otimes n}|Q_{\mathcal{C}_{n}})\ge D_{2}(Q_{Y^{n}|\mathcal{C}_{n}}\|P_{Y}^{\otimes n}|Q_{\mathcal{C}_{n}})\dotgeq\e^{-n\gamma(1)}.
\]
Combining the two lower bounds above yields that $D_{q}(Q_{Y^{n}|\mathcal{C}_{n}}\|P_{Y}^{\otimes n}|Q_{\mathcal{C}_{n}})\dotgeq\e^{-n\min\{\gamma(1),\gamma(q-1)\}}.$

\section{\label{sec:Proof-of-Theorem-exp}Proof of Theorem \ref{thm:exponent}}

To evaluate the performance of the typical set codes, we substitute
$Q_{X}\leftarrow Q_{X^{n}}:=P_{X}^{\otimes n}(\cdot|\mathcal{T}_{\epsilon}^{(n)}(P_{X})),P_{Y|X}\leftarrow P_{Y|X}^{\otimes n},P_{Y}\leftarrow P_{Y}^{\otimes n}$
into Lemma \ref{lem:oneshot}, and by Remark \ref{rem:simplebound},
obtain that for $q\in[2,\infty)$, 
\begin{align}
 & \e^{(q-1)D_{q}(Q_{Y^{n}|\mathcal{C}_{n}}\|P_{Y}^{\otimes n}|Q_{\mathcal{C}_{n}})}\nonumber \\
 & \le\sum_{t=1}^{\tilde{q}-1}\big\{ tS(\tilde{q},t)\Gamma(q-t)+S(\tilde{q},t)\Gamma(\tilde{q}-t)\big\}+\tilde{q}\Gamma(1)+\Gamma(0),\label{eq:-20-2}
\end{align}
where 
\[
\Gamma(s):=e^{-n\gamma_{n}(s)+(q-s-1)D_{\infty}(Q_{Y^{n}}\|P_{Y}^{\otimes n})}
\]
with 
\[
\gamma_{n}(s):=s(R-\frac{1}{n}D_{s+1}(P_{Y|X}^{\otimes n}\|P_{Y}^{\otimes n}|Q_{X^{n}})).
\]

Observe that 
\begin{align}
D_{\infty}(Q_{Y^{n}}\|P_{Y}^{\otimes n}) & \leq D_{\infty}(Q_{X^{n}}\|P_{X}^{\otimes n})\\
 & =\log\frac{1}{P_{X}^{\otimes n}(\mathcal{T}_{\epsilon})}\label{eq:-51-1}\\
 & \leq\frac{1}{P_{X}^{\otimes n}(\mathcal{T}_{\epsilon})}-1\\
 & \doteq P_{X}^{\otimes n}((\mathcal{T}_{\epsilon})^{c}),\label{eq:-58-1}
\end{align}
where $(\mathcal{T}_{\epsilon})^{c}:=\mathcal{X}^{n}\backslash\mathcal{T}_{\epsilon}$.
Now we bound $P_{X}^{\otimes n}((\mathcal{T}_{\epsilon})^{c})$ using
the Chernoff bound in Lemma \ref{lem:chernoff} as 
\begin{align}
P_{X}^{\otimes n}((\mathcal{T}_{\epsilon})^{c}) & \leq2\left|\mathcal{X}\right|\e^{-\frac{\epsilon^{2}nP_{\min}}{3}},\label{eq:-59-1}
\end{align}
where recall that $P_{\min}=\min_{x:P_{X}(x)>0}P_{X}(x)$. Substituting
\eqref{eq:-59-1} into \eqref{eq:-58-1}, we obtain 
\begin{equation}
D_{\infty}(Q_{Y^{n}}\|P_{Y}^{\otimes n})\dotleq2\left|\mathcal{X}\right|\e^{-\frac{\epsilon^{2}nP_{\min}}{3}}.\label{eq:-17}
\end{equation}

On the other hand, 
\begin{align*}
 & \e^{sD_{1+s}(P_{Y|X}^{\otimes n}\|P_{Y}^{\otimes n}|Q_{X^{n}})}\\
 & =\mathbb{E}_{Q_{X^{n}}}\mathbb{E}_{P_{Y|X}^{\otimes n}}[(\frac{P_{Y|X}^{\otimes n}}{P_{Y}^{\otimes n}})^{s}]\\
 & =\mathbb{E}_{Q_{X^{n}}}[e^{\sum_{i=1}^{n}\log\mathbb{E}_{P_{Y|X=X_{i}}}(\frac{P_{Y|X=X_{i}}}{P_{Y}})^{s}}]\\
 & \le e^{n(1+\epsilon)\mathbb{E}_{P_{X}}\log\mathbb{E}_{P_{Y|X}}(\frac{P_{Y|X}}{P_{Y}})^{s}}.
\end{align*}
That is, 
\[
\frac{1}{n}D_{1+s}(P_{Y|X}^{\otimes n}\|P_{Y}^{\otimes n}|Q_{X^{n}})\le(1+\epsilon)\mathbb{E}_{P_{X}}[D_{1+s}(P_{Y|X}\|P_{Y})],
\]
and hence, 
\[
\limsup_{n\to\infty}\gamma_{n}(s)\le\gamma(s,\epsilon),
\]
where recall that 
\[
\gamma(s,\epsilon)=s(R-(1+\epsilon)\mathbb{E}_{P_{X}}[D_{1+s}(P_{Y|X}\|P_{Y})]).
\]

Substituting these into \eqref{eq:-20-2}, we have that 
\begin{align*}
 & (q-1)D_{q}(Q_{Y^{n}|\mathcal{C}_{n}}\|P_{Y}^{\otimes n}|Q_{\mathcal{C}_{n}})\\
 & \le\log[e^{(q-1)D_{\infty}(Q_{Y^{n}}\|P_{Y}^{\otimes n})}+\tilde{q}\Gamma(1)+\sum_{t=1}^{\tilde{q}-1}tS(\tilde{q},t)\Gamma(q-t)+S(\tilde{q},t)\Gamma(\tilde{q}-t)]\\
 & \le e^{(q-1)D_{\infty}(Q_{Y^{n}}\|P_{Y}^{\otimes n})}-1+\tilde{q}\Gamma(1)+\sum_{t=1}^{\tilde{q}-1}tS(\tilde{q},t)\Gamma(q-t)+S(\tilde{q},t)\Gamma(\tilde{q}-t)\\
 & \dotleq\max\{\e^{-\frac{\epsilon^{2}nP_{\min}}{3}},\max_{s\in[\tilde{q}-1]\cup([1:\tilde{q}-1]+\hat{q})}\Gamma(s)\}\\
 & \doteq\max\{\e^{-\frac{\epsilon^{2}nP_{\min}}{3}},e^{-n\min\{\gamma(1,\epsilon),\gamma(q-1,\epsilon)\}}\}\\
 & =e^{-n\min\{\frac{\epsilon^{2}P_{\min}}{3},\gamma(1,\epsilon),\gamma(q-1,\epsilon)\}}.
\end{align*}

\section{\label{sec:Proof-of-Theorem-exp-2}Proof of Theorem \ref{thm:exponent-1}}

By the continuity of $\hat{R}_{q}(P_{Y|X},P_{Y})$ in $q\ge1$ and
the monotonicity of the Rényi divergence in its order, we need to
consider the case $q>1$.

By assumption, $R>\hat{R}_{q}(P_{Y|X},P_{Y})$. From the proof of
Theorem \ref{thm:ResolvabilityRates}, for any $R\ge\hat{R}_{q}(P_{Y|X},P_{Y})$,
it holds that 
\begin{equation}
\eta(R):=\max_{Q_{Y}}\eta(R,Q_{Y})\le0,\label{eq:-43}
\end{equation}
where 
\[
\eta(R,Q_{Y}):=\min_{Q_{X|Y}:I_{Q}(X;Y)\le R}D(Q_{Y|X}\|P_{Y|X}|Q_{X})-q'D(Q_{Y}\|P_{Y}).
\]
On the other hand, by setting $Q_{Y}=P_{Y}$, it is easy to see that
$\eta(R)\ge0.$ So, for all $R\ge\hat{R}_{q}(P_{Y|X},P_{Y})$, it
holds that $\eta(R)=0$, and the maximum in \eqref{eq:-43} is attained
by $Q_{Y}=P_{Y}$. We now make the following claim. 
\begin{claim}
\label{claim:For-,-the} For $R>\hat{R}_{q}(P_{Y|X},P_{Y})$, it holds
that $\eta(R,Q_{Y})\le-\epsilon_{\delta'}$ for all $Q_{Y}\notin\mathcal{B}_{\delta'/2}(P_{Y})$,
where $\epsilon_{\delta'}>0$ is a term vanishing as\footnote{Here $\delta'$ does not denote the Hölder's conjugate of $\delta$.}
$\delta'\downarrow0$ and $\mathcal{B}_{\delta'/2}(P_{Y})=\{Q_{Y}:\|Q_{Y}-P_{Y}\|_{\mathrm{TV}}\le\delta'/2\}$
denotes a ball under the TV distance. In particular, for $R>\hat{R}_{q}(P_{Y|X},P_{Y})$,
the optimization at the LHS of \eqref{eq:-43} is uniquely attained
by $Q_{Y}=P_{Y}$. 
\end{claim}
\begin{IEEEproof}[Proof of Claim \ref{claim:For-,-the}]
We now prove the above claim. By setting $Q_{X|Y}=P_{X|Y}$, we observe
that for each $Q_{Y}$, the distribution $Q_{XY}=Q_{Y}P_{X|Y}$ satisfies
\begin{align*}
 & D(Q_{Y|X}\|P_{Y|X}|Q_{X})-q'D(Q_{Y}\|P_{Y})\\
 & =D(Q_{XY}\|P_{XY})-D(Q_{X}\|P_{X})-q'D(Q_{Y}\|P_{Y})\\
 & =(1-q')D(Q_{Y}\|P_{Y})-D(Q_{X}\|P_{X})\\
 & \le(1-q')D(Q_{Y}\|P_{Y}).
\end{align*}
By Pinsker's inequality, for all $Q_{Y}\notin\mathcal{B}_{\delta'/2}(P_{Y})$,
\begin{align*}
 & D(Q_{Y|X}\|P_{Y|X}|Q_{X})-q'D(Q_{Y}\|P_{Y})\le(1-q')\delta'^{2}/2.
\end{align*}

The condition in \eqref{eq:-43} implies that for all $Q_{Y}$, $\eta(\hat{R}_{q}(P_{Y|X},P_{Y}),Q_{Y})\le0,$
i.e., for all $Q_{Y}$, there is $Q_{X|Y}^{*}$ such that $I_{Q^{*}}(X;Y)\le\hat{R}_{q}(P_{Y|X},P_{Y})$
and $D(Q_{Y|X}^{*}\|P_{Y|X}|Q_{X}^{*})\le q'D(Q_{Y}\|P_{Y})$. We
now define a new conditional distribution 
\[
Q_{X|Y}^{(\lambda)}=(1-\lambda)Q_{X|Y}^{*}+\lambda P_{X|Y},
\]
where $\lambda\in(0,1)$. We denote $Q_{XY}^{(\lambda)}:=Q_{Y}Q_{X|Y}^{(\lambda)}$,
$Q_{XY}^{*}:=Q_{Y}Q_{X|Y}^{*}$, and $Q_{XY}:=Q_{Y}P_{X|Y}$.

By the convexity of $I_{Q}(X;Y)$ in $Q_{X|Y}$ (for given $Q_{Y}$),
it holds that $Q^{(\lambda)}$ satisfies 
\begin{align*}
I_{Q^{(\lambda)}}(X;Y) & \le(1-\lambda)I_{Q^{*}}(X;Y)+\lambda I_{Q}(X;Y)\\
 & \le(1-\lambda)\hat{R}_{q}(P_{Y|X},P_{Y})+\lambda\log|\mathcal{X}|.
\end{align*}
That is, for $R>\hat{R}_{q}(P_{Y|X},P_{Y})$, there is some $\lambda\in(0,1)$
(close to $0$) such that $Q^{(\lambda)}$ satisfies $I_{Q^{(\lambda)}}(X;Y)<R$.
By the convexity of $D(Q\|P)$ in $(Q,P)$, it holds that $Q^{(\lambda)}$
satisfies 
\begin{align*}
D(Q_{X|Y}^{(\lambda)}\|P_{Y|X}|Q_{X}^{(\lambda)}) & \le(1-\lambda)D(Q_{Y|X}^{*}\|P_{Y|X}|Q_{X}^{*})+\lambda D(Q_{Y|X}\|P_{Y|X}|Q_{X})\\
 & \le(1-\lambda)q'D(Q_{Y}\|P_{Y})+\lambda[q'D(Q_{Y}\|P_{Y})+(1-q')\delta'^{2}/2]\\
 & =q'D(Q_{Y}\|P_{Y})+\lambda(1-q')\delta'^{2}/2.
\end{align*}
Hence, $\eta(R,Q_{Y})\le\lambda(1-q')\delta'^{2}/2.$ This completes
the proof of the claim above. 
\end{IEEEproof}
We now start to prove Theorem \ref{thm:exponent-1}. We consider
the typical code constructed in Section \ref{subsec:Strong-Packing-Covering-Lemma}
of size $e^{nR}(1-e^{-n\delta})$, which is denoted as $\tilde{\mathcal{C}}$.
Here $R>I(X;Y)$, since $\hat{R}_{q}(P_{Y|X},P_{Y})\ge I(X;Y)$ for
$q>1$. From Lemma \ref{lem:B-3}, there is a realization $\tilde{c}$
of $\tilde{\mathcal{C}}$ satisfying $\mathcal{B}(\epsilon,\delta')$
for some $\delta>\delta'>0$. That is, 
\begin{align}
 & \Big|\frac{\sum_{m\in[e^{nR}(1-e^{-n\delta})]}\theta_{m}(y^{n})}{e^{nR}(1-e^{-n\delta})\mu}-1\Big|\leq e^{-n\epsilon},\;\forall y^{n}\in\mathcal{T}_{\delta'}^{(n)}(P_{Y}),
\end{align}
where 
\[
\theta_{m}(y^{n}):=P_{Y|X}^{\otimes n}(y^{n}|X^{n}(m))\bone\{(X^{n}(m),y^{n})\in\mathcal{T}_{\delta}^{(n)}(P_{XY})\}
\]
and 
\begin{align*}
\mu & :=\mathbb{E}_{\tilde{\mathcal{C}}}[\theta_{m}(y^{n})]\in P_{Y}^{\otimes n}(y^{n})(1\pm e^{-n\epsilon'})
\end{align*}
for some $\epsilon'>0$. So, 
\begin{align}
 & \frac{\sum_{m\in[e^{nR}(1-e^{-n\delta})]}\theta_{m}(y^{n})}{e^{nR}(1-e^{-n\delta})}\ge(1-e^{-n\epsilon''})P_{Y}^{\otimes n}(y^{n}),\;\forall y^{n}\in\mathcal{T}_{\delta'}^{(n)}(P_{Y}),
\end{align}
for some $\epsilon''>0$.

We now consider the code constructed in Section \ref{subsec:Upper-Bound-in}
of size $e^{n(R-\delta)}$, which is denoted as $\hat{\mathcal{C}}$.
Then, \eqref{eq:-73} still holds for some realization $\hat{c}$
of $\hat{\mathcal{C}}$ with $R'=R-\delta-o_{n}(1)$. 

The final code used here is $c=\tilde{c}\cup\hat{c}$. The rate of
this code is $R$. Observe that for this code $c$, 
\begin{align*}
\e^{sD_{1+s}(P_{Y}^{\otimes n}\|Q_{Y^{n}})} & =\sum_{y^{n}}P^{1+s}(y^{n})Q^{-s}(y^{n})=\Sigma_{1,n}+\Sigma_{2,n},
\end{align*}
where 
\begin{align*}
\Sigma_{1,n} & =\sum_{y^{n}\in\mathcal{T}_{\delta'}^{(n)}}P^{1+s}(y^{n})(\sum_{m}e^{-nR}P(y^{n}|x^{n}(m)))^{-s},\\
\Sigma_{2,n} & =\sum_{y^{n}\notin\mathcal{T}_{\delta'}^{(n)}}P^{1+s}(y^{n})(\sum_{m}e^{-nR}P(y^{n}|x^{n}(m)))^{-s}.
\end{align*}
We estimate the two sums above. For the first sum, we observe that
\begin{align*}
\Sigma_{1,n} & \le\sum_{y^{n}\in\mathcal{T}_{\delta'}^{(n)}}P^{1+s}(y^{n})(e^{-nR}\sum_{m\in[e^{nR}(1-e^{-n\delta})]}P(y^{n}|x^{n}(m))\bone\{(x^{n}(m),y^{n})\in\mathcal{T}_{\delta}^{(n)}\})^{-s}\\
 & \le\sum_{y^{n}\in\mathcal{T}_{\delta'}^{(n)}}P^{1+s}(y^{n})((1-e^{-n\delta})P(y^{n})(1-e^{-n\epsilon''}))^{-s}\\
 & =\sum_{y^{n}\in\mathcal{T}_{\delta'}^{(n)}}P(y^{n})(1-e^{-n\epsilon''})^{-s}(1-e^{-n\delta})^{-s}\\
 & =P(\mathcal{T}_{\delta'}^{(n)})(1-e^{-n\epsilon''})^{-s}(1-e^{-n\delta})^{-s}\\
 & \to1\textrm{ exponentially fast.}
\end{align*}

For the second sum, following steps similar to proof steps in \eqref{eq:-44}-\eqref{eq:-46},
we observe that 
\begin{align*}
\lim_{n\to\infty}\frac{1}{ns}\log\Sigma_{2,n} & \le\lim_{n\to\infty}\max_{T_{Y}\notin\mathcal{B}_{\delta'/2}}\min_{T_{X|Y}:I_{T}(X;Y)\le R-5\epsilon}D(T_{Y|X}\|P_{Y|X}|T_{X})-q'D(T_{Y}\|P_{Y})+\epsilon.
\end{align*}
By Lemma \ref{lemma:continuity} again and letting $\epsilon\downarrow0$
and then by Claim \ref{claim:For-,-the}, we obtain that for $R-\delta>\hat{R}_{q}(P_{Y|X},P_{Y})$
(with $\delta$ chosen sufficiently small), 

\begin{align*}
\lim_{n\to\infty}\frac{1}{ns}\log\Sigma_{2,n} & \le\max_{T_{Y}\notin\mathcal{B}_{\delta'/2}}\min_{Q_{X|Y}:I_{Q}(X;Y)\le R}D(Q_{Y|X}\|P_{Y|X}|Q_{X})-q'D(Q_{Y}\|P_{Y})\\
 & \le-\epsilon_{\delta'}.
\end{align*}
That is, $\Sigma_{2,n}\to0$ exponentially fast.

Therefore, given $\delta$ and $\delta'$, $\e^{sD_{1+s}(P_{Y}^{\otimes n}\|Q_{Y^{n}})}\le1+e^{-n(\epsilon'''+o_{n}(1))}$
for some $\epsilon'''>0$, i.e., for $q>1$, $D_{q}(P_{Y}^{\otimes n}\|Q_{Y^{n}})\le\frac{1}{q-1}e^{-n(\epsilon'''+o_{n}(1))}.$ 

\section{\label{sec:Proof-of-Corollary-asymptotics}Proof of Corollary \ref{cor:asymptotics}}

Statement 1: For $r\ge1$, define 
\[
\varphi_{r}(\alpha):=\min_{Q_{XY}:D(Q_{X}\|P_{X})=\alpha}D(Q_{XY}\|P_{XY})-\frac{D(Q_{Y}\|P_{Y})}{r}.
\]
By Theorem \ref{thm:Resolvability}, we obtain that for $q\ge1$,
\begin{align}
 & \lim_{n\to\infty}\frac{1}{n}\inf_{f:[e^{nR}]\to\mathcal{X}^{n}}D_{q}(Q_{Y^{n}}\|P_{Y}^{\otimes n})\nonumber \\
 & =\min_{\alpha\in[0,1]}\max\{D_{q}(\epsilon)-R,-q'(\varphi_{q'}(\alpha)-\alpha)\}\nonumber \\
 & =\max\{D_{q}(\epsilon)-R,\min_{\alpha\in[0,1]}-q'(\varphi_{q'}(\alpha)-\alpha)\}\nonumber \\
 & =[D_{q}(\epsilon\|1/2)-R]^{+},\label{eq:-32-1}
\end{align}
where  \eqref{eq:-32-1} follows since for any $Q_{X}$ such that
$D(Q_{X}\|P_{X})=\alpha$, it holds that 
\begin{align*}
\varphi_{q'}(\alpha) & \le D(Q_{X}P_{Y|X}\|P_{XY})-\frac{D(Q_{X}\circ P_{Y|X}\|P_{Y})}{q'}\\
 & \le D(Q_{X}\|P_{X})\le\alpha,
\end{align*}
and moreover, $\varphi_{q'}(0)=0$. 

Statement 2: For $0<q<1$, by the dual formula in Proposition \ref{prop:The-expressions-in},
\begin{align}
 & \lim_{n\to\infty}\frac{1}{n}\inf_{f:[e^{nR}]\to\mathcal{X}^{n}}D_{q}(Q_{Y^{n}}\|P_{Y}^{\otimes n})\nonumber \\
 & =\max_{\lambda\in[0,1]}-(\lambda-q')\log\min_{b\in[0,1]}\text{\ensuremath{\max}}\{g(t),g(\bar{t})\}+1-\lambda R\nonumber \\
 & =\max_{\lambda\in[0,1]}-(\lambda-q')\log g(1/2)+1-\lambda R\label{eq:-99}\\
 & =\max_{\lambda\in[0,1]}\lambda\left(1-R-H_{\frac{-q'}{\lambda-q'}}(\epsilon)\right),\nonumber 
\end{align}
where $g(t):=t^{\frac{\lambda-1}{\lambda-q'}}\epsilon^{\frac{-q'}{\lambda-q'}}+\bar{t}^{\frac{\lambda-1}{\lambda-q'}}\bar{\epsilon}^{\frac{-q'}{\lambda-q'}}$,
and \eqref{eq:-99} follows since $g$ is convex. 

Statement 3 follows similarly.  

\section{\label{sec:Proof-of-Theorem-strongqs}Proof of Theorem \ref{thm:strongqs}}

We only need to prove Statements 1 and 3 for finite $q$, since for
$q=\pm\infty$, Statements 1 and 3 follows by taking limits $q\to\infty$
or $q\to-\infty$.

Define the $q$-stability exponent as 
\begin{equation}
\Upsilon_{q}^{(n)}(\alpha):=\begin{cases}
-\frac{1}{n}\log\inf_{A:P_{X}^{\otimes n}(A)\le e^{-n\alpha}}\Vert P_{X|Y}^{\otimes n}(A|\cdot)\Vert_{q}, & q>1\\
-\frac{1}{n}\log\sup_{A:P_{X}^{\otimes n}(A)\ge e^{-n\alpha}}\Vert P_{X|Y}^{\otimes n}(A|\cdot)\Vert_{q}, & q<1
\end{cases}.\label{eq:qstabilityexponent}
\end{equation}
By considering product sets $A=A_{1}\times A_{2}$, it can be seen
that $n\Upsilon_{q}^{(n)}(\alpha)$ is superadditive in $n$ for $q>1$,
and subadditive in $n$ for $q<1$. So, by Fekete's lemma, 
\[
\Upsilon_{q}^{(\infty)}(\alpha):=\lim_{n\to\infty}\Upsilon_{q}^{(n)}(\alpha)=\begin{cases}
\sup_{n\ge1}\Upsilon_{q}^{(n)}(\alpha), & q>1\\
\inf_{n\ge1}\Upsilon_{q}^{(n)}(\alpha), & q<1
\end{cases},
\]
which means that to derive the dimension-free bounds, we only need
to focus on the asymptotic case.

\subsection{Case of $q>0$}

We use the method of types. Since the number of types is polynomial
in $n$, given any $n$ and $A:=A_{n}$, there is a type $T_{X}:=T_{X}^{(n)}$
such that 
\[
-\frac{1}{n}\log P_{X}^{\otimes n}(A)=-\frac{1}{n}\log P_{X}^{\otimes n}(A_{T_{X}})+o_{n}(1),
\]
where $A_{T_{X}}=A\cap\mathcal{T}_{T_{X}}$, and $o_{n}(1)$ is a
term vanishing as $n\to\infty$ uniformly for all sequences of $A_{n}$.

On the other hand, for $q>0$, 
\begin{align*}
 & \frac{1}{n}D_{q}(Q_{Y^{n}}\|P_{Y}^{\otimes n})\\
 & =\frac{1}{n(q-1)}\log\big[\sum_{y^{n}}(\sum_{x^{n}\in A}Q(x^{n})P(y^{n}|x^{n}))^{q}P(y^{n})^{1-q}\big]\\
 & \geq\frac{1}{n(q-1)}\log\big[\sum_{y^{n}}(\sum_{x^{n}\in A_{T_{X}}}Q(x^{n})P(y^{n}|x^{n}))^{q}P(y^{n})^{1-q}\big]\\
 & =\frac{1}{n(q-1)}\log\big[\sum_{y^{n}}(\sum_{x^{n}\in A_{T_{X}}}\frac{P(x^{n})}{P(A)}P(y^{n}|x^{n}))^{q}P(y^{n})^{1-q}\big]\\
 & \approx\frac{1}{n(q-1)}\log\big[\sum_{y^{n}}(\sum_{x^{n}\in A_{T_{X}}}\frac{P(x^{n})}{P(A_{T_{X}})}P(y^{n}|x^{n}))^{q}P(y^{n})^{1-q}\big]\\
 & =\frac{1}{n}D_{q}(P_{Y^{n}|X^{n}\in A_{T_{X}}}\|P_{Y}^{\otimes n})+q'o_{n}(1).
\end{align*}

Therefore, without changing the asymptotics of the minimum $q$-stability,
we can restrict $A$ to a subset of a type class. For $A\subseteq\mathcal{T}_{T_{X}}$,
it holds that $P_{X}^{\otimes n}(\cdot|A)$ is a uniform distribution,
and moreover, 
\[
\alpha\le-\frac{1}{n}\log P_{X}^{\otimes n}(A)=H(T_{X},P_{X})-R,
\]
where $R=\frac{1}{n}\log|A|$. Hence, the minimum $q$-stability problem
for $q>0$ is equivalent to the Rényi resolvability problem with the
coding function restricted to some $f:[e^{nR}]\to\mathcal{T}_{T_{X}}$
for some type $T_{X}$. By Theorem \ref{thm:Resolvability-3}, the
desired bounds follow.

\subsection{\label{subsec:Case-of}Case of $q<0$}

We divide the proof into two parts. We first prove the asymptotic
sharpness of the inequality in Statement 3, which is coined as the
``achievability part''. We then prove the inequality itself in Statement
3, which is coined as the ``converse part''.

Achievability: We first consider the case $q\in(-\infty,0)$. For
each $n$-type $T_{X}$, let $A_{T_{X}}:=\{X^{n}(m)\}$ be a set of
random sequences $X^{n}(m)$ that are drawn independently for different
$m$'s and according to the same distribution $\Unif(\mathcal{T}_{T_{X}})$,
such that
\begin{align*}
P_{X}^{\otimes n}(A_{T_{X}}) & =\begin{cases}
P_{X}^{\otimes n}(\mathcal{T}_{T_{X}}), & P_{X}^{\otimes n}(\mathcal{T}_{T_{X}})<e^{-n\alpha}\\
e^{-n(\alpha+o_{n}(1))}, & P_{X}^{\otimes n}(\mathcal{T}_{T_{X}})\ge e^{-n\alpha},H(T_{X},P_{X})\ge\alpha\\
0, & H(T_{X},P_{X})<\alpha
\end{cases}\\
 & =\begin{cases}
e^{-n(\max\{\alpha,D(T_{X}\|P_{X})\}+o_{n}(1))}, & H(T_{X},P_{X})\ge\alpha\\
0, & H(T_{X},P_{X})<\alpha
\end{cases}.
\end{align*}
Then, $P_{X}^{\otimes n}(A)=e^{-n(\alpha+o_{n}(1))}$, where $A:=\bigcup_{T_{X}}\mathcal{T}_{T_{X}}$.
 Roughly speaking, $A$ contains the whole type class $\mathcal{T}_{T_{X}}$
for all $T_{X}$ such that $D(T_{X}\|P_{X})>\alpha$, and contains
a random subset of $\mathcal{T}_{T_{X}}$ having probability $e^{-n(\alpha+o_{n}(1))}$
for all $T_{X}$ such that $D(T_{X}\|P_{X})\le\alpha$. 

Denote $Q_{X^{n}}:=P_{X}^{\otimes n}(\cdot|A)$. That is, 
\[
Q_{X^{n}}=\sum_{T_{X}}Q_{X^{n}}(\mathcal{T}_{T_{X}})\Unif(A_{T_{X}}),
\]
where $Q_{X^{n}}(\mathcal{T}_{T_{X}})=\frac{P_{X}^{\otimes n}(A_{T_{X}})}{P_{X}^{\otimes n}(A)}=e^{-n(\alpha_{T_{X}}+o_{n}(1))}$
with 
\begin{equation}
\alpha_{T_{X}}:=[D(T_{X}\|P_{X})-\alpha]^{+}.\label{eq:-103}
\end{equation}

Denote $R_{T_{X}}=\frac{1}{n}\log|A_{T_{X}}|$, which must satisfy
$0\le R_{T_{X}}\le H_{T}(X)$ for $T_{X}$ such that $|A_{T_{X}}|\ge1$,
i.e., $H(T_{X},P_{X})\ge\alpha$. Then, for such $T_{X}$, $-\frac{1}{n}\log P_{X}^{\otimes n}(A_{T_{X}})=H(T_{X},P_{X})-R_{T_{X}},$
and 
\begin{align*}
R_{T_{X}} & =H(T_{X},P_{X})-\max\{\alpha,D(T_{X}\|P_{X})\}+o_{n}(1)\\
 & =H_{T}(X)-[\alpha-D(T_{X}\|P_{X})]^{+}+o_{n}(1).
\end{align*}

Denote $s=-q$. Then, it holds that 
\begin{align*}
 & \e^{sD_{1+s}(P_{Y}^{\otimes n}\|Q_{Y^{n}})}\\
 & =\sum_{y^{n}}P^{1+s}(y^{n})(\sum_{x^{n}}Q(x^{n})P(y^{n}|x^{n}))^{-s}\\
 & \dotleq\sum_{T_{Y}}\sum_{y^{n}\in\mathcal{T}_{T_{Y}}}e^{(1+s)n\sum T_{Y}\log P_{Y}}(\sum_{T_{X|Y}:H(T_{X},P_{X})\ge\alpha}\sum_{x^{n}\in\mathcal{T}_{T_{X|Y}}(y^{n})\cap A_{T_{X}}}e^{-n(\alpha_{T_{X}}+R_{T_{X}})}e^{n\sum T_{XY}\log P_{Y|X}})^{-s}\\
 & \dotleq\max_{T_{Y}}\min_{T_{X|Y}:H(T_{X},P_{X})\ge\alpha}\sum_{y^{n}\in\mathcal{T}_{T_{Y}}}e^{sn(\alpha_{T_{X}}+R_{T_{X}})+(1+s)n\sum T_{Y}\log P_{Y}}e^{-sn\sum T_{XY}\log P_{Y|X}}|\mathcal{T}_{T_{X|Y}}(y^{n})\cap A_{T_{X}}|^{-s}\\
 & \dotleq\max_{T_{Y}}\min_{T_{X|Y}:H(T_{X},P_{X})\ge\alpha}e^{nH_{T}(Y)+sn(\alpha_{T_{X}}+R_{T_{X}})+(1+s)n\sum T_{Y}\log P_{Y}}\\
 & \qquad\qquad\times e^{-sn\sum T_{XY}\log P_{Y|X}}\begin{cases}
e^{-sn(R_{T_{X}}-I_{T}(X;Y))} & I_{T}(X;Y)\le R_{T_{X}}-\epsilon\\
\infty & I_{T}(X;Y)>R_{T_{X}}-\epsilon
\end{cases},
\end{align*}
where the last inequality follows by Lemma \ref{lem:B-1}.

Therefore, 
\begin{align*}
\frac{1}{n}D_{1+s}(P_{Y}^{\otimes n}\|Q_{Y^{n}}) & \apprle\max_{T_{Y}}\min_{T_{X|Y}:H(T_{X},P_{X})\ge\alpha,\,I_{T}(X;Y)\le R_{T_{X}}-\epsilon}D(T_{Y|X}\|P_{Y|X}|T_{X})-\frac{1+s}{s}D(T_{Y}\|P_{Y})+\alpha_{T_{X}}\\
 & =\max_{T_{Y}}\min_{\substack{T_{X|Y}:H(T_{X},P_{X})\ge\alpha,\\
I_{T}(X;Y)\le\min\{-\alpha-\sum T_{X}\log P_{X},H_{T}(X)\}-\epsilon
}
}D(T_{Y|X}\|P_{Y|X}|T_{X})-\frac{1}{q'}D(T_{Y}\|P_{Y})+[D(T_{X}\|P_{X})-\alpha]^{+}.
\end{align*}
Letting $n\to\infty$ and $\epsilon\downarrow0$ yields 
\begin{align*}
\limsup_{n\to\infty}\frac{1}{n}D_{1+s}(P_{Y}^{\otimes n}\|Q_{Y^{n}}) & \le\max_{Q_{Y}}\min_{\substack{Q_{X|Y}:I_{Q}(X;Y)\le\\
-\alpha-\sum Q_{X}\log P_{X}
}
}D(Q_{Y|X}\|P_{Y|X}|Q_{X})-\frac{1}{q'}D(Q_{Y}\|P_{Y})+[D(Q_{X}\|P_{X})-\alpha]^{+}.
\end{align*}
Combined with the relationship between $q$-stability and Rényi resolvability,
this implies the asymptotic sharpness of the inequality in Statement
3 for $q\in(-\infty,0)$.

The desired asymptotic sharpness for $q=-\infty$ follows in a similar
way.

Converse: We only consider the case $q\in(-\infty,0)$, since the
case $q=-\infty$ follows by taking limits. For this case, we follows
proof steps similar to those in Appendix \eqref{subsec:Lower-Bound-in}. 

Let $A$ be a subset of $\mathcal{X}^{n}$. We write $A=\bigcup_{T_{X}}A_{T_{X}}$,
where $A_{T_{X}}:=A\cap\mathcal{T}_{T_{X}}$. Denote $R_{T_{X}}=\frac{1}{n}\log|A_{T_{X}}|$
(which is equal to $-\infty$ if $|A_{T_{X}}|=0$). Then, for such
$T_{X}$, $-\frac{1}{n}\log P_{X}^{\otimes n}(A_{T_{X}})=H(T_{X},P_{X})-R_{T_{X}}.$
Since $P_{X}^{\otimes n}(A_{T_{X}})\le e^{-n\max\{\alpha,D(T_{X}\|P_{X})\}},$
it holds that 
\begin{align*}
R_{T_{X}} & \le H(T_{X},P_{X})-\max\{\alpha,D(T_{X}\|P_{X})\}\\
 & =H_{T}(X)-[\alpha-D(T_{X}\|P_{X})]^{+}.
\end{align*}

Denote $Q_{X^{n}}=P_{X}^{\otimes n}(\cdot|A),$ and it holds that
$Q(A_{T_{X}})=\frac{P_{X}^{\otimes n}(A_{T_{X}})}{P_{X}^{\otimes n}(A)}\le e^{n(\alpha-D(T_{X}\|P_{X}))}$
and $Q(A_{T_{X}})\le1,$ i.e., 
\[
Q(A_{T_{X}})\le e^{-n[D(T_{X}\|P_{X})-\alpha]^{+}}.
\]

Similarly to the definition in \eqref{eq:-6}, we define
\begin{align*}
B_{T_{Y}} & :=\bigcup_{T_{X|Y}:I_{T}(X;Y)>R_{T_{X}}+\epsilon}\bigcup_{x^{n}\in A_{T_{X}}}\mathcal{T}_{T_{Y|X}}(x^{n})\\
 & =\{y^{n}:B_{y^{n}}\neq\emptyset\},
\end{align*}
where 
\begin{align*}
B_{y^{n}} & :=\bigcup_{T_{X|Y}:I_{T}(X;Y)>R_{T_{X}}+\epsilon}\mathcal{T}_{T_{X|Y}}(y^{n})\cap A\\
 & =\{x^{n}\in A:\exists T_{X|Y},I_{T}(X;Y)>R_{T_{X}}+\epsilon,(x^{n},y^{n})\in\mathcal{T}_{T_{XY}}\}.
\end{align*}
By the argument below \eqref{eq:-6}, the set $B_{T_{Y}}$ is exponentially
smaller than $\mathcal{T}_{T_{Y}}$. The set $\mathcal{T}_{T_{Y}}\backslash B_{T_{Y}}$
contains only the sequences $y^{n}$ such that $B_{y^{n}}$ is empty.
Equivalently, for each $y^{n}\in\mathcal{T}_{T_{Y}}\backslash B_{T_{Y}}$,
all sequences $x^{n}$ in $A$ together with $y^{n}$ having joint
type $T_{XY}$ such that $I_{T}(X;Y)\le R_{T_{X}}+\epsilon$. 

Denote $s=-q$. Analogizing \eqref{eq:-104}-\eqref{eq:-105}, one
can obtain that 
\begin{align*}
\e^{sD_{1+s}(P_{Y}^{\otimes n}\|Q_{Y^{n}})} & \dotgeq\max_{T_{Y}}\min_{T_{X|Y}:I_{T}(X;Y)\le R_{T_{X}}+\epsilon}e^{snD(T_{Y|X}\|P_{Y|X}|T_{X})-(1+s)nD(T_{Y}\|P_{Y})}Q(A_{T_{X}})^{-s},
\end{align*}
Therefore, 
\begin{align*}
\frac{1}{n}D_{1+s}(P_{Y}^{\otimes n}\|Q_{Y^{n}}) & \apprge\max_{T_{Y}}\min_{T_{X|Y}:I_{T}(X;Y)\le R_{T_{X}}+\epsilon}D(T_{Y|X}\|P_{Y|X}|T_{X})-\frac{1+s}{s}D(T_{Y}\|P_{Y})-\frac{1}{n}\log Q(A_{T_{X}})\\
 & \ge\max_{T_{Y}}\min_{\substack{T_{X|Y}:I_{T}(X;Y)\le\\
H_{T}(X)-[\alpha-D(T_{X}\|P_{X})]^{+}+\epsilon
}
}D(T_{Y|X}\|P_{Y|X}|T_{X})-\frac{1}{q'}D(T_{Y}\|P_{Y})+[D(T_{X}\|P_{X})-\alpha]^{+}.
\end{align*}
The constraint in the last line is equivalent to $I_{T}(X;Y)\le\min\{H_{T}(X),H(T_{X},P_{X})-\alpha\}+\epsilon$,
i.e., $I_{T}(X;Y)\le H(T_{X},P_{X})-\alpha+\epsilon.$ Letting $n\to\infty$
and $\epsilon\downarrow0$ yields the desired bound. 

\section{\label{sec:Proof-of-Theorem-IT}Proof of Theorem \ref{thm:ITcharacterization}}

Before proving Theorem \ref{thm:ITcharacterization}, we first introduce
the following two lemmas. The OT divergence, admits the following
dual formula. Denote $C_{\mathrm{b}}(\mathcal{X})$ as the set of
continuous bounded real-valued functions on $\mathcal{X}$. 
\begin{lem}[Divergence-Inner Product Duality \cite{leonard2001minimization,leonard2013survey}]
\label{lem:OT-divergence} Let $\calX,\calY$ be Polish spaces. For
probability measures $Q_{X},Q_{Y}$ and a $\sigma$-finite nonnegative
measure $\mu_{XY}$, the infimum in the definition of $\D(Q_{X},Q_{Y}\|\mu_{XY})$
given in \eqref{eq:mathbbD} is attained\footnote{If the infimum is equal to $+\infty$, then $D(Q_{XY}\|\mu_{XY})=+\infty$
for all $Q_{XY}\in\Pi(Q_{X},Q_{Y})$}. Moreover, for $(\phi,\psi)\in C_{\mathrm{b}}(\mathcal{X})\times C_{\mathrm{b}}(\mathcal{Y})$
and $Q_{X},Q_{Y}$, it holds that 
\begin{align}
\D(Q_{X},Q_{Y}\|\mu_{XY}) & =\sup_{\phi,\psi}\int\phi\mathrm{d}Q_{X}+\int\psi\mathrm{d}Q_{Y}-\log\int e^{\phi+\psi}\mathrm{d}\mu_{XY},\label{eq:-127}\\
-\log\int e^{\phi+\psi}\mathrm{d}\mu_{XY} & =\!\inf_{Q_{X},Q_{Y}}\D(Q_{X},Q_{Y}\|\mu_{XY})-\int\phi\mathrm{d}Q_{X}-\int\psi\mathrm{d}Q_{Y},\label{eq:-54}
\end{align}
where the supremum above is taken over all functions $(\phi,\psi)\in C_{\mathrm{b}}(\mathcal{X})\times C_{\mathrm{b}}(\mathcal{Y})$
and the infimum above is taken over all probability measures $Q_{X},Q_{Y}$. 
\end{lem}
The duality in \eqref{eq:-127}, due to Léonard \cite{leonard2001minimization,leonard2013survey},
is a consequence of a general minimax theorem, and the duality in
\eqref{eq:-54} is a consequence of information projection \cite{csiszar1975divergence}.
This duality in fact implies the information-theoretic characterization
of BL (or hypercontractivity) inequalities and their reverse. 

By using the duality above, we obtain the following  duality between
the relative entropy and the ``$q$-norm''. 
\begin{lem}[Divergence-Norm Duality]
\label{lem:duality} Let $\calX,\calY$ be Polish spaces. For a probability
measure $Q_{X}$ and a $\sigma$-finite nonnegative measure $\mu_{XY}$,
it holds that 
\begin{align}
\sup_{\phi}-\log\|\mu_{X|Y}(e^{\phi})\|_{L^{q}(\mu_{Y})}+\int\phi\mathrm{d}Q_{X} & =\begin{cases}
\inf_{Q_{Y}}\kappa_{q}(Q_{X},Q_{Y}), & q>0\\
\sup_{Q_{Y}}\kappa_{q}(Q_{X},Q_{Y}), & q<0
\end{cases},\label{eq:-55}\\
-\log\|\mu_{X|Y}(e^{\phi})\|_{L^{q}(\mu_{Y})} & =\begin{cases}
\inf_{Q_{X},Q_{Y}}\kappa_{q}(Q_{X},Q_{Y})-\int\phi\mathrm{d}Q_{X}, & q>0\\
\sup_{Q_{Y}}\inf_{Q_{X}}\kappa_{q}(Q_{X},Q_{Y})-\int\phi\mathrm{d}Q_{X}, & q<0
\end{cases},\label{eq:-56}
\end{align}
where 
\begin{align}
\kappa_{q}(Q_{X},Q_{Y}) & :=\D(Q_{X},Q_{Y}\|\mu_{XY})-\frac{1}{q'}D(Q_{Y}\|\mu_{Y}).\label{eq:kappapq}
\end{align}
\end{lem}
\begin{IEEEproof}[Proof of Lemma \ref{lem:duality}]
We first prove \eqref{eq:-55}. For $q>1$, 
\begin{align}
 & \inf_{Q_{Y}}\D(Q_{X},Q_{Y}\|\mu_{XY})-\frac{1}{q'}D(Q_{Y}\|\mu_{Y})\nonumber \\
 & =\inf_{Q_{Y|X},S_{Y}}D(Q_{XY}\|\mu_{XY})+\frac{1}{q'}D(Q_{Y}\|S_{Y})-\frac{1}{q'}D(Q_{Y}\|\mu_{Y})\nonumber \\
 & =\inf_{R_{XY},S_{Y}}\sup_{\phi}D(R_{XY}\|\mu_{XY})+\frac{1}{q'}D(R_{Y}\|S_{Y})-\frac{1}{q'}D(R_{Y}\|\mu_{Y})+\int\phi\mathrm{d}Q_{X}-\int\phi\mathrm{d}R_{X}\nonumber \\
 & =\sup_{\phi}\inf_{S_{Y}}-\log\int(\frac{\d S_{Y}}{\d\mu_{Y}})^{\frac{1}{q'}}e^{\phi}\mathrm{d}\mu_{XY}+\int\phi\mathrm{d}Q_{X}\label{eq:-52}\\
 & =\sup_{\phi}\int\phi\mathrm{d}Q_{X}-\log\|\mu_{X|Y}(e^{\phi})\|_{L^{q}(\mu_{Y})},\nonumber 
\end{align}
where we swap $\inf_{R_{XY},S_{Y}}$ and $\sup_{\phi}$ in \eqref{eq:-52}
by using the minimax theorem (e.g., \cite[Theorem 2.10.2]{zalinescu2002convex})
since the objective function can be written as 
\[
\frac{1}{q}D(R_{XY}\|\mu_{XY})+\frac{1}{q'}D(R_{X|Y}\|\mu_{X|Y}|R_{Y})+\frac{1}{q'}D(R_{Y}\|S_{Y})+\int\phi\mathrm{d}Q_{X}-\int\phi\mathrm{d}R_{X}
\]
which is jointly convex in $(R_{XY},S_{Y})$ and linear in $\phi$. 

For $0<q<1$, by Theorem \ref{lem:OT-divergence}, we obtain that
\begin{align}
 & \inf_{Q_{Y}}\D(Q_{X},Q_{Y}\|\mu_{XY})-\frac{1}{q'}D(Q_{Y}\|\mu_{Y})\nonumber \\
 & =\inf_{Q_{Y}}\sup_{\phi,\psi}\int\phi\mathrm{d}Q_{X}+\int\psi\mathrm{d}Q_{Y}-\log\int e^{\phi+\psi}\mathrm{d}\mu_{XY}-\frac{1}{q'}D(Q_{Y}\|\mu_{Y})\nonumber \\
 & =\sup_{\phi,\psi}\int\phi\mathrm{d}Q_{X}-\log\int e^{\phi+\psi}\mathrm{d}\mu_{XY}+\frac{1}{q'}\log\int e^{q'\psi}\mathrm{d}\mu_{Y}\label{eq:-53}\\
 & =\sup_{\phi}\int\phi\mathrm{d}Q_{X}-\log\|\mu_{X|Y}(e^{\phi})\|_{L^{q}(\mu_{Y})},\nonumber 
\end{align}
where \eqref{eq:-53} follows by the minimax theorem. 

Similarly, for $q<0$, by Theorem \ref{lem:OT-divergence} again,
we obtain that 
\begin{align*}
 & \sup_{Q_{Y}}\D(Q_{X},Q_{Y}\|\mu_{XY})-\frac{1}{q'}D(Q_{Y}\|\mu_{Y})\\
 & =\sup_{Q_{Y}}\sup_{\phi,\psi}\int\phi\mathrm{d}Q_{X}+\int\psi\mathrm{d}Q_{Y}-\log\int e^{\phi+\psi}\mathrm{d}\mu_{XY}-\frac{1}{q'}D(Q_{Y}\|\mu_{Y})\\
 & =\sup_{\phi,\psi}\int\phi\mathrm{d}Q_{X}-\log\int e^{\phi+\psi}\mathrm{d}\mu_{XY}+\frac{1}{q'}\log\int e^{q'\psi}\mathrm{d}\mu_{Y}\\
 & =\sup_{\phi}\int\phi\mathrm{d}Q_{X}-\log\|\mu_{X|Y}(e^{\phi})\|_{L^{q}(\mu_{Y})}.
\end{align*}

The duality in \eqref{eq:-56} follows by the information projection
theory \cite{csiszar1975divergence} and also can be easily proven
by the nonnegativity of the relative entropy \cite{csiszar2003information}. 
\end{IEEEproof}
We now start to prove Theorem \ref{thm:ITcharacterization}. We may
assume, by homogeneity, that $\Vert f\Vert_{p}=1$. Then, without
loss of generality, we can write $f^{p}=\frac{\mathrm{d}Q_{X}}{\mathrm{d}P_{X}}$
for some probability measures $Q_{X}\ll P_{X}$. Moreover, we require
$f<\infty$. Hence, $Q_{X}\ll\gg P_{X}$ if $p<0$. By Lemma \ref{lem:duality},
\begin{align*}
-\log\Vert P_{X|Y}^{\otimes n}(f)\Vert_{q} & =\begin{cases}
\inf_{R_{XY}}\xi_{p,q}(R_{XY},Q_{X}), & q>0\\
\sup_{R_{Y}}\inf_{R_{X|Y}}\xi_{p,q}(R_{XY},Q_{X}), & q<0
\end{cases},
\end{align*}
where 
\begin{align*}
\xi_{p,q}(R_{XY},Q_{X}) & :=D(R_{XY}\|P_{XY})-\frac{1}{q'}D(R_{Y}\|P_{Y})\\
 & \qquad+\frac{1}{p}D(R_{X}\|Q_{X})-\frac{1}{p}D(R_{X}\|P_{X})\\
 & =D(R_{X|Y}\|P_{X|Y}|R_{Y})+\frac{1}{q}D(R_{Y}\|P_{Y})\\
 & \qquad-\frac{1}{p}\mathbb{E}_{R_{X}}[\log\frac{\mathrm{d}Q_{X}}{\mathrm{d}P_{X}}].
\end{align*}
Therefore, 
\begin{align*}
\underline{\Gamma}_{p,q} & =\inf_{Q_{X}}\begin{cases}
\inf_{R_{XY}}\xi_{p,q}(R_{XY},Q_{X}), & q>0\\
\sup_{R_{Y}}\inf_{R_{X|Y}}\xi_{p,q}(R_{XY},Q_{X}), & q<0
\end{cases},\\
\overline{\Gamma}_{p,q} & =\sup_{Q_{X}}\begin{cases}
\inf_{R_{XY}}\xi_{p,q}(R_{XY},Q_{X}), & q>0\\
\sup_{R_{Y}}\inf_{R_{X|Y}}\xi_{p,q}(R_{XY},Q_{X}), & q<0
\end{cases}.
\end{align*}

Proof of $\underline{\Gamma}_{p,q}$: We first consider $\underline{\Gamma}_{p,q}(X;Y)$.
We divide the proof into several parts.

Case of $q>0$: For the case $q>0$, swapping two infimizations, we
obtain 
\[
\underline{\Gamma}_{p,q}=\begin{cases}
\inf_{R_{X},R_{Y}}\theta_{p,q'}(R_{X},R_{Y}), & p,q>0\\
{\displaystyle -\infty} & p<0<q
\end{cases}.
\]

Case of $q<0<p$: Observe that for $q<0$, $\xi_{p,q}(R_{XY},Q_{X})$
is convex in $R_{X|Y}$, convex in $Q_{X}$, and concave in $R_{Y}$.
Define 
\[
\hat{\xi}_{p,q}(R_{Y},Q_{X}):=\inf_{R_{X|Y}}\xi_{p,q}(R_{XY},Q_{X}).
\]
To determine the convexity and concavity of $\hat{\xi}_{p,q}$, we
require the following two facts. 
\begin{fact}
The pointwise infimum of a set of concave functions is still concave.
Let $f:(x,y)\in A\times B\mapsto f(x,y)\in\mathbb{R}$ be a function
that is concave in $y$, where $B$ is a convex set. Define $g(y):=\inf_{x\in A}f(x,y)$.
Then, $g$ is concave. 
\end{fact}
\begin{fact}
Let $f:A\times B\to\mathbb{R}$ be a convex function, where $A,B$
are two convex sets. Define $g(y):=\inf_{x\in A}f(x,y)$. Then, $g$
is convex as well. 
\end{fact}
By Fact 1, $\hat{\xi}_{p,q}(R_{Y},Q_{X})$ is concave in $R_{Y}$.
By Fact 2, $\hat{\xi}_{p,q}(R_{Y},Q_{X})$ is convex in $Q_{X}$.
By the minimax theorem, 
\begin{align*}
\underline{\Gamma}_{p,q} & =\inf_{Q_{X}}\sup_{R_{Y}}\hat{\xi}_{p,q}(R_{Y},Q_{X})\\
 & =\sup_{R_{Y}}\inf_{Q_{X}}\hat{\xi}_{p,q}(R_{Y},Q_{X})\\
 & =\sup_{R_{Y}}\inf_{R_{X|Y}}\inf_{Q_{X}}\xi_{p,q}(R_{XY},Q_{X})\\
 & =\sup_{R_{Y}}\inf_{R_{X|Y}}D(R_{XY}\|P_{XY})-\frac{1}{q'}D(R_{Y}\|P_{Y})-\frac{1}{p}D(R_{X}\|P_{X}).
\end{align*}
In particular, when $p>1$, $\underline{\Gamma}_{p,q}=0.$ 

Case of $p<0,q<0$: If there is some set $A$ such that $0<P_{X}(A)<1$
and $P_{Y}\{y:P_{X|Y}(A|y)=1\}=0$, then $\underline{\Gamma}_{p,q}=-\infty.$ 

Proof of $\overline{\Gamma}_{p,q}$: We next consider $\overline{\Gamma}_{p,q}$.
We divide the proof into several parts.

Case of $q<1$: For $q<1$, $\overline{\Gamma}_{p,q}$ is just the
reverse BL exponent whose information-theoretic characterization was
already given in \cite{kamath2015reverse,beigi2016equivalent,liu2016brascamp}.
 This information-theoretic characterization is also implied by Lemmas
\ref{lem:OT-divergence} and \ref{lem:duality}.

Case of $q>1,p<0$: For the case $p<0<q$, $\xi_{p,q}(R_{XY},Q_{X})$
is convex in $R_{XY}$ and concave in $Q_{X}$. By the minimax theorem,
\begin{align*}
\overline{\Gamma}_{p,q} & =\inf_{R_{XY}}\sup_{Q_{X}}\xi_{p,q}(R_{XY},Q_{X})\\
 & =\inf_{R_{XY}}D(R_{XY}\|P_{XY})-\frac{1}{q'}D(R_{Y}\|P_{Y})-\frac{1}{p}D(R_{X}\|P_{X})\\
 & =0.
\end{align*}

Case of $q>1,0<p<1$: For $q>1$, $\overline{\Gamma}_{p,q}=\sup_{Q_{X}}\inf_{R_{XY}}\xi_{p,q}(R_{XY},Q_{X}).$
From this formula, we obtain that 
\begin{align*}
\overline{\Gamma}_{p,q} & \ge\inf_{R_{XY}}\xi_{p,q}(R_{XY},P_{X})\\
 & =\inf_{R_{XY}}D(R_{XY}\|P_{XY})-\frac{1}{q'}D(R_{Y}\|P_{Y})\\
 & \ge0.
\end{align*}
On the other hand, by setting $R_{X}$ to $Q_{X}$, we obtain 
\begin{align*}
\overline{\Gamma}_{p,q} & =\sup_{Q_{X}}\inf_{R_{XY}}\xi_{p,q}(R_{XY},Q_{X})\\
 & \le\sup_{R_{X}}\inf_{R_{Y|X}}D(R_{XY}\|P_{XY})-\frac{1}{q'}D(R_{Y}\|P_{Y})-\frac{1}{p}D(R_{X}\|P_{X})\\
 & =\sup_{R_{X}}\inf_{R_{Y}}\theta_{p,q'}(R_{X},R_{Y}).
\end{align*}
By further setting $R_{Y|X}$ to $P_{Y|X}$, we obtain 
\begin{align*}
\overline{\Gamma}_{p,q} & \le\sup_{R_{X}}D(R_{X}P_{Y|X}\|P_{XY})-\frac{1}{q'}D(R_{X}\circ P_{Y|X}\|P_{Y})-\frac{1}{p}D(R_{X}\|P_{X})\\
 & =\sup_{R_{X}}\frac{1}{p'}D(R_{X}\|P_{X})-\frac{1}{q'}D(R_{X}\circ P_{Y|X}\|P_{Y})\\
 & =0.
\end{align*}
So, $\overline{\Gamma}_{p,q}=0$.

Case of $p>1,q>1$: By setting $R_{X}$ to $Q_{X}$, we obtain 
\begin{align*}
\overline{\Gamma}_{p,q} & =\sup_{Q_{X}}\inf_{R_{XY}}\xi_{p,q}(R_{XY},Q_{X})\\
 & \le\gamma:=\sup_{R_{X}}\inf_{R_{Y|X}}D(R_{XY}\|P_{XY})-\frac{1}{q'}D(R_{Y}\|P_{Y})-\frac{1}{p}D(R_{X}\|P_{X})\\
 & =\sup_{R_{X}}\inf_{R_{Y}}\theta_{p,q'}(R_{X},R_{Y}).
\end{align*}

We next prove $\overline{\Gamma}_{p,q}\ge\gamma$. By Lemma \ref{lem:duality},
it holds that 
\begin{align}
 & \inf_{Q_{Y}}\kappa_{q}(Q_{X},Q_{Y})\nonumber \\
 & =\sup_{\phi}-\log\|\mu_{X|Y}(e^{\phi})\|_{L^{q}(\mu_{Y})}+\int\phi\mathrm{d}Q_{X}\label{eq:-58}\\
 & =\sup_{S_{X}}-\log\|\mu_{X|Y}((\frac{\d S_{X}}{\d\mu_{X}})^{\frac{1}{p}})\|_{L^{q}(\mu_{Y})}+\frac{1}{p}\int\log\frac{\d S_{X}}{\d\mu_{X}}\mathrm{d}Q_{X}\label{eq:-57}\\
 & =\sup_{S_{X}}-\log\|\mu_{X|Y}((\frac{\d S_{X}}{\d\mu_{X}})^{\frac{1}{p}})\|_{L^{q}(\mu_{Y})}+\frac{1}{p}D(Q_{X}\|\mu_{X})-\frac{1}{p}D(Q_{X}\|S_{X})\nonumber \\
 & \le\sup_{S_{X}}-\log\|\mu_{X|Y}((\frac{\d S_{X}}{\d\mu_{X}})^{\frac{1}{p}})\|_{L^{q}(\mu_{Y})}+\frac{1}{p}D(Q_{X}\|\mu_{X})\nonumber \\
 & =-\log\inf_{f\ge0}\frac{\|\mu_{X|Y}(f)\|_{L^{q}(\mu_{Y})}}{\|f\|_{L^{p}(\mu_{X})}}+\frac{1}{p}D(Q_{X}\|\mu_{X})\nonumber \\
 & =\overline{\Gamma}_{p,q}+\frac{1}{p}D(Q_{X}\|\mu_{X}),\nonumber 
\end{align}
where $\kappa_{q}$ is defined in \eqref{eq:kappapq}, and without
loss of generality, we set $\phi=\frac{1}{p}\log\frac{\d S_{X}}{\d\mu_{X}}$
in \eqref{eq:-57} since substituting $\phi\leftarrow\phi+a$ with
a constant $a$, the objective function in \eqref{eq:-58} remains
unchanged. Therefore, for any $Q_{X}$, it holds that 
\[
\overline{\Gamma}_{p,q}\ge\inf_{Q_{Y}}\kappa_{q}(Q_{X},Q_{Y})-\frac{1}{p}D(Q_{X}\|\mu_{X}).
\]
Taking supremum over $Q_{X}$ yields that 
\[
\overline{\Gamma}_{p,q}\ge\sup_{Q_{X}}\inf_{Q_{Y}}\kappa_{q}(Q_{X},Q_{Y})-\frac{1}{p}D(Q_{X}\|\mu_{X}).
\]

\section{\label{sec:Proof-of-Theorem}Proof of Theorem \ref{thm:anticontractivity}}

The last two clauses in \eqref{eq:RAH-1} and the last clause in \eqref{eq:FAH-1}
are obvious. So, we only prove the remaining cases in the following. 

\subsection{\label{subsec:Proof-of-}Proof of $\overline{\Gamma}_{p,q}^{(n)}$
for $p,q\ge1$}

Our proof combines Theorem \ref{thm:strongqs} with the layer representation
idea from \cite[Proof of Theorem 1.8]{kirshner2021moment}. Observe
that by the product construction, the optimal exponents $n\overline{\Gamma}_{p,q}^{(n)}$
is superadditive in $n$. So, by Fekete's lemma, $\overline{\Gamma}_{p,q}^{(\infty)}=\sup_{n\ge1}\overline{\Gamma}_{p,q}^{(n)}$,
which means that we only need to focus on the asymptotic case.

We may assume, by homogeneity, that $\left\Vert f\right\Vert _{p}=1$,
which means that $f^{p}\le1/P_{X,\min}^{n}$ with $P_{X,\min}:=\min_{x}P_{X}(x)$.
For sufficiently large $a>0$, the points at which $f^{p}<e^{-na}$
contributes little to $\|f\|_{p}$, and $\|P_{X|Y}^{\otimes n}(f)\|_{q}$,
in the sense that if we set $f$ to be zero at these points (the resulting
function denoted as $f_{a}$), then $\frac{1}{n}\log\|f\|_{p}$ and
$\frac{1}{n}\log\|P_{X|Y}^{\otimes n}(f)\|_{q}$ only change by amounts
of the order of $o_{n}(1)$, where $o_{n}(1)$ denotes a term vanishing
as $n\to\infty$ uniformly over all $f$ with $\|f\|_{p}=1$. This
is because, $f_{a}\le f\le f_{a}+e^{-na/p},$ which implies 
\begin{align*}
\|f_{a}\|_{p}^{p} & \le\|f\|_{p}^{p}\le2^{p-1}(\|f_{a}\|_{p}^{p}+e^{-na}),\\
\|P_{X|Y}^{\otimes n}(f_{a})\|_{q} & \le\|P_{X|Y}^{\otimes n}(f)\|_{q}.
\end{align*}

We next use a layer representation of the function $f$. The set $\{x^{n}:e^{-na}\le f^{p}(x^{n})\le1/P_{X,\min}^{n}\}$
can be partitioned into $m=m_{a,b}$ level sets $A_{1},...,A_{m}$
such that $f^{p}$ varies by a factor of at most $e^{nb}$ in each
level set, where $b>0$. Let $\alpha_{i}:=-\frac{1}{n}\log P_{X}^{\otimes n}(A_{i})$,
and let $\mu_{i}=\frac{1}{n}\log(u_{i})$, where $u_{i}$ is the median
value of $f^{p}$ on $A_{i}$. (If $A_{i}$ is empty then $u_{i}$
can be chosen to be any value within the level set defining $A_{i}$.)
Note that $f^{p}(x^{n})\in[u_{i}e^{-nb},u_{i}e^{nb}]$ on the set
$A_{i}$. Then, 
\begin{align*}
\frac{1}{n}\log\|f\|_{p} & =\frac{1}{n}\log\|f_{a}\|_{p}+o_{n}(1)\\
 & \le\frac{1}{np}\log[\sum_{i\in[m]}P_{X}^{\otimes n}(A_{i})u_{i}]+b+o_{n}(1)\\
 & =\frac{1}{np}\log[\max_{i\in[m]}P_{X}^{\otimes n}(A_{i})u_{i}]+b+o_{n}(1),
\end{align*}
where $o_{n}(1)$ tends to zero as $n\to\infty$ for large enough
but fixed $a$ and any fixed $b$. Given $n$, denote $i^{*}:=\arg\max_{i\in[m]}P_{X}^{\otimes n}(A_{i})u_{i}$.

On the other hand, 
\begin{align}
\frac{1}{n}\log\|P_{X|Y}^{\otimes n}(f)\|_{q} & \geq\frac{1}{n}\log\|P_{X|Y}^{\otimes n}(f_{a})\|_{q}\nonumber \\
 & \geq\frac{1}{nq}\log\mathbb{E}[(\sum_{i\in[m]}P_{X|Y}^{\otimes n}(A_{i}|Y^{n})u_{i}^{1/p})^{q}\cdot e^{-nqb}]\nonumber \\
 & \geq\frac{1}{nq}\log\mathbb{E}[(P_{X|Y}^{\otimes n}(A_{i^{*}}|Y^{n})u_{i^{*}}^{1/p})^{q}]-b\label{eq:-4-4}\\
 & =\frac{1}{nq}\log\mathbb{E}[P_{X|Y}^{\otimes n}(A_{i^{*}}|Y^{n})^{q}]+\frac{1}{np}\log u_{i^{*}}-b.
\end{align}
Therefore, 
\begin{align*}
\frac{1}{n}\log\frac{\|P_{X|Y}^{\otimes n}(f)\|_{q}}{\|f\|_{p}} & \geq\frac{1}{nq}\log\mathbb{E}[P_{X|Y}^{\otimes n}(A_{i^{*}}|Y^{n})^{q}]-\frac{1}{np}\log P_{X}^{\otimes n}(A_{i^{*}})-2b+o_{n}(1).
\end{align*}

Recall the $q$-stability exponent $\Upsilon_{q}^{(n)}(\alpha)$ defined
in \eqref{eq:qstabilityexponent}. Then, 
\begin{align*}
\overline{\Gamma}_{p,q}^{(n)} & \le\sup_{\alpha\ge0}\Upsilon_{q}^{(n)}(\alpha)-\frac{\alpha}{p}-2b+o_{n}(1).
\end{align*}
Letting $n\to\infty$ yields that 
\begin{align*}
\overline{\Gamma}_{p,q}^{(\infty)} & \le\sup_{n\ge1}\sup_{\alpha\ge0}\overline{\Upsilon}^{(n)}(\alpha)-\frac{\alpha}{p}-2b\\
 & =\sup_{\alpha\ge0}\sup_{n\ge1}\overline{\Upsilon}^{(n)}(\alpha)-\frac{\alpha}{p}-2b\\
 & =\sup_{\alpha\ge0}\Upsilon_{q}^{(\infty)}(\alpha)-\frac{\alpha}{p}-2b.
\end{align*}
Then, letting $b\downarrow0$ yields that $\overline{\Gamma}_{p,q}^{(\infty)}\le\sup_{\alpha\ge0}\Upsilon_{q}^{(\infty)}(\alpha)-\frac{\alpha}{p}.$
Substituting the bound on $\Upsilon_{q}^{(\infty)}(\alpha)$ in Theorem
\ref{thm:strongqs} for $q>1$ yields the following bound: 

\begin{align}
\overline{\Gamma}_{p,q}^{(\infty)} & \le\sup_{\alpha\ge0}\max_{Q_{X}:D(Q_{X}\|P_{X})\le\alpha\le H(Q_{X},P_{X})}\min_{Q_{Y|X}}\min\{\hat{\eta}_{q'}(Q_{XY},\alpha),\eta_{q'}(Q_{XY})\}+\frac{\alpha}{p'}.\label{eq:-85-1}
\end{align}

On the other hand, the asymptotic sharpness of this bound can be verified
easily by using the $q$-stability result in Theorem \ref{thm:strongqs}.
So, the equality in \eqref{eq:-85-1} holds. 

The expression at the RHS of \eqref{eq:-85-1} can be rewritten as
\begin{equation}
\max_{Q_{X}}\max_{\alpha:D(Q_{X}\|P_{X})\le\alpha\le H(Q_{X},P_{X})}\min_{Q_{Y|X}}\min\{\hat{\eta}_{q'}(Q_{XY},\alpha),\eta_{q'}(Q_{XY})\}+\frac{\alpha}{p'}.\label{eq:-86}
\end{equation}
We next further simplify this expression. 

If $p\ge q\ge1$, then $1/p'-1/q'\ge0$, which implies that $\min_{Q_{Y|X}}\min\{\hat{\eta}_{q'}(Q_{XY},\alpha),\eta_{q'}(Q_{XY})\}+\frac{\alpha}{p'}$
is increasing in $\alpha$. So, for this case, $\alpha=H(Q_{X},P_{X})$
is a maximizer in \eqref{eq:-86}, which implies that the expression
in \eqref{eq:-86} is equal to 
\begin{align}
 & \max_{Q_{X}}\min_{Q_{Y|X}}\min\{\hat{\eta}_{q'}(Q_{XY},H(Q_{X},P_{X})),\eta_{q'}(Q_{XY})\}+\frac{H(Q_{X},P_{X})}{p'}\nonumber \\
 & =\max_{Q_{X}}\min_{Q_{Y|X}}\min\{D(Q_{Y|X}\|P_{Y|X}|Q_{X})-\frac{1}{q'}D(Q_{Y|X}\|P_{Y}|Q_{X}),\nonumber \\
 & \qquad D(Q_{Y|X}\|P_{Y|X}|Q_{X})-\frac{1}{q'}D(Q_{Y}\|P_{Y})\}+\frac{H(Q_{X},P_{X})}{p'}\nonumber \\
 & =\max_{Q_{X}}\min_{Q_{Y|X}}D(Q_{Y|X}\|P_{Y|X}|Q_{X})-\frac{1}{q'}D(Q_{Y|X}\|P_{Y}|Q_{X})+\frac{H(Q_{X},P_{X})}{p'}\nonumber \\
 & =\max_{Q_{X}}\frac{H(Q_{X},P_{X})}{p'}-\frac{1}{q'}\mathbb{E}_{Q_{X}}[D_{q}(P_{Y|X}\|P_{Y})]\\
 & =\max_{x}\frac{1}{p'}\log\frac{1}{P_{X}(x)}-\frac{1}{q'}D_{q}(P_{Y|X=x}\|P_{Y}).\label{eq:-87}
\end{align}

If $q>p\ge1$, then $1/p'-1/q'<0$, which implies that $\min_{Q_{Y|X}}\hat{\eta}_{q'}(Q_{XY},\alpha)+\frac{\alpha}{p'}$
is decreasing in $\alpha$ and $\min_{Q_{Y|X}}\eta_{q'}(Q_{XY})+\frac{\alpha}{p'}$
is increasing in $\alpha$. Moreover, letting these two quantities
equal, we obtain the solution 
\begin{align}
\alpha^{*} & =H(Q_{X},P_{X})+\min_{Q_{Y|X}}\{q'D(Q_{Y|X}\|P_{Y|X}|Q_{X})-D(Q_{Y|X}\|P_{Y}|Q_{X})\}\nonumber \\
 & \qquad-q'\min_{Q_{Y|X}}\eta_{q'}(Q_{XY})\nonumber \\
 & =H(Q_{X},P_{X})-\mathbb{E}_{Q_{X}}[D_{q}(P_{Y|X}\|P_{Y})]-q'\min_{Q_{Y|X}}\eta_{q'}(Q_{XY}).\label{eq:-93}
\end{align}
It is easy to see that $D(Q_{X}\|P_{X})\le\alpha^{*}\le H(Q_{X},P_{X})$,
which is a feasible solution for the second maximization in \eqref{eq:-86}.
So, $\alpha=\alpha^{*}$ is the optimal solution for the second maximization
in \eqref{eq:-86}, which implies that the expression in \eqref{eq:-86}
is equal to 
\begin{align}
 & \max_{Q_{X}}\min\{\min_{Q_{Y|X}}\hat{\eta}_{q'}(Q_{XY},\alpha^{*}),\min_{Q_{Y|X}}\eta_{q'}(Q_{XY})\}+\frac{\alpha^{*}}{p'}\nonumber \\
 & =\max_{Q_{X}}\frac{\alpha^{*}}{p'}+\min_{Q_{Y|X}}\eta_{q'}(Q_{XY})\nonumber \\
 & =\max_{Q_{X}}\frac{1}{p'}H(Q_{X},P_{X})-\frac{1}{p'}\mathbb{E}_{Q_{X}}[D_{q}(P_{Y|X}\|P_{Y})]\nonumber \\
 & \qquad+(1-\frac{q'}{p'})\min_{Q_{Y|X}}\eta_{q'}(Q_{XY}).\label{eq:-88}
\end{align}

The expressions in \eqref{eq:-87} and \eqref{eq:-88} are the desired
ones. So, the proof for the case $p,q\ge1$ is completed. 

\subsection{Proof of $\underline{\Gamma}_{p,q}^{(n)}$ for $0<p,q<1$}

The proof for this case is the almost same as that for $p,q>1$. In
particular, we can show that 
\begin{align}
\underline{\Gamma}_{p,q}^{(\infty)} & =\inf_{\alpha\ge0}\min_{Q_{XY}:D(Q_{X}\|P_{X})\le\alpha\le H(Q_{X},P_{X})}\max\{\hat{\eta}_{q'}(Q_{XY},\alpha),\eta_{q'}(Q_{XY})\}+\frac{\alpha}{p'}\label{eq:-94}\\
 & =\min_{Q_{XY}}\inf_{D(Q_{X}\|P_{X})\le\alpha\le H(Q_{X},P_{X})}\max\{\hat{\eta}_{q'}(Q_{XY},\alpha),\eta_{q'}(Q_{XY})\}+\frac{\alpha}{p'}.
\end{align}
Similarity as \eqref{eq:-86}-\eqref{eq:-88}, one can observe that
for $0<p\le q<1$, the optimal $\alpha=H(Q_{X},P_{X})$, and for $0<q<p<1$,
the optimal 
\begin{equation}
\alpha=H(Q_{X},P_{X})-I_{Q}(X;Y).\label{eq:-95}
\end{equation}
We then obtain the expressions in the first two clauses in \eqref{eq:RAH-1}.
We omit proof details. 

\subsection{Proof of $\underline{\Gamma}_{p,q}^{(n)}$ for $q<0<p<1$}

We use a proof idea similar to that in Appendix \ref{subsec:Case-of}.

Achievability (Upper Bound on $\underline{\Gamma}_{p,q}^{(\infty)}$):
Recall that 
\begin{align*}
\underline{\Gamma}_{p,q}^{(n)} & =\inf_{Q_{X^{n}}}\frac{1}{np'}D_{p}(Q_{X^{n}}\|P_{X}^{\otimes n})+\frac{1}{n}D_{1-q}(P_{Y}^{\otimes n}\|Q_{Y^{n}}).
\end{align*}
Let $m\ge1$ be an integer which is assumed to be fixed. Let $R(\cdot):\calP(\calX)\times[m]\to[0,\infty)$
be a function such that 
\[
R(Q_{X},i):=R_{Q_{X},i}=\frac{i-1}{m}H_{Q}(X),\forall Q_{X},\forall i\in[m].
\]
Let $\alpha_{\max}:=-\log\min_{x}P(x).$ Let 
\begin{equation}
\alpha_{T_{X},i}:=\frac{1}{-p'}(\alpha_{\max}+R_{T_{X},i}+\sum T_{X}\log P_{X}).\label{eq:-59}
\end{equation}
So, $\alpha_{T_{X},i}\ge0,\forall Q_{X},\forall i\in[m]$, and $\alpha_{T_{X},i}=0$
for $i=1$ and $T_{X}=\delta_{x^{*}}$ with $x^{*}=\arg\min_{x}P(x)$.
Let $Q_{X^{n}}$ be a probability measure given by 
\[
Q_{X^{n}}=\frac{1}{\Lambda}\sum_{i\in[m]}\sum_{T_{X}}e^{-n\alpha_{T_{X},i}}\Unif(A_{T_{X},i}),
\]
where $\Lambda:=\sum_{i\in[m]}\sum_{T_{X}}e^{-n\alpha_{T_{X},i}}=e^{o(n)}$
and $A_{T_{X},i},i\in[m]$ are disjoint random subsets of $\mathcal{T}_{T_{X}}$
with $|A_{T_{X},i}|=e^{nR_{T_{X},i}}$. Here $A_{T_{X},i},i\in[m]$
can be generated in the following way. We first generate a random
codebook which consists of $\sum_{i\in[m]}|A_{T_{X},i}|$ codewords
chosen randomly and independently from $\mathcal{T}_{T_{X}}$. We
then assign the codewords in this codebook to the sets $A_{T_{X},i},i\in[m]$
via partitioning the indices of codewords into $m$ parts of sizes
$|A_{T_{X},i}|,i\in[m]$ in an arbitrary but fixed way. This procedure
is feasible when $n$ is sufficiently large, since $\sum_{i\in[m]}|A_{T_{X},i}|\le|\mathcal{T}_{T_{X}}|$
for sufficiently large $n$.

We then have that 
\begin{align*}
\frac{1}{n}D_{p}(Q_{X^{n}}\|P_{X}^{\otimes n}) & =\frac{1}{n}\frac{1}{p-1}\log\sum_{T_{X}}\sum_{i\in[m]}\sum_{x^{n}\in A_{T_{X},i}}\frac{e^{-pn(\alpha_{T_{X},i}+R_{T_{X},i})}}{e^{(p-1)n\sum_{x}T_{X}\log P_{X}}}\\
 & \approx\frac{1}{p-1}\max_{T_{X},i}R_{T_{X},i}-p(\alpha_{T_{X},i}+R_{T_{X},i})-(p-1)\sum T_{X}\log P_{X}\\
 & =\max_{T_{X},i}-p'\alpha_{T_{X},i}-R_{T_{X},i}-\sum T_{X}\log P_{X}\\
 & =\alpha_{\max},
\end{align*}
where we substitute the expression of $\alpha_{T_{X},i}$ in \eqref{eq:-59}
in the last line. 

Denote $s=-q>0$. Then, it holds that 
\begin{align*}
 & \e^{sD_{1+s}(P_{Y}^{\otimes n}\|Q_{Y^{n}})}\\
 & =\sum_{y^{n}}P^{1+s}(y^{n})(\sum_{x^{n}}Q(x^{n})P(y^{n}|x^{n}))^{-s}\\
 & =\sum_{T_{Y}}\sum_{y^{n}\in\mathcal{T}_{T_{Y}}}e^{(1+s)n\sum T_{Y}\log P_{Y}}(\sum_{i\in[m]}\sum_{T_{X|Y}}\sum_{x^{n}\in\mathcal{T}_{T_{X|Y}}(y^{n})\cap A_{T_{X},i}}e^{-n(\alpha_{T_{X},i}+R_{T_{X},i})}e^{n\sum T_{XY}\log P_{Y|X}})^{-s}\\
 & \dotleq\max_{T_{Y}}\min_{i\in[m]}\min_{T_{X|Y}}\sum_{y^{n}\in\mathcal{T}_{T_{Y}}}e^{(1+s)n\sum T_{Y}\log P_{Y}}(e^{-n(\alpha_{T_{X},i}+R_{T_{X},i})+n\sum T_{XY}\log P_{Y|X}}|\mathcal{T}_{T_{X|Y}}(y^{n})\cap A_{T_{X},i}|)^{-s}\\
 & \le\max_{T_{Y}}\min_{i\in[m]}\min_{T_{X|Y}}\sum_{y^{n}\in\mathcal{T}_{T_{Y}}}e^{sn(\alpha_{T_{X},i}+R_{T_{X},i}-\sum T_{XY}\log P_{Y|X})+(1+s)n\sum T_{Y}\log P_{Y}}\\
 & \qquad\times\begin{cases}
e^{-sn(R_{T_{X},i}-I_{T}(X;Y))} & I_{T}(X;Y)\le R_{T_{X},i}-\epsilon\\
\infty & I_{T}(X;Y)>R_{T_{X},i}-\epsilon
\end{cases}\\
 & \doteq\max_{T_{Y}}\min_{i\in[m]}\min_{T_{X|Y}:I_{T}(X;Y)\le R_{T_{X},i}-\epsilon}e^{nH_{T}(Y)+sn(\alpha_{T_{X},i}+I_{T}(X;Y)-\sum T_{XY}\log P_{Y|X})+(1+s)n\sum T_{Y}\log P_{Y}},
\end{align*}
where the inequality follows by Lemma \ref{lem:B-1}. That is, 
\begin{align}
\frac{1}{n}D_{1+s}(P_{Y}^{\otimes n}\|Q_{Y^{n}}) & \apprle\max_{T_{Y}}\min_{i\in[m]}\min_{T_{X|Y}:I_{T}(X;Y)\le R_{T_{X},i}-\epsilon}D(T_{Y|X}\|P_{Y|X}|T_{X})-\frac{1}{q'}D(T_{Y}\|P_{Y})+\alpha_{T_{X},i}\nonumber \\
 & =\max_{T_{Y}}\min_{i\in[m]}\min_{T_{X|Y}:I_{T}(X;Y)\le R_{T_{X},i}-\epsilon}D(T_{Y|X}\|P_{Y|X}|T_{X})-\frac{1}{q'}D(T_{Y}\|P_{Y})\nonumber \\
 & \qquad\qquad+\frac{1}{-p'}(\alpha_{\max}+R_{T_{X},i}+\sum T_{X}\log P_{X}).\label{eq:-61}
\end{align}

We claim that given any $T_{Y}$, the double minimizations in the
last line above is upper bounded by (in fact approximately equal to)
\begin{align}
 & \min_{T_{X|Y}:I_{T}(X;Y)\le(1-\frac{1}{m})H_{T}(X)-\epsilon}D(T_{Y|X}\|P_{Y|X}|T_{X})-\frac{1}{q'}D(T_{Y}\|P_{Y})\nonumber \\
 & \qquad\qquad+\frac{1}{-p'}(\alpha_{\max}+I_{T}(X;Y)+\frac{1}{m}H_{T}(X)+\epsilon+\sum T_{X}\log P_{X}).\label{eq:-60}
\end{align}

Given $T_{Y}$, let $T_{X|Y}^{*}$ be an optimal conditional distribution
attaining the minimum in \eqref{eq:-60}. It satisfies $I_{T_{Y}T_{X|Y}^{*}}(X;Y)\le R_{T_{X}^{*},m}-\epsilon$.
We can choose $i\in[m]$ in \eqref{eq:-61} such that $I_{T_{Y}T_{X|Y}^{*}}(X;Y)$
is sandwiched between $R_{T_{X}^{*},i-1}-\epsilon$ and $R_{T_{X}^{*},i}-\epsilon$
with $T_{X}^{*}:=T_{Y}\circ T_{X|Y}^{*}$. Such $i$ together with
$T_{X|Y}^{*}$ forms a feasible solution to \eqref{eq:-61}. Using
this feasible solution, we obtain the upper bound in \eqref{eq:-60}.
Hence, the claim follows. 

For a fixed $\epsilon$, choosing $m$ sufficiently large, we obtain
a simpler upper bound: 
\begin{align}
 & \min_{T_{X|Y}:H_{T}(X|Y)\ge2\epsilon}D(T_{Y|X}\|P_{Y|X}|T_{X})-\frac{1}{q'}D(T_{Y}\|P_{Y})\nonumber \\
 & \qquad\qquad+\frac{1}{-p'}(\alpha_{\max}+I_{T}(X;Y)+2\epsilon+\sum T_{X}\log P_{X}).\label{eq:-60-3}
\end{align}

Therefore, 
\begin{align*}
\overline{\Gamma}_{p,q}^{(n)} & \apprle\frac{1}{p'}\alpha_{\max}+\max_{T_{Y}}\min_{T_{X|Y}:H_{T}(X|Y)\ge2\epsilon}D(T_{Y|X}\|P_{Y|X}|T_{X})-\frac{1}{q'}D(T_{Y}\|P_{Y})\\
 & \qquad\qquad+\frac{1}{-p'}(\alpha_{\max}+I_{T}(X;Y)+\sum T_{X}\log P_{X}+2\epsilon)\\
 & =\max_{T_{Y}}\min_{T_{X|Y}:H_{T}(X|Y)\ge2\epsilon}D(T_{Y|X}\|P_{Y|X}|T_{X})-\frac{1}{q'}D(T_{Y}\|P_{Y})\\
 & \qquad\qquad+\frac{1}{-p'}(I_{T}(X;Y)+\sum T_{X}\log P_{X}+2\epsilon).
\end{align*}
Letting $n\to\infty$ yields 
\begin{align*}
\overline{\Gamma}_{p,q}^{(\infty)} & \le\max_{Q_{Y}}\min_{Q_{X|Y}:H_{Q}(X|Y)\ge2\epsilon}\chi(Q_{XY})+\frac{2\epsilon}{-p'},
\end{align*}
where 
\begin{align*}
\chi(Q_{XY}) & :=D(Q_{Y|X}\|P_{Y|X}|Q_{X})-\frac{1}{q'}D(Q_{Y}\|P_{Y})\\
 & \qquad\qquad+\frac{1}{-p'}(I_{Q}(X;Y)+\sum Q_{X}\log P_{X}).
\end{align*}

We now remove $\epsilon$ in the upper bound given above. Observe
that $Q_{X|Y}\mapsto\chi(Q_{XY})$ and $Q_{X|Y}\mapsto H_{Q}(X|Y)$
are respectively convex and concave. So, $\min_{Q_{X|Y}:H_{Q}(X|Y)\ge2\epsilon}\chi(Q_{XY})$
is nondecreasing and  convex in $\epsilon\ge0$. Since pointwise maximum
of a family of convex functions is convex, $\max_{Q_{Y}}\min_{Q_{X|Y}:H_{Q}(X|Y)\ge2\epsilon}\chi(Q_{XY})$
is nondecreasing and convex in $\epsilon\ge0$, which implies the
continuity of this function at $\epsilon=0$. Hence, we obtain the
desired bound, i.e., $\overline{\Gamma}_{p,q}^{(\infty)}\le\max_{Q_{Y}}\min_{Q_{X|Y}}\chi(Q_{XY}).$ 

Converse (Lower Bound on $\underline{\Gamma}_{p,q}^{(\infty)}$):
Here we combine the ideas in Appendices \ref{subsec:Proof-of-} and
\ref{subsec:Case-of}.   Let $f$ be a nonnegative function maximizing
$\|P_{X|Y}^{\otimes n}(f)\|_{q}/\left\Vert f\right\Vert _{p}$. We
may assume, by homogeneity, that $\left\Vert f\right\Vert _{p}=1$.
On the other hand, $\|P_{X|Y}^{\otimes n}(f)\|_{q}\ge1$ since this
lower bound corresponds to the $1$-valued constant function. The
condition $\left\Vert f\right\Vert _{p}=1$ means that $f^{p}\le P_{X,\min}^{-n}$
with $P_{X,\min}:=\min_{x}P_{X}(x)$, and moreover, there is a sequence
$x^{n}$ such that $f^{p}(x^{n})\ge1$. The condition $\|P_{X|Y}^{\otimes n}(f)\|_{q}\ge1$
means that $P_{X|Y}^{\otimes n}(f)^{q}\le P_{Y,\min}^{-n}$ with $P_{Y,\min}:=\min_{y}P_{Y}(y)$.
Here $P_{X,\min}>0,P_{Y,\min}>0$. 

For sufficiently large $a>0$, the points at which $f^{p}<e^{-na}$
contributes little to $\|f\|_{p}$, and $\|P_{X|Y}^{\otimes n}(f)\|_{q}$,
in the sense that if we set $f$ to be zero at these points (the resulting
function denoted as $f_{a}$), then $\frac{1}{n}\log\|f\|_{p}$ and
$\frac{1}{n}\log\|P_{X|Y}^{\otimes n}(f)\|_{q}$ only change by amounts
of the order of $o_{n}(1)$, where $o_{n}(1)$ denotes a term vanishing
as $n\to\infty$ uniformly over all $f$ with $\|f\|_{p}=1$. This
is because, $f_{a}\le f\le f_{a}+e^{-na/p},$ which implies 
\begin{align}
\|f_{a}\|_{p}^{p} & \le\|f\|_{p}^{p}\le\|f_{a}\|_{p}^{p}+e^{-na},\label{eq:-67}\\
P_{X|Y}^{\otimes n}(f_{a}) & \le P_{X|Y}^{\otimes n}(f)\le P_{X|Y}^{\otimes n}(f_{a})+e^{-na/p}.\label{eq:-66}
\end{align}
Since $P_{X|Y}^{\otimes n}(f)\ge P_{Y,\min}^{-n/q}$, we know that
$e^{-na/p}$ is exponentially smaller than $P_{X|Y}^{\otimes n}(f_{a})(y^{n})$
for every $y^{n}$, when $a$ is chosen such that $e^{-a/p}\le P_{Y,\min}^{-1/q}$.
Hence, \eqref{eq:-66} implies that 
\begin{align}
\|P_{X|Y}^{\otimes n}(f_{a})\|_{q} & \le\|P_{X|Y}^{\otimes n}(f)\|_{q}\le2\|P_{X|Y}^{\otimes n}(f_{a})\|_{q}.\label{eq:-68}
\end{align}
By \eqref{eq:-67} and \eqref{eq:-68}, the asymptotic exponents for
$\|P_{X|Y}^{\otimes n}(f)\|_{q}/\left\Vert f\right\Vert _{p}$ and
$\|P_{X|Y}^{\otimes n}(f_{a})\|_{q}/\left\Vert f_{a}\right\Vert _{p}$
remain the same. So, we only need to focus on nonnegative functions
$f$ taking values in $\{0\}\cup[e^{-na/p},P_{X,\min}^{-n/p}]$ 
for a sufficiently large but fixed $a$. 

We next use a layer representation. For a function $f^{p}$ taking
values in $\{0\}\cup[e^{-na},P_{X,\min}^{-n}]$, we partition its
support  into $m=m_{a,b}$ level sets $A_{1},...,A_{m}$ such that
$f^{p}$ varies by a factor of at most $e^{nb}$ in each level set,
where $b>0$. Denote $A=\bigcup_{i\in[m]}A_{i}$. Denote $A_{T_{X},i}:=A_{i}\cap\mathcal{T}_{T_{X}}$,
$R_{T_{X},i}:=\frac{1}{n}\log|A_{T_{X},i}|$, and $\alpha_{T_{X},i}:=-\frac{1}{n}\log P(A_{T_{X},i})$,
$i\in[m]$.  Denote 
\begin{align*}
B_{T_{Y},i} & :=\bigcup_{T_{X|Y}:I_{T}(X;Y)>R_{T_{X},i}+\epsilon}\bigcup_{x^{n}\in A_{T_{X},i}}\mathcal{T}_{T_{Y|X}}(x^{n})\\
 & =\{y^{n}:B_{y^{n},i}\neq\emptyset\},
\end{align*}
where 
\begin{align*}
B_{y^{n},i} & :=\bigcup_{T_{X|Y}:I_{T}(X;Y)>R_{T_{X}}+\epsilon}\mathcal{T}_{T_{X|Y}}(y^{n})\cap A_{i}\\
 & =\{x^{n}\in A_{i}:\exists T_{X|Y},I_{T}(X;Y)>R_{T_{X},i}+\epsilon,(x^{n},y^{n})\in\mathcal{T}_{T_{XY}}\}.
\end{align*}
Denote $B_{T_{Y}}=\bigcup_{i\in[m]}B_{T_{Y},i}$ and $B_{y^{n}}=\bigcup_{i\in[m]}B_{y^{n},i}$.

By the argument below \eqref{eq:-6}, the set $B_{T_{Y}}$ is exponentially
smaller than $\mathcal{T}_{T_{Y}}$. The set $\mathcal{T}_{T_{Y}}\backslash B_{T_{Y}}$
contains only the sequences $y^{n}$ such that $B_{y^{n}}$ is empty.
 In other words, for each $y^{n}\in\mathcal{T}_{T_{Y}}\backslash B_{T_{Y}}$,
all sequences $x^{n}$ in $A$ together with $y^{n}$ having joint
type $T_{XY}$ such that $I_{T}(X;Y)\le\max_{i\in[m]}R_{T_{X},i}+\epsilon$.

Let $Q_{X^{n}}$ be such that $\frac{Q_{X^{n}}}{P_{X}^{\otimes n}}\propto f$.
On one hand, 
\begin{align}
\frac{1}{n}D_{p}(Q_{X^{n}}\|P_{X}^{\otimes n}) & \le\frac{1}{n}\frac{1}{p-1}\log\sum_{i\in[m]}\sum_{T_{X}}\sum_{x^{n}\in A_{T_{X},i}}\frac{e^{-pn(\alpha_{T_{X},i}+R_{T_{X},i})}e^{-nb}}{e^{(p-1)n\sum_{x}T_{X}\log P_{X}}}\nonumber \\
 & \approx\frac{1}{p-1}\max_{T_{X},i}\{R_{T_{X},i}-p(\alpha_{T_{X},i}+R_{T_{X},i})-(p-1)\sum T_{X}\log P_{X}\}+\frac{b}{1-p}\nonumber \\
 & =\min_{i\in[m]}\min_{T_{X}}\{-p'\alpha_{T_{X},i}-R_{T_{X},i}-\sum T_{X}\log P_{X}\}+\frac{b}{1-p}.\label{eq:-69}
\end{align}

On the other hand, denoting $s=-q>0$ and analogizing \eqref{eq:-104}-\eqref{eq:-105},
one can obtain that 
\begin{align*}
 & \e^{sD_{1+s}(P_{Y}^{\otimes n}\|Q_{Y^{n}})}\\
 & \dotgeq\max_{T_{Y}}\min_{i\in[m]}\min_{T_{X|Y}:I_{T}(X;Y)\le R_{T_{X},i}+\epsilon}e^{snD(T_{Y|X}\|P_{Y|X}|T_{X})-(1+s)nD(T_{Y}\|P_{Y})}Q(A_{T_{X},i})^{-s},
\end{align*}
Therefore, 
\begin{align}
\frac{1}{n}D_{1-q}(P_{Y}^{\otimes n}\|Q_{Y^{n}}) & \ge\max_{T_{Y}}\min_{i\in[m]}\min_{T_{X|Y}:I_{T}(X;Y)\le R_{T_{X},i}+\epsilon}D(T_{Y|X}\|P_{Y|X}|T_{X})-\frac{1}{q'}D(T_{Y}\|P_{Y})+\alpha_{T_{X},i}.\label{eq:-70}
\end{align}

Combining \eqref{eq:-69} and \eqref{eq:-70}
\begin{align}
\underline{\Gamma}_{p,q}^{(n)} & \ge\frac{1}{np'}D_{p}(Q_{X^{n}}\|P_{X}^{\otimes n})+\frac{1}{n}D_{1-q}(P_{Y}^{\otimes n}\|Q_{Y^{n}})\nonumber \\
 & \apprge\max_{T_{Y}}\min_{i\in[m]}\min_{T_{X|Y}:I_{T}(X;Y)\le R_{T_{X},i}+\epsilon}\{D(T_{Y|X}\|P_{Y|X}|T_{X})-\frac{1}{q'}D(T_{Y}\|P_{Y})+\alpha_{T_{X},i}\}\nonumber \\
 & \qquad\qquad-\min_{i\in[m]}\min_{T_{X}}\{\alpha_{T_{X},i}+\frac{1}{p'}R_{T_{X},i}+\frac{1}{p'}\sum T_{X}\log P_{X}\}-\frac{b}{p}\nonumber \\
 & \ge\max_{T_{Y}}\min_{i\in[m]}\min_{T_{X|Y}:I_{T}(X;Y)\le R_{T_{X},i}+\epsilon}D(T_{Y|X}\|P_{Y|X}|T_{X})-\frac{1}{q'}D(T_{Y}\|P_{Y})-\frac{1}{p'}(R_{T_{X},i}+\sum T_{X}\log P_{X})-\frac{b}{p}\label{eq:-71}\\
 & \ge\max_{T_{Y}}\min_{T_{X|Y}}D(T_{Y|X}\|P_{Y|X}|T_{X})-\frac{1}{q'}D(T_{Y}\|P_{Y})-\frac{1}{p'}(I_{T}(X;Y)-\epsilon+\sum T_{X}\log P_{X})-\frac{b}{p}.\nonumber 
\end{align}
Invoking Lemma \ref{lem:minequality} and letting $\epsilon\downarrow0,b\downarrow0$
yields the desired bound. 

\section{\label{sec:Proof-of-Corollary}Proof of Corollary \ref{cor:binary}}

The first clause in \eqref{eq:FAH} and the first clause in \eqref{eq:RAH}
follow from Theorem \ref{thm:anticontractivity}. 

Using \eqref{eq:q0p1}, we have that for $q<0<p<1$, 
\begin{align*}
\underline{\Gamma}_{p,q}^{(\infty)} & =\min_{S_{X}}-\frac{1}{q}\log\sum_{y}P_{Y}(\sum_{x}P_{X|Y}^{p}\frac{S_{X}}{P_{X}})^{q/p}\\
 & =\min_{a\in[0,1]}\frac{1}{q}-\frac{1}{p}-\frac{1}{q}\log[(\epsilon^{p}a+(1-\epsilon)^{p}(1-a))^{q/p}+((1-\epsilon)^{p}a+\epsilon^{p}(1-a))^{q/p}]\\
 & =-\frac{1}{q}\log[(\epsilon^{p}+(1-\epsilon)^{p})^{q/p}]\\
 & =-\frac{1}{p}\log[\epsilon^{p}+(1-\epsilon)^{p}]\\
 & =\frac{1}{p'}H_{p}(\epsilon).
\end{align*}
Similarly, using \eqref{eq:q0p1} again, it can be proven that for
$0<q<p<1$, $\underline{\Gamma}_{p,q}^{(\infty)}=\frac{1}{p'}H_{p}(\epsilon).$
This completes the proof of the second clause in \eqref{eq:RAH}. 

We lastly prove the second clause in \eqref{eq:FAH}. Using \eqref{eq:-85}
and denoting $S_{Y}=\Bern(b)$, we have that for $1\le p<q$, 
\begin{align*}
\overline{\Gamma}_{p,q}^{(\infty)} & =\min_{S_{Y}}\max_{x}\frac{1}{p'}\log\frac{1}{P_{X}(x)}-\frac{1}{p'}D_{q}(P_{Y|X=x}\|P_{Y})\\
 & \qquad-(1-\frac{q'}{p'})\log\mathbb{E}_{P_{Y|X=x}}[(\frac{S_{Y}}{P_{Y}})^{1/q'}]\\
 & =\frac{1}{p'}-\frac{1}{p'}(1-H_{q}(\epsilon))\\
 & \qquad-(1-\frac{q'}{p'})\log\max_{b\in[0,1]}\min\{(1-\epsilon)(2(1-b))^{1/q'}+\epsilon(2b)^{1/q'},\\
 & \qquad\qquad\epsilon(2(1-b))^{1/q'}+(1-\epsilon)(2b)^{1/q'}\}\\
 & =\frac{1}{p'}H_{q}(\epsilon)-(1-\frac{q'}{p'})\log\max_{b\in[0,1/2]}\{\epsilon(2(1-b))^{1/q'}+(1-\epsilon)(2b)^{1/q'}\}\\
 & =\frac{1}{p'}H_{q}(\epsilon),
\end{align*}
where the optimal $b$ is $1/2$. 

The above evaluations of the second clause in \eqref{eq:RAH} and
the second clause in \eqref{eq:FAH} can be also done by using the
primal expressions in Theorem \ref{thm:anticontractivity}.  The
optimality of the random codes given in Corollary \ref{cor:binary}
follows by invoking the connections between Rényi resolvability, noise
stability, and anti-contractivity (especially \eqref{eq:-82} and
\eqref{eq:Stability-Resolvability}) and Remark \ref{rem:iid}. 

\section*{Acknowledgements}

The author would like to thank Professor Alexander Barg for pointing
out some details in \cite{pathegama2023smoothing}.

 \bibliographystyle{unsrt}
\bibliography{ref}

\begin{thebibliography}{10}

\bibitem{Han93}
T.~S. Han and S.~Verd\'{u}.
\newblock Approximation theory of output statistics.
\newblock {\em IEEE Transactions on Information Theory}, 39(3):752--772, 1993.

\bibitem{Hayashi06}
M.~Hayashi.
\newblock General nonasymptotic and asymptotic formulas in channel
  resolvability and identification capacity and their application to the
  wiretap channel.
\newblock {\em IEEE Transactions on Information Theory}, 52(4):1562--1575,
  2006.

\bibitem{Hayashi11}
M.~Hayashi.
\newblock Exponential decreasing rate of leaked information in universal random
  privacy amplification.
\newblock {\em IEEE Transactions on Information Theory}, 57(6):3989--4001,
  2011.

\bibitem{yu2019renyi}
L.~Yu and V.~Y.~F. Tan.
\newblock R{\'e}nyi resolvability and its applications to the wiretap channel.
\newblock {\em IEEE Transactions on Information Theory}, 65(3):1862--1897,
  2019.

\bibitem{samorodnitsky2022some}
A.~Samorodnitsky.
\newblock On some properties of random and pseudorandom codes.
\newblock {\em arXiv preprint arXiv:2206.05135}, 2022.

\bibitem{pathegama2023smoothing}
M.~Pathegama and A.~Barg.
\newblock Smoothing of binary codes, uniform distributions, and applications.
\newblock {\em Entropy}, 25(11):1515, 2023.

\bibitem{Liu}
J.~Liu, P.~Cuff, and S.~Verd{\'u}.
\newblock {$E_{\gamma}$}-resolvability.
\newblock {\em IEEE Transactions on Information Theory}, 63(5):2629--2658,
  2017.

\bibitem{yu2019simulation}
L.~Yu and V.~Y.~F. Tan.
\newblock Simulation of random variables under {R\'enyi} divergence measures of
  all orders.
\newblock {\em IEEE Transactions on Information Theory}, 65(6):3349--3383, Jun
  2019.

\bibitem{yu2018asymptotic}
L.~Yu and V.~Y.~F. Tan.
\newblock Asymptotic coupling and its applications in information theory.
\newblock {\em IEEE Transactions on Information Theory}, 65(3):1321--1344,
  2018.

\bibitem{WynerCI}
A.~D. Wyner.
\newblock The common information of two dependent random variables.
\newblock {\em IEEE Transactions on Information Theory}, 21(2):163--179, Mar
  1975.

\bibitem{YuTan2018}
L.~Yu and V.~Y.~F. Tan.
\newblock Wyner's common information under {R}\'enyi divergence measures.
\newblock {\em IEEE Transactions on Information Theory}, 64(5):3616--3623, May
  2018.

\bibitem{yu2020corrections}
L.~Yu and V.~Y.~F. Tan.
\newblock {Corrections to ``Wyner's common information under R{\'e}nyi
  divergence measures''}.
\newblock {\em IEEE Transactions on Information Theory}, 66(4):2599--2608,
  2020.

\bibitem{yagli2019exact}
S.~Yagli and P.~Cuff.
\newblock Exact exponent for soft covering.
\newblock {\em IEEE Transactions on Information Theory}, 65(10):6234--6262,
  2019.

\bibitem{Parizi}
M.~B. Parizi, E.~Telatar, and N.~Merhav.
\newblock Exact random coding secrecy exponents for the wiretap channel.
\newblock {\em IEEE Transactions on Information Theory}, 63(1):509--531, 2017.

\bibitem{yu2021strong}
L.~Yu.
\newblock Strong {Brascamp--Lieb} inequalities.
\newblock {\em ArXiv e-prints, arXiv:2102.06935}, 2021.

\bibitem{gacs1973common}
P.~G{\'a}cs and J.~K{\"o}rner.
\newblock Common information is far less than mutual information.
\newblock {\em Problems of Control and Information Theory}, 2(2):149--162,
  1973.

\bibitem{witsenhausen1975sequences}
Hans~S Witsenhausen.
\newblock On sequences of pairs of dependent random variables.
\newblock {\em SIAM Journal on Applied Mathematics}, 28(1):100--113, 1975.

\bibitem{yu2021non}
L.~Yu and V.~Y.~F. Tan.
\newblock On non-interactive simulation of binary random variables.
\newblock {\em IEEE Transactions on Information Theory}, 67(4):2528--2538,
  2021.

\bibitem{kahn1988influence}
J.~Kahn, G.~Kalai, and N.~Linial.
\newblock The influence of variables on {Boolean} functions.
\newblock In {\em IEEE Symposium on Foundations of Computer Science (FOCS)},
  pages 68--80, 1988.

\bibitem{mossel2006non}
E.~Mossel, R.~O'Donnell, O.~Regev, J.~E. Steif, and B.~Sudakov.
\newblock Non-interactive correlation distillation, inhomogeneous {Markov}
  chains, and the reverse {Bonami-Beckner} inequality.
\newblock {\em Israel Journal of Mathematics}, 154(1):299--336, 2006.

\bibitem{ODonnell14analysisof}
R.~O'Donnell.
\newblock {\em Analysis of {Boolean} Functions}.
\newblock Cambridge University Press, 2014.

\bibitem{kamath2016non}
S.~Kamath and V.~Anantharam.
\newblock On non-interactive simulation of joint distributions.
\newblock {\em IEEE Transactions on Information Theory}, 62(6):3419--3435,
  2016.

\bibitem{ordentlich2020note}
O.~Ordentlich, Y.~Polyanskiy, and O.~Shayevitz.
\newblock A note on the probability of rectangles for correlated binary
  strings.
\newblock {\em IEEE Transactions on Information Theory}, 66(11):7878--7886,
  2020.

\bibitem{kirshner2021moment}
N.~Kirshner and A.~Samorodnitsky.
\newblock A moment ratio bound for polynomials and some extremal properties of
  {Krawchouk} polynomials and {Hamming} spheres.
\newblock {\em IEEE Transactions on Information Theory}, 67(6):3509--3541,
  2021.

\bibitem{borell1985geometric}
C.~Borell.
\newblock Geometric bounds on the {Ornstein--Uhlenbeck} velocity process.
\newblock {\em Probability Theory and Related Fields}, 70(1):1--13, 1985.

\bibitem{kindler2015remarks}
G.~Kindler, R.~O'Donnell, and D.~Witmer.
\newblock Remarks on the most informative function conjecture at fixed mean.
\newblock {\em ArXiv e-prints, arXiv:1506.03167}, 2015.

\bibitem{mossel2005coin}
E.~Mossel and R.~O'Donnell.
\newblock Coin flipping from a cosmic source: {On} error correction of truly
  random bits.
\newblock {\em Random Structures \& Algorithms}, 26(4):418--436, 2005.

\bibitem{courtade2014boolean}
T.~A. Courtade and G.~R. Kumar.
\newblock Which {Boolean} functions maximize mutual information on noisy
  inputs?
\newblock {\em IEEE Transactions on Information Theory}, 60(8):4515--4525,
  2014.

\bibitem{li2020boolean}
J.~Li and M.~M{\'e}dard.
\newblock Boolean functions: noise stability, non-interactive correlation
  distillation, and mutual information.
\newblock {\em IEEE Transactions on Information Theory}, 67(2):778--789, 2020.

\bibitem{yu2023phi}
L.~Yu.
\newblock On the {$\Phi$}-stability and related conjectures.
\newblock {\em Probability Theory and Related Fields}, 186:1045--1080, 2023.

\bibitem{samorodnitsky2016entropy}
A.~Samorodnitsky.
\newblock On the entropy of a noisy function.
\newblock {\em IEEE Transactions on Information Theory}, 62(10):5446--5464,
  2016.

\bibitem{yu2022common}
L.~Yu and V.~Y.~F. Tan.
\newblock Common information, noise stability, and their extensions.
\newblock {\em Foundations and Trends in Communications and Information
  Theory}, 19(2):107--389, 2022.

\bibitem{bonami1968ensembles}
A.~Bonami.
\newblock Ensembles {$\Lambda (p)$} dans le dual de {$D^{\infty}$}.
\newblock In {\em Annales de l'institut Fourier}, volume~18, pages 193--204,
  1968.

\bibitem{kiener1969uber}
K.~Kiener.
\newblock {\em Uber Produkte von quadratisch integrierbaren Funktionen
  endlicher Vielfalt}.
\newblock PhD thesis, PhD thesis, Dissertation, Universit{\"a}t Innsbruck,
  1969.

\bibitem{schreiber1969fermeture}
M.~Schreiber.
\newblock Fermeture en probabilit{\'e} de certains sous-espaces d'un espace
  {$L^2$}.
\newblock {\em Zeitschrift f{\"u}r Wahrscheinlichkeitstheorie und Verwandte
  Gebiete}, 14(1):36--48, 1969.

\bibitem{bonami1970etude}
A.~Bonami.
\newblock {\'E}tude des coefficients de {Fourier} des fonctions de {$ L^{p}(G)
  $}.
\newblock In {\em Annales de l'institut Fourier}, volume~20, pages 335--402,
  1970.

\bibitem{beckner1975inequalities}
W.~Beckner.
\newblock Inequalities in {F}ourier analysis.
\newblock {\em Annals of Mathematics}, pages 159--182, 1975.

\bibitem{gross1975logarithmic}
L.~Gross.
\newblock Logarithmic {Sobolev} inequalities.
\newblock {\em American Journal of Mathematics}, 97(4):1061--1083, 1975.

\bibitem{ahlswede1976spreading}
R.~Ahlswede and P.~G{\'a}cs.
\newblock Spreading of sets in product spaces and hypercontraction of the
  {Markov} operator.
\newblock {\em Annals of Probability}, pages 925--939, 1976.

\bibitem{borell1982positivity}
C.~Borell.
\newblock Positivity improving operators and hypercontractivity.
\newblock {\em Mathematische Zeitschrift}, 180(3):225--234, 1982.

\bibitem{bakry1994hypercontractivite}
D.~Bakry.
\newblock L'hypercontractivit{\'e} et son utilisation en th{\'e}orie des
  semigroupes.
\newblock In {\em Lectures on Probability Theory}, pages 1--114. Springer,
  1994.

\bibitem{mossel2013reverse}
E.~Mossel, K.~Oleszkiewicz, and A.~Sen.
\newblock On reverse hypercontractivity.
\newblock {\em Geometric and Functional Analysis}, 23(3):1062--1097, 2013.

\bibitem{carlen2009subadditivity}
E.~A. Carlen and D.~Cordero-Erausquin.
\newblock Subadditivity of the entropy and its relation to {Brascamp--Lieb}
  type inequalities.
\newblock {\em Geometric and Functional Analysis}, 19(2):373--405, 2009.

\bibitem{nair2014equivalent}
C.~Nair.
\newblock Equivalent formulations of hypercontractivity using information
  measures.
\newblock In {\em International Zurich Seminar (IZS) Workshop}, 2014.

\bibitem{kamath2015reverse}
S.~Kamath.
\newblock Reverse hypercontractivity using information measures.
\newblock In {\em Allerton Conference on Communication, Control, and
  Computing}, pages 627--633, Monticello, Illinois, USA, 2015.

\bibitem{beigi2016equivalent}
S.~Beigi and C.~Nair.
\newblock Equivalent characterization of reverse {Brascamp--Lieb-type}
  inequalities using information measures.
\newblock In {\em IEEE International Symposium on Information Theory (ISIT)},
  pages 1038--1042, 2016.

\bibitem{liu2016brascamp}
J.~Liu, T.~A. Courtade, P.~Cuff, and S.~Verd{\'u}.
\newblock {Brascamp-Lieb} inequality and its reverse: An information theoretic
  view.
\newblock In {\em 2016 IEEE International Symposium on Information Theory
  (ISIT)}, pages 1048--1052. IEEE, 2016.

\bibitem{liu2018forward}
J.~Liu, T.~A. Courtade, P.~W. Cuff, and S.~Verd{\'u}.
\newblock A forward-reverse brascamp-lieb inequality: Entropic duality and
  gaussian optimality.
\newblock {\em Entropy}, 20(6):418, 2018.

\bibitem{Csi97}
I.~Csisz\'{a}r and J.~{K\"{o}rner}.
\newblock {\em Information Theory: Coding Theorems for Discrete Memoryless
  Systems}.
\newblock Cambridge University Press, 2011.

\bibitem{Erven}
T.~{van Erven} and P.~Harremo\"es.
\newblock {R{\'e}nyi} divergence and {Kullback-Leibler} divergence.
\newblock {\em IEEE Transactions on Information Theory}, 60(7):3797--3820,
  2014.

\bibitem{cuff13}
P.~Cuff.
\newblock Distributed channel synthesis.
\newblock {\em IEEE Transactions on Information Theory}, 59(11):7071--7096,
  2013.

\bibitem{li2021boolean}
J.~Li and M.~M{\'e}dard.
\newblock Boolean functions: noise stability, non-interactive correlation
  distillation, and mutual information.
\newblock {\em IEEE Transactions on Information Theory}, 67(2):778--789, 2021.

\bibitem{raginsky2013logarithmic}
M.~Raginsky.
\newblock Logarithmic {Sobolev} inequalities and strong data processing
  theorems for discrete channels.
\newblock In {\em IEEE International Symposium on Information Theory (ISIT)},
  pages 419--423, 2013.

\bibitem{Dembo}
A.~Dembo and O.~Zeitouni.
\newblock {\em Large Deviations Techniques and Applications}.
\newblock Springer, 2nd edition, 1998.

\bibitem{gallagerIT}
R.~G. Gallager.
\newblock {\em {Information Theory and Reliable Communication}}.
\newblock Wiley, New York, 1968.

\bibitem{Tan11_IT}
V.~Y.~F. Tan, A.~Anandkumar, L.~Tong, and A.~S. Willsky.
\newblock A large-deviation analysis for the maximum likelihood learning of
  {M}arkov tree structures.
\newblock {\em IEEE Transactions on Information Theory}, 57(3):1714--35, Mar
  2011.

\bibitem{YuTan2020b}
L.~Yu and V.~Y.~F. Tan.
\newblock Exact channel synthesis.
\newblock {\em IEEE Transactions on Information Theory}, 66(5):2299--2818, May
  2020.

\bibitem{schieler2016henchman}
C.~Schieler and P.~Cuff.
\newblock The henchman problem: Measuring secrecy by the minimum distortion in
  a list.
\newblock {\em IEEE Transactions on Information Theory}, 62(6):3436--3450,
  2016.

\bibitem{elgamal}
A.~{El~Gamal} and Y.-H. Kim.
\newblock {\em Network Information Theory}.
\newblock Cambridge University Press, Cambridge, U.K., 2012.

\bibitem{Mitzenmacher}
M.~Mitzenmacher and E.~Upfal.
\newblock {\em Probability and Computing: Randomized Algorithms and
  Probabilistic Analysis}.
\newblock Cambridge university press, 2005.

\bibitem{boucheron2013concentration}
S.~Boucheron, G.~Lugosi, and P.~Massart.
\newblock {\em Concentration Inequalities: A Nonasymptotic Theory of
  Independence}.
\newblock Oxford University Press, 2013.

\bibitem{kavian2023statistics}
M.~Kavian, M.~M. Mojahedian, M.~H. Yassaee, M.~Mirmohseni, and M.~R. Aref.
\newblock Statistics of random binning based on {Tsallis} divergence.
\newblock {\em arXiv preprint arXiv:2304.12606}, 2023.

\bibitem{leonard2001minimization}
C.~L{\'e}onard.
\newblock Minimization of energy functionals applied to some inverse problems.
\newblock {\em Applied mathematics and optimization}, 44(3):273--297, 2001.

\bibitem{leonard2013survey}
C.~L{\'e}onard.
\newblock A survey of the {Schr\"odinger} problem and some of its connections
  with optimal transport.
\newblock {\em arXiv preprint arXiv:1308.0215}, 2013.

\bibitem{csiszar1975divergence}
I.~Csisz{\'a}r.
\newblock I-divergence geometry of probability distributions and minimization
  problems.
\newblock {\em The annals of probability}, pages 146--158, 1975.

\bibitem{zalinescu2002convex}
C.~Zalinescu.
\newblock {\em Convex analysis in general vector spaces}.
\newblock World scientific, 2002.

\bibitem{csiszar2003information}
I.~Csisz{\'a}r and F.~Matus.
\newblock Information projections revisited.
\newblock {\em IEEE Transactions on Information Theory}, 49(6):1474--1490,
  2003.

\end{thebibliography}

%\begin{IEEEbiographynophoto}{Lei Yu}
%(Member, IEEE) received the B.E. and Ph.D. degrees in electronic
%engineering from the University of Science and Technology of China
%(USTC) in 2010 and 2015, respectively. From 2015 to 2020, he worked
%as a Post-Doctoral Researcher at the USTC, National University of
%Singapore, and University of California at Berkeley. He is currently
%an Associate Professor at the School of Statistics and Data Science,
%LPMC, KLMDASR, and LEBPS, Nankai University, China. His research interests
%lie in the intersection of probability theory, information theory,
%and combinatorics. 
%\end{IEEEbiographynophoto}
\end{document}